\documentclass[preprint,10pt,columnsep=15pt, frontmatter]{elsarticle}

\usepackage[english]{babel}
\usepackage[utf8]{inputenc}
\usepackage{float}

\usepackage{lineno}
\usepackage{overpic}
 \usepackage{transparent}
\usepackage{amsthm}
\usepackage{mathtools}
\usepackage{physics}
\usepackage{xcolor}
\usepackage{graphicx}
\usepackage[left=23mm,right=13mm,top=35mm,columnsep=15pt]{geometry} 
\usepackage{adjustbox}
\usepackage{placeins}
\usepackage[T1]{fontenc}
\usepackage{lipsum}
\usepackage{csquotes}

\biboptions{numbers,sort&compress}

\usepackage[pdftex, pdftitle={Article}, pdfauthor={Author},colorlinks = true,allcolors = blue]{hyperref} 

\title{Alignment of the CLAS12 central hybrid tracker with a Kalman Filter}

\begin{document}
\begin{frontmatter}

\journal{Nuclear Instrumentation and Methods A}

\newcommand*{\CSUDH}{California State University, Dominguez Hills, Carson, CA 90747}
\newcommand*{\CSUDHindex}{1}
\newcommand*{\CANISIUS}{Canisius College, Buffalo, NY}
\newcommand*{\CANISIUSindex}{2}
\newcommand*{\SACLAY}{IRFU, CEA, Universit\'{e} Paris-Saclay, F-91191 Gif-sur-Yvette, France}
\newcommand*{\SACLAYindex}{3}
\newcommand*{\CNU}{Christopher Newport University, Newport News, Virginia 23606}
\newcommand*{\CNUindex}{4}
\newcommand*{\UCONN}{University of Connecticut, Storrs, Connecticut 06269}
\newcommand*{\UCONNindex}{5}
\newcommand*{\DUQUESNE}{Duquesne University, Pittsburgh, PA 15282 USA }
\newcommand*{\DUQUESNEindex}{6}
\newcommand*{\FU}{Fairfield University, Fairfield CT 06824}
\newcommand*{\FUindex}{7}
\newcommand*{\FERRARAU}{Universit\`a di Ferrara, 44121 Ferrara, Italy}
\newcommand*{\FERRARAUindex}{8}
\newcommand*{\FIU}{Florida International University, Miami, Florida 33199}
\newcommand*{\FIUindex}{9}
\newcommand*{\GWUI}{The George Washington University, Washington, DC 20052}
\newcommand*{\GWUIindex}{10}
\newcommand*{\GSIFFN}{GSI Helmholtzzentrum fur Schwerionenforschung GmbH, D-64291 Darmstadt, Germany}
\newcommand*{\GSIFFNindex}{11}
\newcommand*{\INFNFE}{INFN, Sezione di Ferrara, 44100 Ferrara, Italy}
\newcommand*{\INFNFEindex}{12}
\newcommand*{\INFNFR}{INFN, Laboratori Nazionali di Frascati, 00044 Frascati, Italy}
\newcommand*{\INFNFRindex}{13}
\newcommand*{\INFNCAT}{INFN, Sezione di Catania, 95123 Catania, Italy}
\newcommand*{\INFNCATindex}{14}
\newcommand*{\INFNGE}{INFN, Sezione di Genova, 16146 Genova, Italy}
\newcommand*{\INFNGEindex}{15}
\newcommand*{\INFNRO}{INFN, Sezione di Roma Tor Vergata, 00133 Rome, Italy}
\newcommand*{\INFNROindex}{16}
\newcommand*{\INFNTUR}{INFN, Sezione di Torino, 10125 Torino, Italy}
\newcommand*{\INFNTURindex}{17}
\newcommand*{\INFNPAV}{INFN, Sezione di Pavia, 27100 Pavia, Italy}
\newcommand*{\INFNPAVindex}{18}
\newcommand*{\ORSAY}{Universit\'{e} Paris-Saclay, CNRS/IN2P3, IJCLab, 91405 Orsay, France}
\newcommand*{\ORSAYindex}{19}
\newcommand*{\Juelich}{Institute fur Kernphysik (Juelich), Juelich, Germany}
\newcommand*{\Juelichindex}{20}
\newcommand*{\JMU}{James Madison University, Harrisonburg, Virginia 22807}
\newcommand*{\JMUindex}{21}
\newcommand*{\KNU}{Kyungpook National University, Daegu 41566, Republic of Korea}
\newcommand*{\KNUindex}{22}
\newcommand*{\LAMAR}{Lamar University, 4400 MLK Blvd, PO Box 10046, Beaumont, Texas 77710}
\newcommand*{\LAMARindex}{23}
\newcommand*{\MIT}{Massachusetts Institute of Technology, Cambridge, Massachusetts  02139-4307}
\newcommand*{\MITindex}{24}
\newcommand*{\MISS}{Mississippi State University, Mississippi State, MS 39762-5167}
\newcommand*{\MISSindex}{25}
\newcommand*{\ITEP}{National Research Centre Kurchatov Institute - ITEP, Moscow, 117259, Russia}
\newcommand*{\ITEPindex}{26}
\newcommand*{\UNH}{University of New Hampshire, Durham, New Hampshire 03824-3568}
\newcommand*{\UNHindex}{27}
\newcommand*{\NMSU}{New Mexico State University, PO Box 30001, Las Cruces, NM 88003, USA}
\newcommand*{\NMSUindex}{28}
\newcommand*{\OHIOU}{Ohio University, Athens, Ohio  45701}
\newcommand*{\OHIOUindex}{29}
\newcommand*{\ODU}{Old Dominion University, Norfolk, Virginia 23529}
\newcommand*{\ODUindex}{30}
\newcommand*{\JLUGiessen}{II Physikalisches Institut der Universitaet Giessen, 35392 Giessen, Germany}
\newcommand*{\JLUGiessenindex}{31}
\newcommand*{\URICH}{University of Richmond, Richmond, Virginia 23173}
\newcommand*{\URICHindex}{32}
\newcommand*{\ROMAII}{universit\`a di Roma Tor Vergata, 00133 Rome Italy}
\newcommand*{\ROMAIIindex}{33}
\newcommand*{\MSU}{Skobeltsyn Institute of Nuclear Physics, Lomonosov Moscow State University, 119234 Moscow, Russia}
\newcommand*{\MSUindex}{34}
\newcommand*{\SCAROLINA}{University of South Carolina, Columbia, South Carolina 29208}
\newcommand*{\SCAROLINAindex}{35}
\newcommand*{\TEMPLE}{Temple University,  Philadelphia, PA 19122 }
\newcommand*{\TEMPLEindex}{36}
\newcommand*{\JLAB}{Thomas Jefferson National Accelerator Facility (Jefferson Lab), Newport News, Virginia 23606}
\newcommand*{\JLABindex}{37}
\newcommand*{\UTFSM}{Universidad T\'{e}cnica Federico Santa Mar\'{i}a, Casilla 110-V Valpara\'{i}so, Chile}
\newcommand*{\UTFSMindex}{38}
\newcommand*{\INSUBRIA}{Universit\`{a} degli Studi dell'Insubria, 22100 Como, Italy}
\newcommand*{\INSUBRIAindex}{39}
\newcommand*{\BRESCIA}{Universit\`{a} degli Studi di Brescia, 25123 Brescia, Italy}
\newcommand*{\BRESCIAindex}{40}
\newcommand*{\MESSINA}{Università degli Studi di Messina, 98166 Messina, Italy}
\newcommand*{\MESSINAindex}{41}
\newcommand*{\GENOVA}{Universit\`{a} di Genova, Dipartimento di Fisica, 16146 Genova, Italy}
\newcommand*{\GENOVAindex}{42}
\newcommand*{\UCR}{University of California Riverside, 900 University Avenue, Riverside, CA 92521, USA}
\newcommand*{\UCRindex}{43}
\newcommand*{\GLASGOW}{University of Glasgow, Glasgow G12 8QQ, United Kingdom}
\newcommand*{\GLASGOWindex}{44}
\newcommand*{\YORK}{University of York, York YO10 5DD, United Kingdom}
\newcommand*{\YORKindex}{45}
\newcommand*{\WM}{College of William and Mary, Williamsburg, Virginia 23187-8795}
\newcommand*{\WMindex}{46}
\newcommand*{\YEREVAN}{Yerevan Physics Institute, 375036 Yerevan, Armenia}
\newcommand*{\YEREVANindex}{47}


 \author[toUCR]{S.J.~Paul}
 \author[toUCR]{A.~Peck}
 \author[toUCR,toJLAB]{M.~Arratia}
 \ead[Correspondence email address: ]{miguel.arratia@ucr.edu}
 \author[toJLAB]{Y.~Gotra}
 \author[toJLAB]{V.~Ziegler}
 \author[toINFNGE]{R.~De~Vita}
 \author[toIRFU]{F.~Boss\`u}
 \author[toIRFU]{M.~Defurne}

\author[toTEMPLE]{H.~Atac}
\author[toWM]{C.~Ayerbe~Gayoso}
% ...

% ...
\address[toINFNGE]{\INFNGE} 
% ...
\address[toTEMPLE]{\TEMPLE} 
 \address[toJLAB]{\JLAB} 
% ...
\address[toUCR]{\UCR} 
% ...
\address[toWM]{\WM} 
% ...

\date{\today} 
\begin{abstract}
    Several factors can contribute to the difficulty of aligning the sensors of tracking detectors, including a large number of modules, multiple types of detector technologies, and non-linear strip patterns on the sensors.  All three of these factors apply to the CLAS12 CVT, which is a hybrid detector consisting of planar silicon sensors with non-parallel strips, and cylindrical micromegas sensors with longitudinal and arc-shaped strips located within a 5~T superconducting solenoid. To align this detector, we used the Kalman Alignment Algorithm, which accounts for correlations between the alignment parameters without requiring the time-consuming inversion of large matrices.  This is the first time that this algorithm has been adapted for use with hybrid technologies, non-parallel strips, and curved sensors. We present the results for the first alignment of the CLAS12 CVT using straight tracks from cosmic rays and from a target with the magnetic field turned off.  After running this procedure, we achieved alignment at the level of 10~$\mu$m, and the widths of the residual spectra were greatly reduced.  These results attest to the flexibility of this algorithm and its applicability to future use in the CLAS12 CVT and other hybrid or curved trackers, such as those proposed for the future Electron-Ion Collider. 
\end{abstract}

\end{frontmatter}
\twocolumn

\section{Introduction}\label{sec:introduction}

Aligning a tracking detector is a non-trivial task, which can involve large numbers of degrees of freedom.  Various algorithms have been developed for this task, such as HIP \cite{Karimaki:2006az} and MillePede \cite{Blobel:2006yh}.  The Kalman Alignment Algorithm (KAA) \cite{Widl:2006mz,Widl:2008aqa}, which is based on the Kalman-filter algorithm, was first implemented to align the CMS silicon tracker \cite{Widl_2010}, and we use it to align the CEBAF Large Acceptance Spectrometer (CLAS12) Central Vertex Tracker (CVT) \cite{Burkert:2020akg,ANTONIOLI2020163701,ACKER2020163423} in Hall~B at Jefferson Lab.

These algorithms take the fitted tracks, reconstructed from misaligned detector data, and a model of the dependence of the residuals of the track fit to the alignment and track parameters.  Here, the residuals are the differences between the measurements along the track and the values interpolated from the track fit.
  The goal of these algorithms is then to find the values of the alignment parameters that minimize the sum of squares of the residuals ({\it i.e.} the track fit $\chi^2$). 

When choosing an alignment algorithm, two important factors are the computational speed and biases in the results.  One drawback to the MillePede algorithm is that it requires the inversion of a large matrix, typically of rank $N_{\rm align}\cross N_{\rm align}$, where $N_{\rm align}$ is the number of alignment parameters, which can be time-consuming.  
The Hits and Impact Points (HIP) algorithm is similar to MillePede, except that it forces the analogous matrix to be block-diagonal (and thus much faster to invert) at the cost of ignoring the dependence of the residuals on the track parameters (which MillePede and the KAA take into account).  Because this dependence is ignored, the correlations between alignment parameters for one module and those of another module are not accounted for and can introduce biases in the results.  The KAA overcomes both of these problems.  Like MillePede, it corrects for the biases caused by the track-parameter dependence of the residuals, but the KAA does so in a manner that avoids the inversion of large matrices.  

The results obtained with the HIP, KAA, and Millepede algorithms for the CMS inner tracker were compared to one another in Ref.~\cite{Widl_2010}.  The tracking-residual distributions obtained with the three algorithms were all centered within a few $\mu$m of zero and had comparable RMS values to one another (about 300~$\mu$m). 

One important difference between CMS and the CLAS12 CVT is that the strips in the sensors in CMS are straight and parallel\footnote{The strips in one sensor of CMS are not necessarily parallel to those in an another sensor, since there is a stereo angle between nearby sensors.}, whereas the CLAS12 CVT has both non-parallel strips within the same sensor and sensors that are curved.  These two features cause the tracking residuals to depend non-linearly on the alignment parameters.  The HIP, KAA, and Millepede algorithms all approximate the relationship between these residuals and the alignment parameters as linear, causing such algorithms to converge at non-optimal values for the alignment parameters.   

A solution to this issue was used for the LHCb VELO (silicon VErtex LOcator), which consists of silicon sensors with azimuthally curved and radial strips rigidly mounted on half-disks \cite{Antunes-Nobrega:630827}.  To determine the relative alignment of the radial and azimuthal sensors, they used multiple iterations of a fast, specialized algorithm that is similar to HIP, and refitted the tracks between iterations with the alignment parameters obtained from the previous iteration \cite{VIRET2008157}.  From this, they obtained an alignment precision on the level of 1.3~$\mu$m between the radial and azimuthal sensors of each half disk.

In this work, we use the KAA to align the CLAS12 CVT using a multiple-iteration approach similar to Ref.~\cite{VIRET2008157}. The CLAS12 CVT presents two new challenges for the KAA that were not applicable when it was first implemented for CMS: the CVT is a hybrid of two different types of sensor technology, silicon and micromegas, while CMS is a fully silicon tracker, and the CVT includes curved sensors, while the sensors at CMS are flat.  Thus, the alignment of the CVT using the KAA is a test of the versatility and flexibility of the algorithm for diverse detectors.

Details of the CLAS12 CVT are given in Sec.~\ref{sec:CVT}.  We then describe  the KAA in Sec.~\ref{sec:KAA}.  Section \ref{sec:data} describes the datasets used for alignment. In Sec.~\ref{sec:strategy}, we describe the procedure for running the KAA for the CLAS12 CVT. We then present the results for the data in Sec.~\ref{sec:results} and we conclude in Sec.~\ref{sec:conclusions}.

\section{The CLAS12 Central Vertex Tracker}
\label{sec:CVT}
\begin{figure}[]
    \centering
    \includegraphics[width=1.1\columnwidth,trim=7cm 0 0 0,clip]{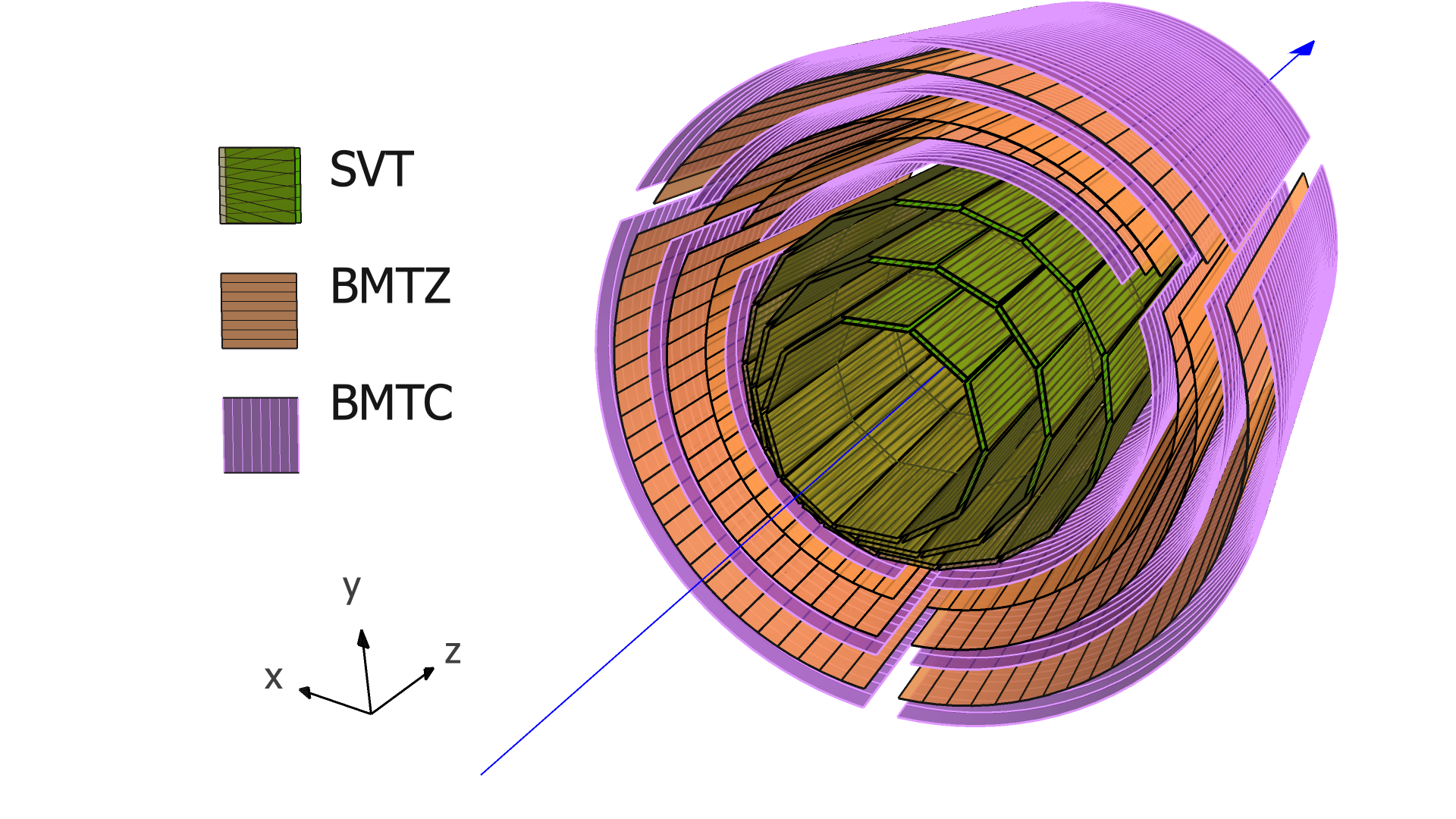}
    \caption{Rendering of the CLAS12 CVT, which consists of three double layers of SVT (inner, leaf-green) and six layers of BMT, with Z layers in orange and C layers in purple.  The blue line represents the beamline.  The lines within the sensors indicate every 32nd strip.}
    \label{fig:cvt}
\end{figure}

The CLAS12 CVT, which covers the polar-angle\footnote{Throughout this paper, the lab-frame coordinates are defined as follows:  $z$ is along the beam direction, $y$ is the up direction, and $x$ is to the left when looking at the detector from upstream.} range $35^\circ<\theta<125^\circ$, is shown in Fig.~\ref{fig:cvt}. It consists of three regions of double-sided Silicon Vertex Tracker (SVT) modules \cite{ANTONIOLI2020163701} and six layers of the Barrel Micromegas Tracker (BMT) \cite{ACKER2020163423}.  

\begin{figure}
    \centering
    \includegraphics[width=\columnwidth, trim=10cm 0 0 0, clip]{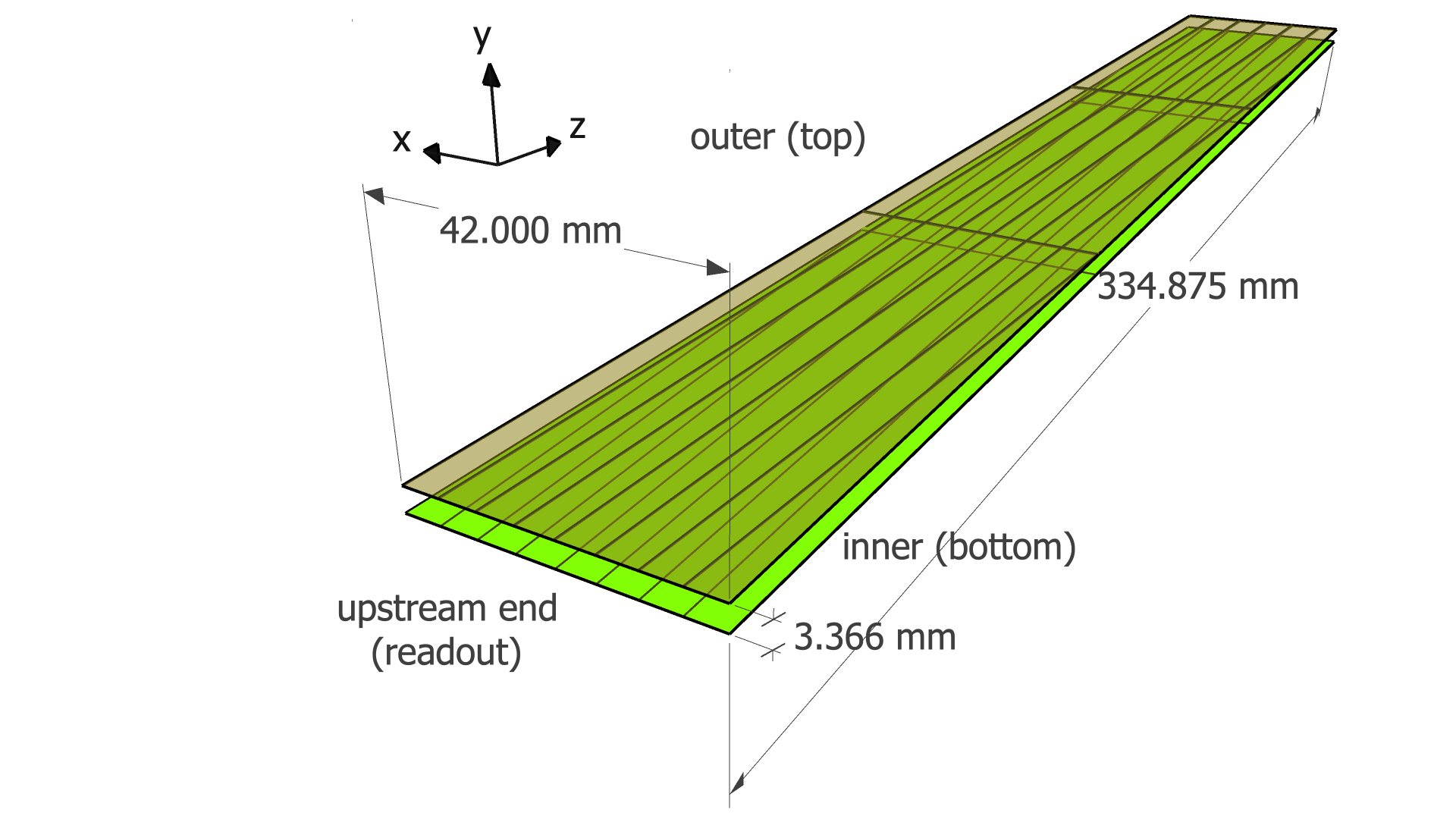}
    \includegraphics[width=\columnwidth]{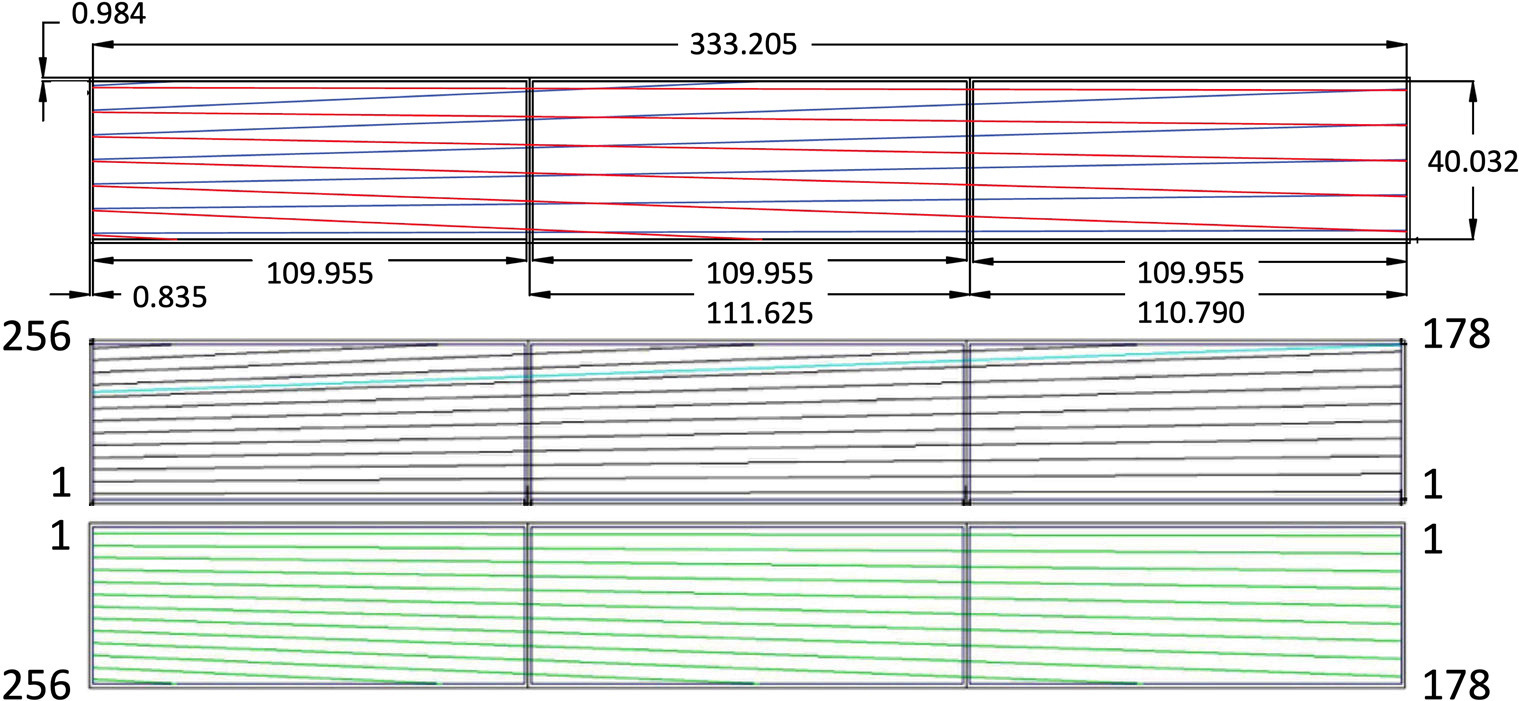}
    \caption{Top: 3D Rendering of one of the SVT sector modules.  The inner (outer) sensor of the module is shown in green (yellow).  Every 32nd strip is shown for both sensors as lines on the sensors.  Bottom (from Ref.~\cite{ANTONIOLI2020163701}): Sensor strip layout. The upstream end, which has the readout, is on the left side. Strip numbers are indicated. Dimensions are in mm.}
    \label{fig:svt_module}
\end{figure}

The SVT regions are arranged in concentric polygons with 10, 14, and 18 sectors in the inner, middle, and outer pairs of layers\footnote{Since the pairs of layers have different numbers of modules, the sectors in one double layer do not line up with those in the other double layers, with the exception of the top and bottom sector in each double layer.}.  The geometry of the SVT is summarized in Table~\ref{tab:SVT}.  Within each pair of layers, each sector is a separate module, consisting of one sensor on each of the two layers, 
separated radially by 3.162~mm.  The sensor consists of three daisy-chained silicon microstrip detectors and has 256 strips. Each detector is 320~$\mu$m thick, 42.00~mm wide, and 111.63~mm long.  A rendering of the geometry of the SVT module is shown in Fig.~\ref{fig:svt_module}. 

At the upstream end of the sensor planes, where the strips connect to the readout, they have 156~$\mu$m pitch, but they fan out, with the angle of the strip relative to longitudinal direction of the sensor increasing linearly from 0$^\circ$ at the first strip to 3$^\circ$ at the last strip.  The two sensors in each module are mounted back-to-back, so that the first strip of one sensor corresponds with the last strip of the other and vice versa. This geometry allows measurements of the longitudinal hit positions due to the 3$^\circ$ stereo angle between the two sensors on each module.  

\begin{table}[]
    \centering
    \begin{tabular}{c|ccc}
        Layer & Radius (mm) & Pitch ($\mu$m) & Sectors  \\
        \hline
        1 & 65.29 & 156-224 & 10 \\
        2 & 68.77 & 156-224 & 10 \\
        3 & 92.89 & 156-224 & 14 \\
        4 & 96.37 & 156-224 & 14 \\
        5 & 120.32 & 156-224 & 18 \\
        6 & 123.80 & 156-224 & 18 \\
    \end{tabular}
    \caption{Summary of parameters of each SVT layer.  The radii given are the nominal values for the perpendicular distance between the midplane of the SVT backing structure and the beamline.  The pitch varies from 156~$\mu$m at the upstream end to about 224~$\mu$m at the downstream end.}
    \label{tab:SVT}
\end{table}

The BMT is divided azimuthally into three sectors, each of which consists of six cylindrical arc layers.  There are two types of sensors:  Z-type (layers 2, 3, and 5), in which the strips are (nominally) parallel to the beamline and measure the azimuthal position of the particle's trajectory, and C-type (layers 1, 4, and 6) in which the strips curve azimuthally around the beamline and measure the longitudinal position of the particle's trajectory (which is also used to measure the polar angle of the trajectory).  Throughout this paper, we refer to the Z layers as the BMTZ and the C layers as the BMTC.  The radii, pitches,  and strip orientations for each layer are given in Table~\ref{tab:BMT}.

\begin{table}[]
    \centering
    \begin{tabular}{c|ccc}
        Layer & Radius (mm) & Pitch
        ($\mu$m) & Strip orientation \\
        \hline
        1 & 146.15 & 330–860 & C \\
        2 & 161.15 & 487 & Z \\
        3 & 176.15 & 536 & Z \\
        4 & 191.15 & 340–770 & C \\
        5 & 206.15 & 529 & Z \\
        6 & 221.15 & 330–670 & C \\
    \end{tabular}
    \caption{Summary of parameters of each BMT layer. The pitches of the C layers vary from strip to strip, with wider strips towards the front and back, and narrower strips near the center.}
    \label{tab:BMT}
\end{table}

\section{Methodology}
\label{sec:KAA}
We used the KAA, which is described in detail in Refs.~\cite{Widl:2006mz,Widl:2008aqa}.  Here we present a summary of the main features of the algorithm and detail the specific implementation to the CLAS12 case.  We note here that our method relies on straight tracks to obtain the alignment parameters and was validated with both straight and curved tracks.

A Kalman filter is an algorithm that uses an ordered sequence of measurements and produces estimates of unknown parameters that converge upon more precise values than those obtained from a single measurement.
Like any other Kalman-filter algorithm, the KAA begins with an estimate of the parameters to be fitted and a matrix of the covariances among these parameters.   It then loops through the measurements in the input sample and updates the values of the parameters and their covariance matrix after each measurement.  In the case of the KAA, the parameters to be fitted are the alignment parameters, and the measurements are fitted tracks and the tracking residuals thereof\footnote{This is analogous to the Kalman-filter track-fitting algorithm, where the parameters of a single track are fitted, and the individual measurements are the hits and/or clusters along the track.}.  As more tracks are processed, the uncertainties on the alignment parameters (that is, the square roots of the diagonal elements of the covariance matrix) decrease, and the alignment parameters converge to more precise values.

In the KAA, the deviations of each sensor and module from their nominal positions  are represented by the column vector $\mathbf d$. 
The KAA requires a preliminary estimate of $\mathbf d$ and its covariance matrix $\mathbf{D}$, and a set of several matrices for each track.  These matrices, which are summarized below, model the track residuals for each measurement in the track, their dependence on the alignment and track parameters, and the expected resolution on these residuals and are summarized below.  
The alignment values and its covariance matrix are updated sequentially for every track in the sample of input events.

Straight-line tracks in the CVT are represented by their direction, $\hat u$, and a point on the line, $\vec x_{\rm ref}$.   Unless otherwise noted, all coordinates are given in the lab frame.  We use the following track parameters: the distance of closest approach of the track to the beamline, $d_0$, the azimuthal angle of the track direction, $\phi_0$, the longitudinal position of the track's point of closest approach, $z_0$, and the tangent of the track's dip angle, $t_0$.  Expressed in terms of these parameters, $\vec x_{\rm ref}$ and $\hat u$ are:

\begin{equation}
   \vec x_{\rm ref} = (-d_0\sin\phi_0+x_b,\, d_0\cos\phi_0+y_b,\,z_0)
   \label{eq:xref}
\end{equation}
and 
\begin{equation}
    \hat u = \frac{(\cos\phi_0,\,\sin\phi_0,\, t_0)}{\sqrt{1+t_0^2}},
   \label{eq:u}
\end{equation}
where $(x_b, y_b)$ is the beam position.  

\begin{figure*}
    \centering
    \includegraphics[width=0.33\textwidth]{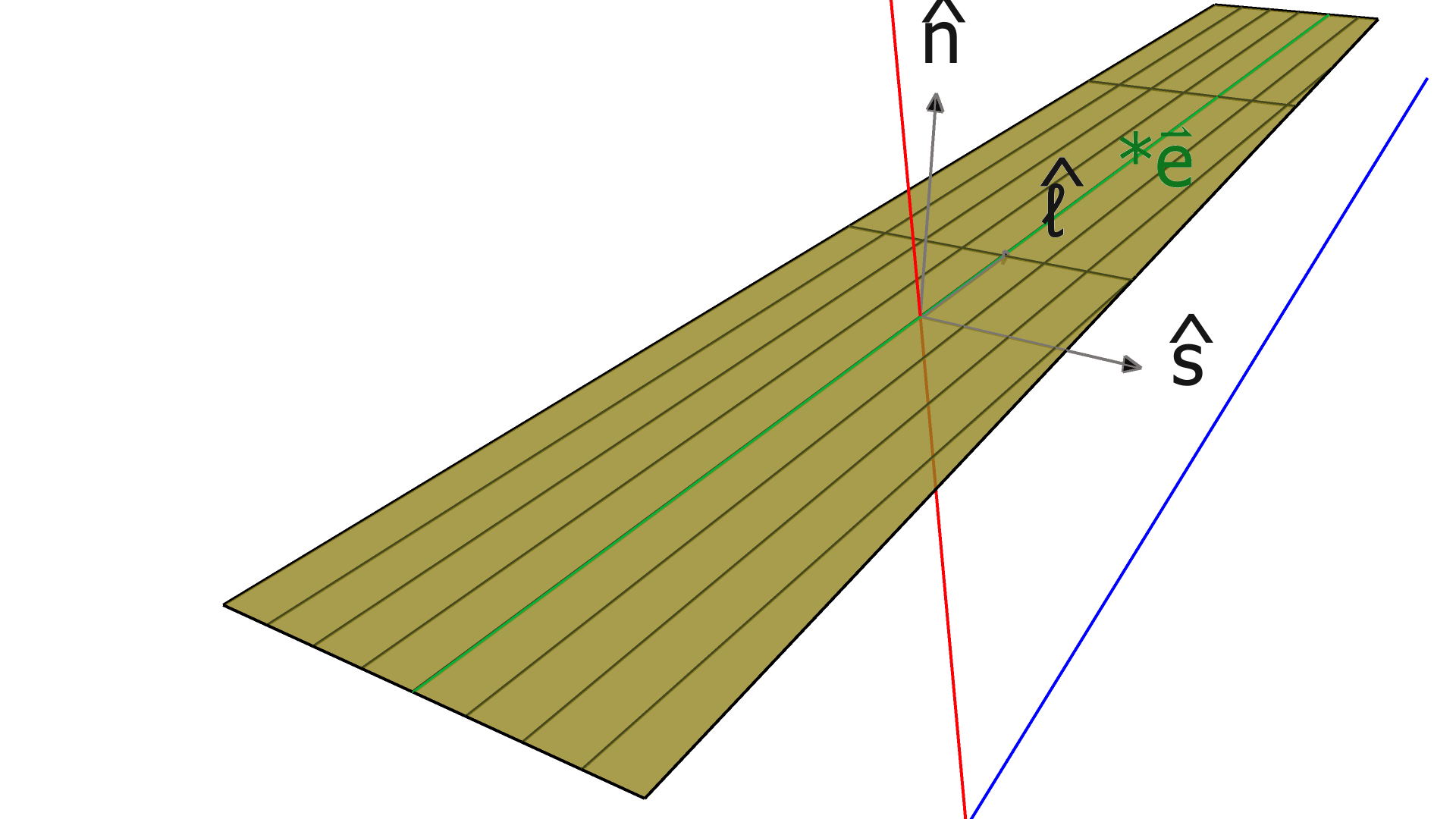}\includegraphics[width=0.33\textwidth]{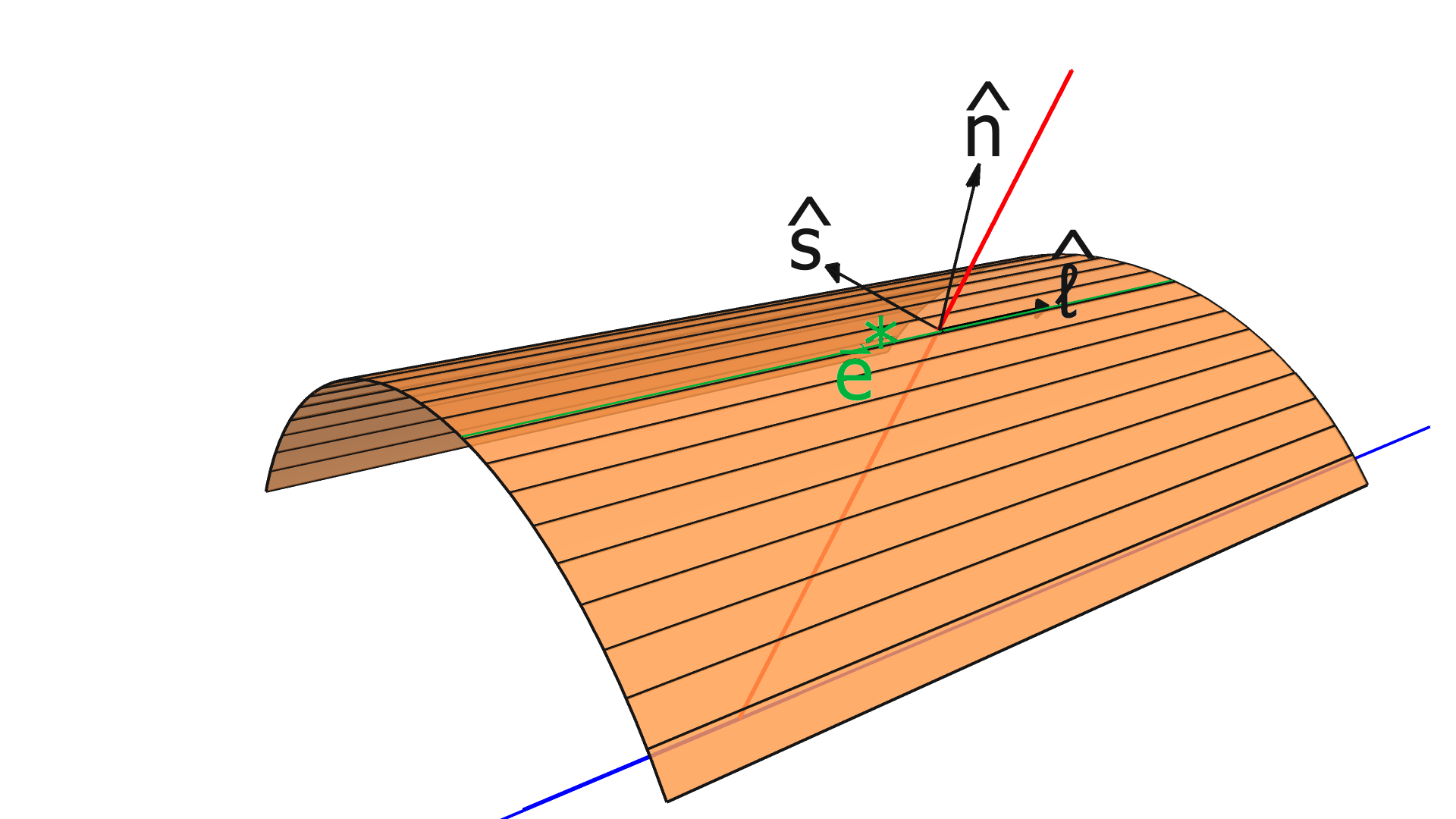}\includegraphics[width=0.33\textwidth]{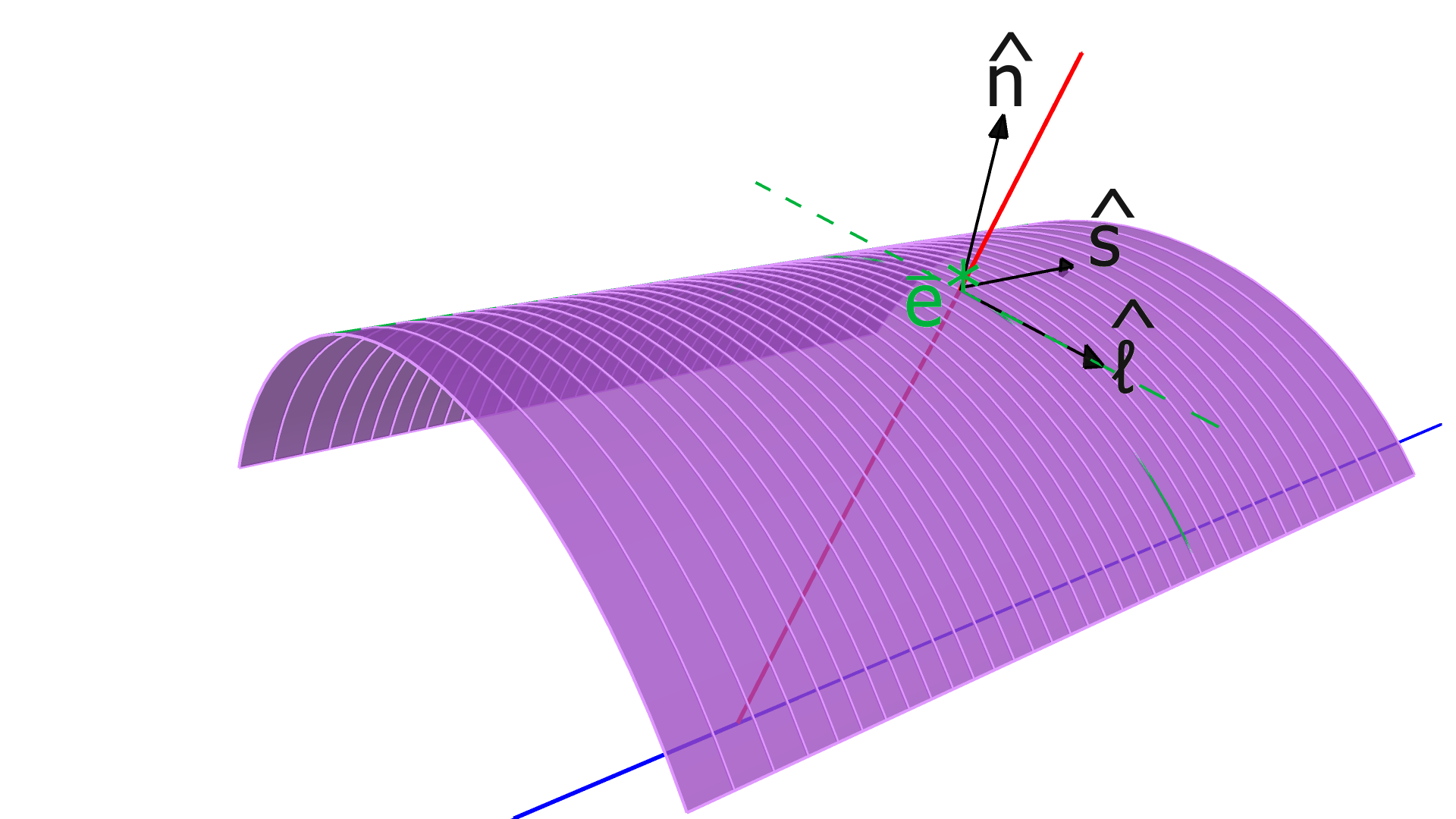}
    \caption{Illustrations of the vectors $\hat s$, $\hat n$, and $\hat \ell$ for the SVT (left), BMTZ (middle), and BMTC (right).  The beamline and the reference trajectory are shown in blue and red, respectively.  The struck strip is shown as a solid green line or arc.  For the BMTC, the tangent line to the struck strip is shown as a green dashed line.  A point on the line, $\vec e$, is indicated by an asterisk (for the SVT and BMTZ, we chose one of the endpoints of the strip; for the BMTC, we used the tangent point).}
    \label{fig:snl}
\end{figure*}

In the CVT, each measurement corresponds to a contiguous cluster of hits in one of the SVT or BMT layers.  
We represented these clusters as line segments connecting the centroids\footnote{weighted by the reconstructed energy deposited in the strip} of the endpoints of the strips on one end of the sensor to the centroid of the endpoints of the strips on the other end.  Notice the direction of each line segment in the
lab frame is not necessarily parallel to a particular strip.  We defined the vector $\vec e$ to be the coordinates (in the lab frame) of a point on this line segment (arbitrarily, we chose the midpoint), and $\hat\ell$  to be the direction of this line, \textit{i.e.}, the direction of the lines connecting the centroids of the endpoints of the
strips on each end of the sensor\footnote{For the SVT, which has non-parallel strips, this is the weighted average of the directions of the strips in the cluster}.  
We also defined the unit vector $\hat n$ as the unit normal vector to the sensor, and $\hat s=\hat n\times\hat \ell$, which we call the ``measurement direction'', as shown in Fig.~\ref{fig:snl}.

For the BMTC, each strip is an arc, 
therefore we analogously constructed a ``centroid'' arc using the centroids of two endpoints and centers of the individual strip's arcs.  We then extrapolated the track to the BMTC layer, and find the line that is tangent to the arc at the same azimuthal position as the extrapolation point (right panel of Fig.~\ref{fig:snl}).  The vectors $\vec e$ and $\hat\ell$ are then defined as a point on this line (we chose the tangent point) and the direction of the line respectively.  The measurement direction, $\hat s$, is defined to be along the BMTC layer's cylindrical axis, and $\hat n$ is normal to the sensor at the extrapolated azimuthal position.  

Using these representations of the track and its clusters, we then determined the matrices needed for the KAA's input.  The first two matrices are a column vector of the 1D measurements along the track, $\mathbf m$, and another column vector $\mathbf c$ of the expected values for each measurement based on a track fit performed in reconstruction, which is made using the the Kalman Filter algorithm~\cite{FRUHWIRTH1987444}. The tracking-residuals column vector, $\mathbf{r}$, is defined as their difference, $\mathbf{m}-\mathbf{c}$.  

We calculated the element of the column vectors $\mathbf c$, $\mathbf m$, and $\mathbf r$ corresponding to the $i^{\rm th}$ measurement along the track using the following formulas:
\begin{equation}
\label{eq:c}
    c_i = \hat s\cdot\left(\vec x_{\rm ref} + \hat u \frac{\hat n\cdot(\vec e-\vec x_{\rm ref})}{\hat u\cdot\hat n}\right),
\end{equation} 
\begin{equation}
\label{eq:m}
    m_i = \hat s\cdot \vec e,
\end{equation}
and 
\begin{align}
\label{eq:r}
    r_i &= m_i-c_i,\\
        &=\hat s\cdot\left(\vec e-\vec x_{\rm ref} - \hat u \frac{\hat n\cdot(\vec e-\vec x_{\rm ref})}{\hat u\cdot\hat n}\right)\\
        &=\vec s\,'\cdot(\vec e-\vec x_{\rm ref}), \label{eq:rformula}
\end{align} 
where 
\begin{equation}
\vec s\,' = \hat s - \frac{\hat s\cdot \hat u}{\hat u\cdot\hat n}\hat n.
\label{eq:sp}
\end{equation}
Eq.~\ref{eq:rformula} is equivalent to the distance along the measurement direction, $\hat s$, between the centroid line of the cluster of hits on the sensor and the extrapolated position where the track intersects the sensor.  

The next matrix in the input, $\mathbf{V}$, represents the stochastic part of the measurement.  The elements of $\mathbf V$ can be expressed as the expectation value of the product of the residuals for two (not necessarily distinct) measurements in a track,
\begin{equation}
    V_{ij} = \langle r_i r_j \rangle,
\end{equation}
where $i$ and $j$ are the indices of the two measurements within the track. 

In models where the residuals in one sensor are uncorrelated with those in the other sensors (as is assumed in this work), this matrix is diagonal, where each element is the square of the resolution for the corresponding measurement in the track.  We used the spacial resolutions that are calculated in the CLAS12 reconstruction package \cite{ZIEGLER2020163472}.  For the SVT, where the strips get wider further downstream, the width is calculated at the longitudinal position of the intersection of the clusters in a stereo pair.

The dependence of the residuals on the alignment parameters and on the track parameters are modeled linearly by the alignment-derivative matrix, $\mathbf{A}$, and the track-derivative matrix $\mathbf{B}$.  The elements of $\mathbf{A}$ are defined by

\begin{equation}
     A_{ij} = \frac{\partial r_i}{\partial x_j},
     \label{eq:Adef}
\end{equation}
where $r_i$ is the residual of the $i^{\rm th}$ measurement in the track and $x_j$ is the $j^{\rm th}$ alignment parameter.  In this work, we assume that every module is a rigid body, and therefore  consider only rotation and translation, but not deformations within any module.  For three rotation variables and three translation variables per module,  $\mathbf{A}$ has dimension $n_{\rm meas}\times 6n_{\rm mod}$, where $n_{\rm meas}$ is the number of measurements (clusters) in the track, and $n_{\rm mod}$ is the total number of modules to be aligned.

The elements of $\mathrm{B}$ are likewise defined as 

\begin{equation}
\label{eq:Bdef}
    B_{ij} = \frac{\partial r_i}{\partial t_j},   
\end{equation}
where $r_i$ is the residual of the $i^{\rm th}$ measurement in the track and $t_j$ is the $j^{\rm th}$ track parameter.  Since four parameters define a straight track, $\mathbf{B}$ has dimension $n_{\rm meas}\times 4$.  

In our implementation, the elements of the alignment-derivative matrix, $\mathbf{A}$, are
\begin{equation}
\label{eq:AT}
A_{i,\vec T} = \vec s\,'
\end{equation}
and 
\begin{equation}
\label{eq:AR}
A_{i,\vec R} = -\vec s\,'\times \left(\vec x_{\rm ref}+\left(\frac{\vec n \cdot (\vec e-\vec x_{\rm ref})}{\hat u\cdot\hat n}\right) \hat u\right).
\end{equation}
The $\vec T$ and $\vec R$ vectors represent the groups of indices corresponding to the translation and rotation parameters of the module which the $i^{\rm th}$ measurement in the track takes place in.    

The elements of the track-derivative matrix, $\mathbf{B}$, are 
\begin{align}
    \label{eq:Bd}
    B_{i,d_0} =& -\vec s\,'\cdot (-\sin\phi_0,\,\cos\phi_0,\,0)\\
      B_{i,\phi_0} =& -\vec s\,'\cdot\left(\frac{\hat n\cdot(\vec e-\vec x_{\rm ref})}{\hat u\cdot\hat n\sqrt{1+t_0^2}}(-\sin\phi_0,\,\cos\phi_0,\,0)\right.\\\nonumber
      &\hspace{2.5cm}\left.-d_0(\cos\phi_0,\,\sin\phi_0,\,0)\vphantom{\frac12}\right)\\
    B_{i,z_0} =& -s'_z\\
    B_{i,t_0} =& -s'_z\frac{\hat n\cdot(\vec e-\vec x_{\rm ref})}{\hat u\cdot\hat n}.\label{eq:Bt}
\end{align}

Equation~\ref{eq:AT} was obtained by taking the derivative of the formula for the residuals (Eq.~\ref{eq:rformula}) with respect to $\vec e$.  To obtain Eq.~\ref{eq:AR}, we took the derivative of Eq.~\ref{eq:rformula} with respect to an infinitesimal rotation $d\vec R$ of the sensor:  $\hat n\rightarrow \hat n + d\vec R\cross \hat n$, and likewise for $\hat s$, $\hat \ell$, and $\vec e$.  The track is not rotated, so the vectors $\hat u$ and $\vec x_{\rm ref}$ are not rotated.  

Eqs.~\ref{eq:Bd}-\ref{eq:Bt} were obtained by taking the derivative of Eq.~\ref{eq:rformula} with respect to the track parameters $d_0$, $\phi_0$, $z_0$, and $t_0$, using the definitions of $\vec x_{\rm ref}$ and $\hat u$ in Eqs.~\ref{eq:xref} and \ref{eq:u}.

The degrees of freedom corresponding to the matrix elements of $\mathbf{A}$ and $\mathbf{B}$ are illustrated in Figs.~\ref{fig:deriv_SVT}, \ref{fig:deriv_BMTZ}, and \ref{fig:deriv_BMTC} for the SVT, BMTZ, and BMTC, respectively.  
\newcommand{\customwidth}{0.39\textwidth}
\begin{figure*}
    \centering
    \includegraphics[width=\customwidth]{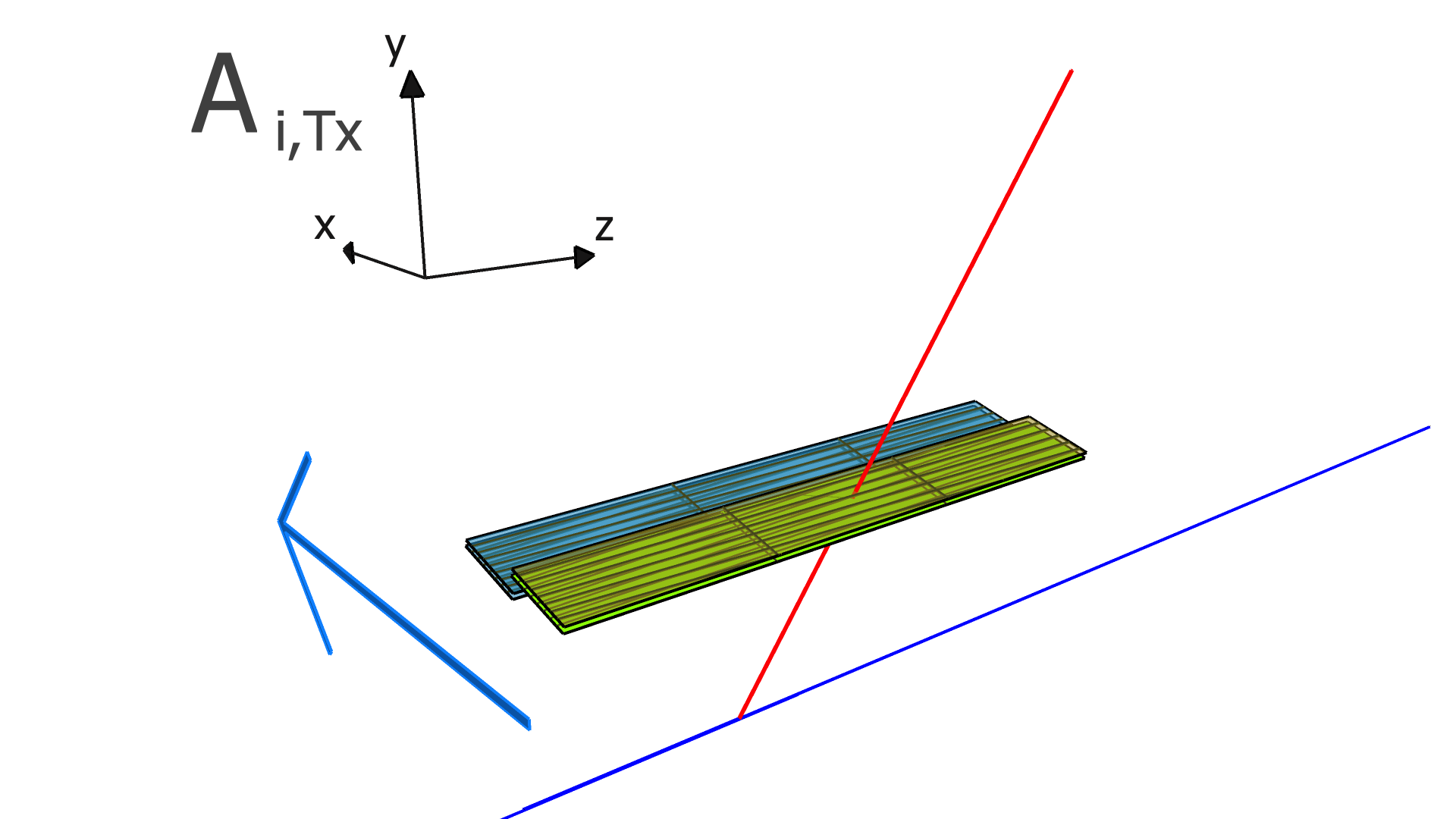}\includegraphics[width=\customwidth]{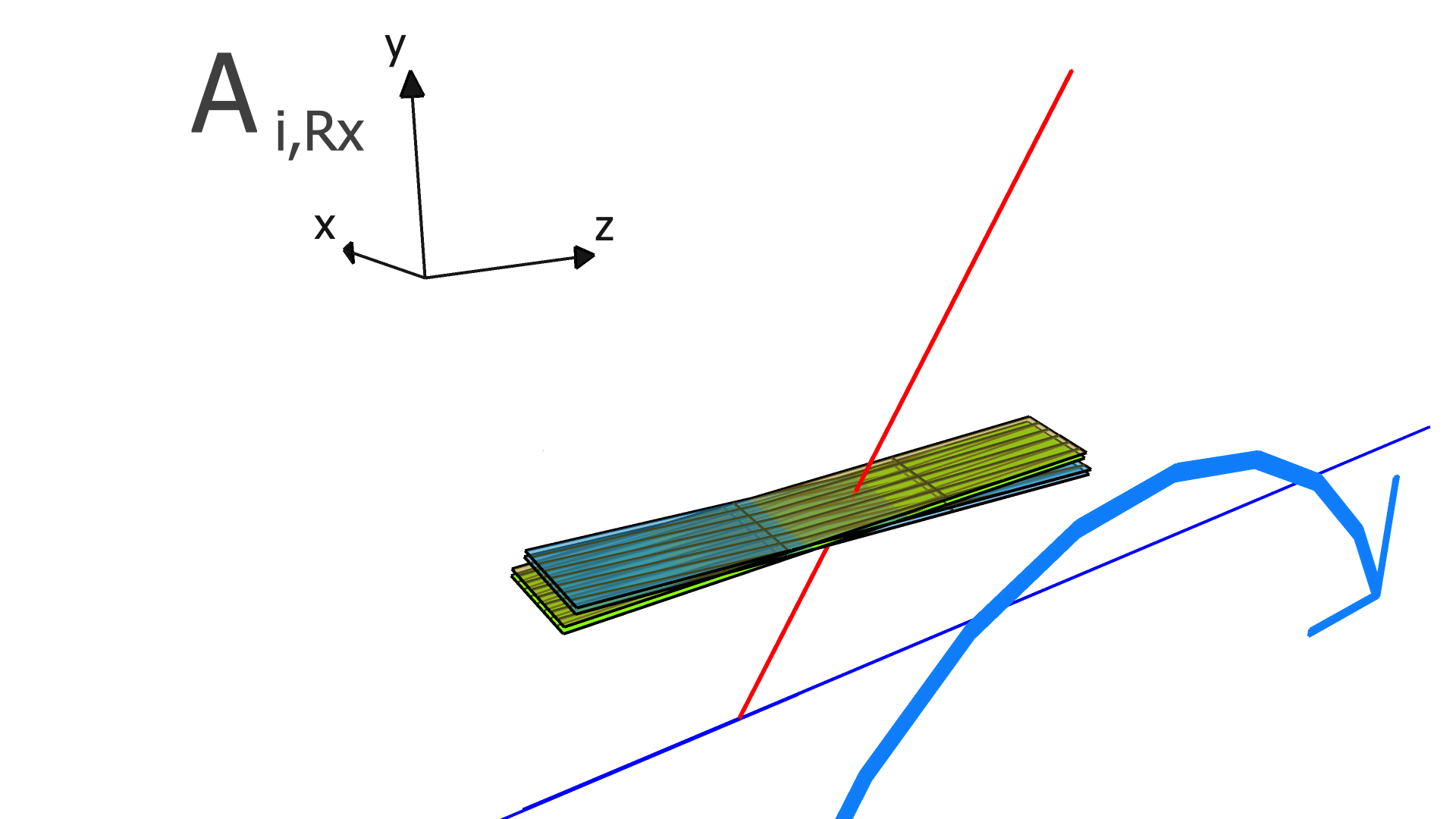}
    
    \includegraphics[width=\customwidth]{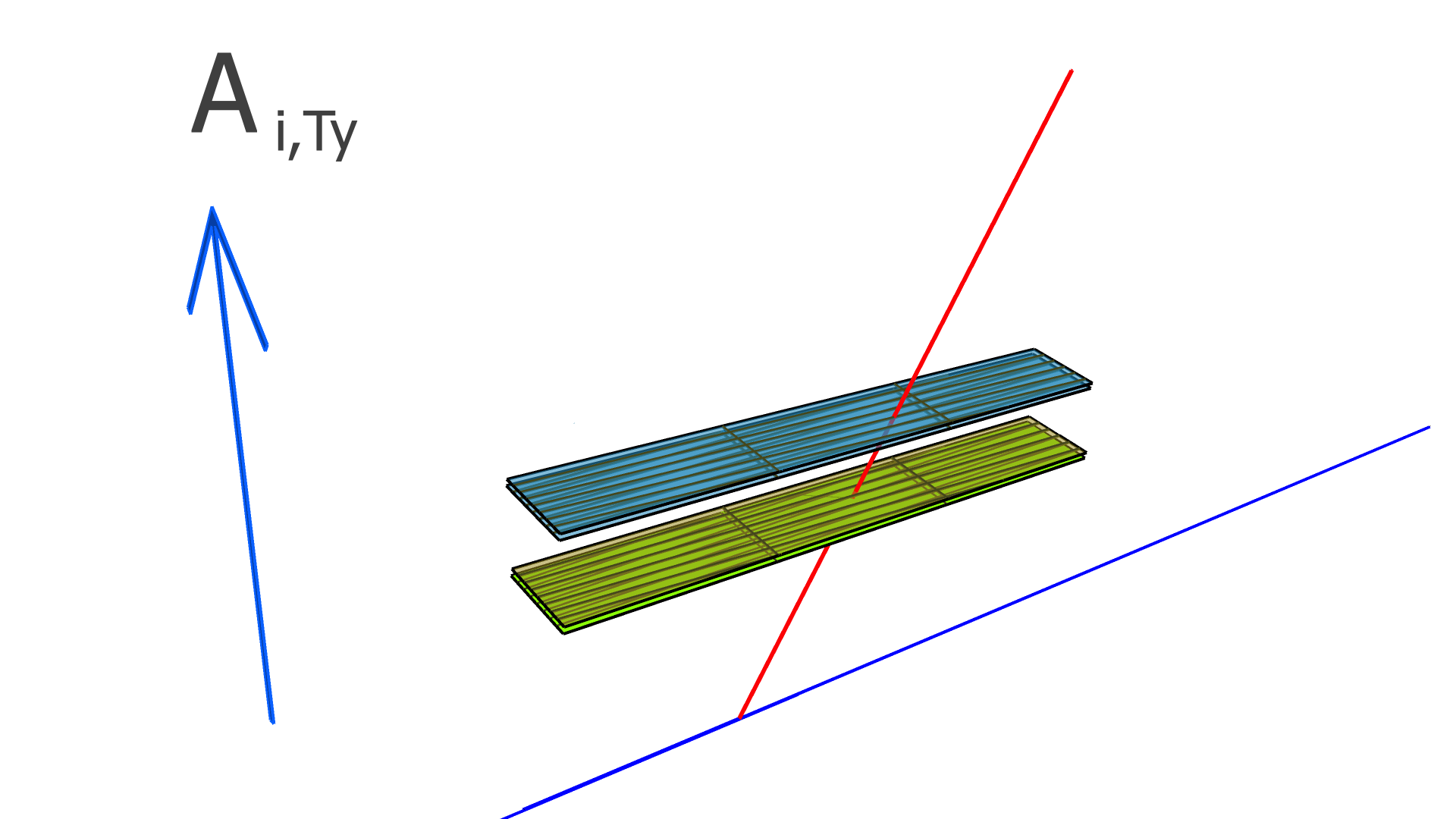}\includegraphics[width=\customwidth]{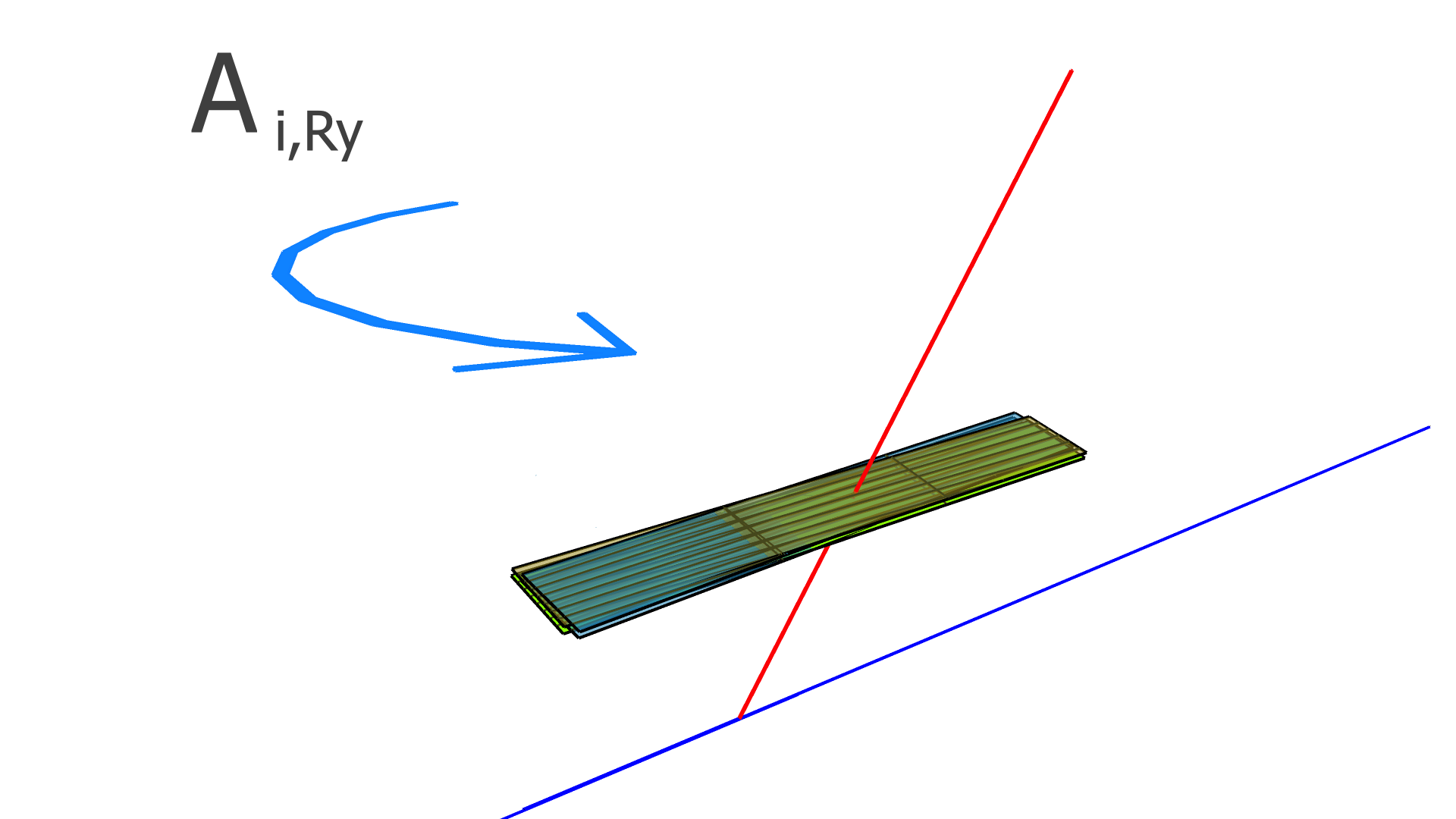}
    
    \includegraphics[width=\customwidth]{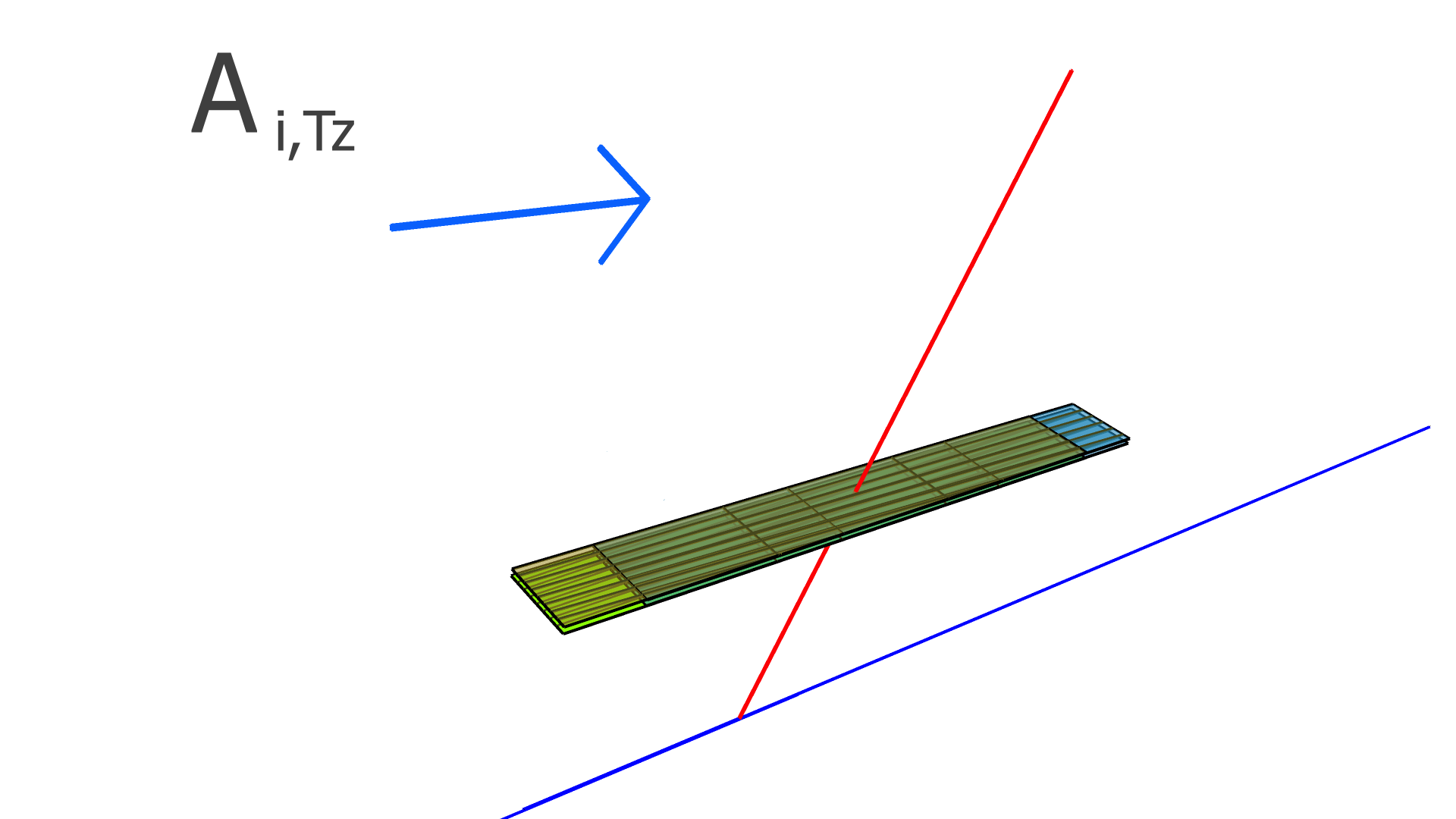}\includegraphics[width=\customwidth]{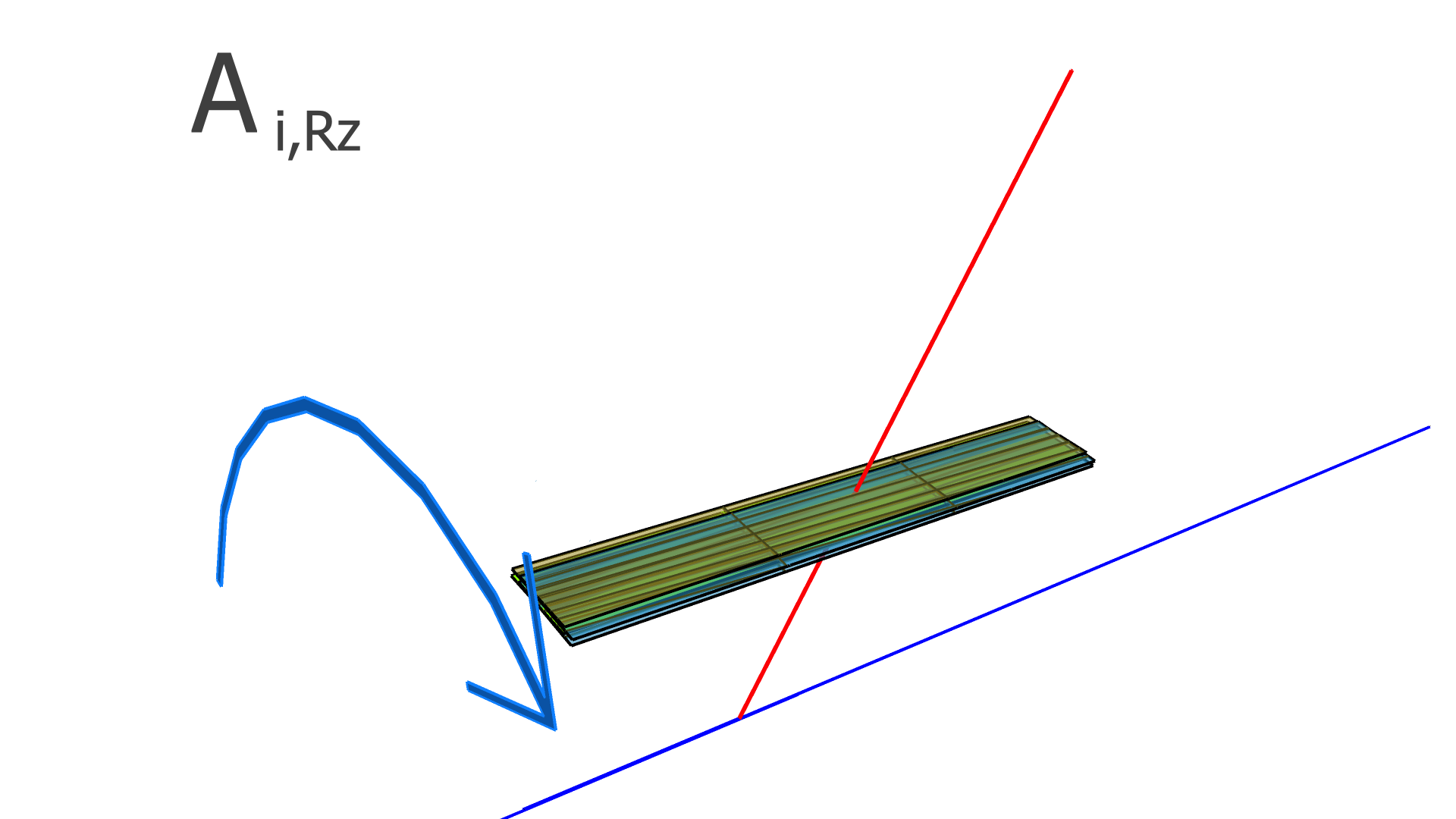}
    
    \includegraphics[width=\customwidth]{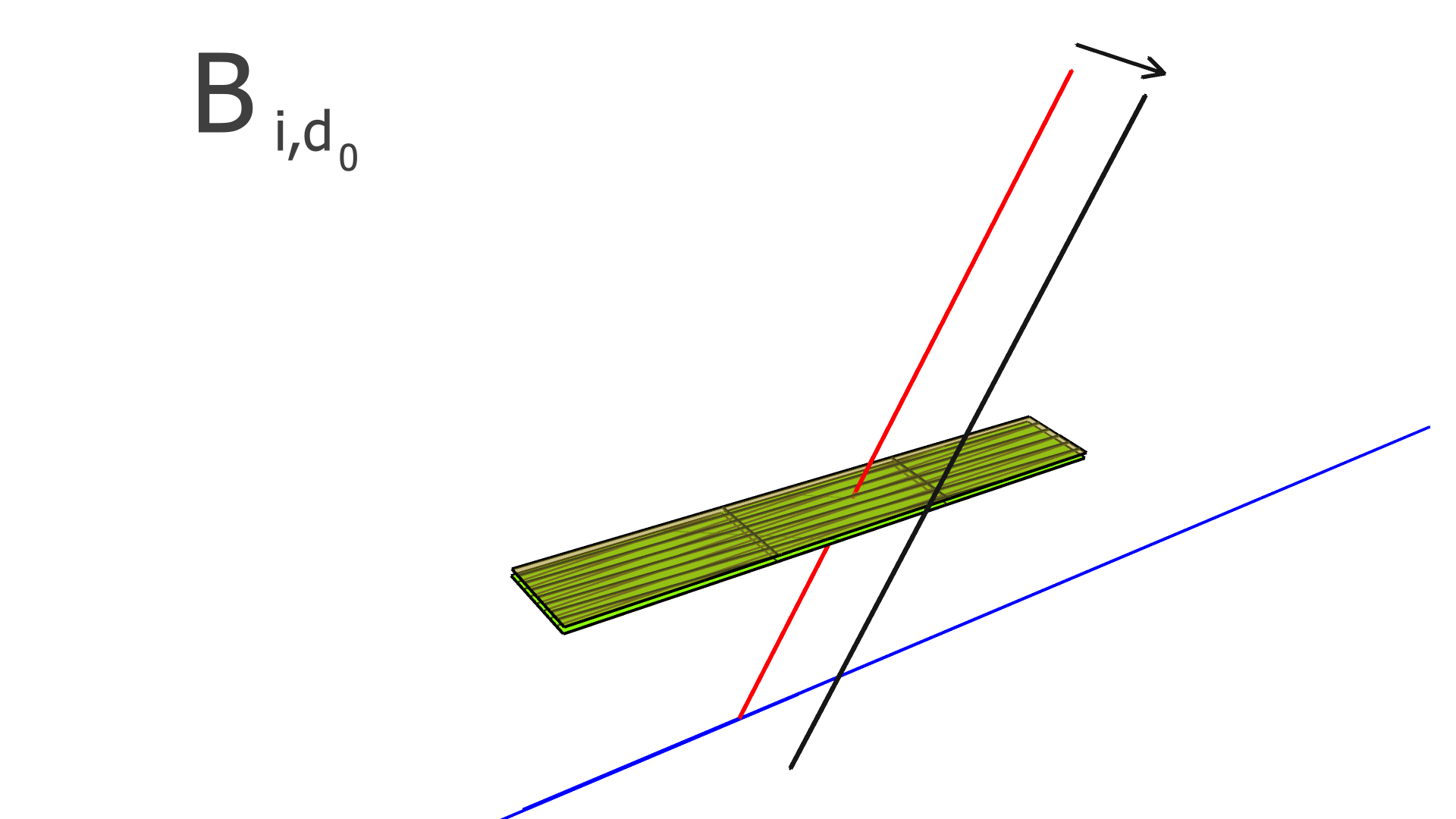}\includegraphics[width=\customwidth]{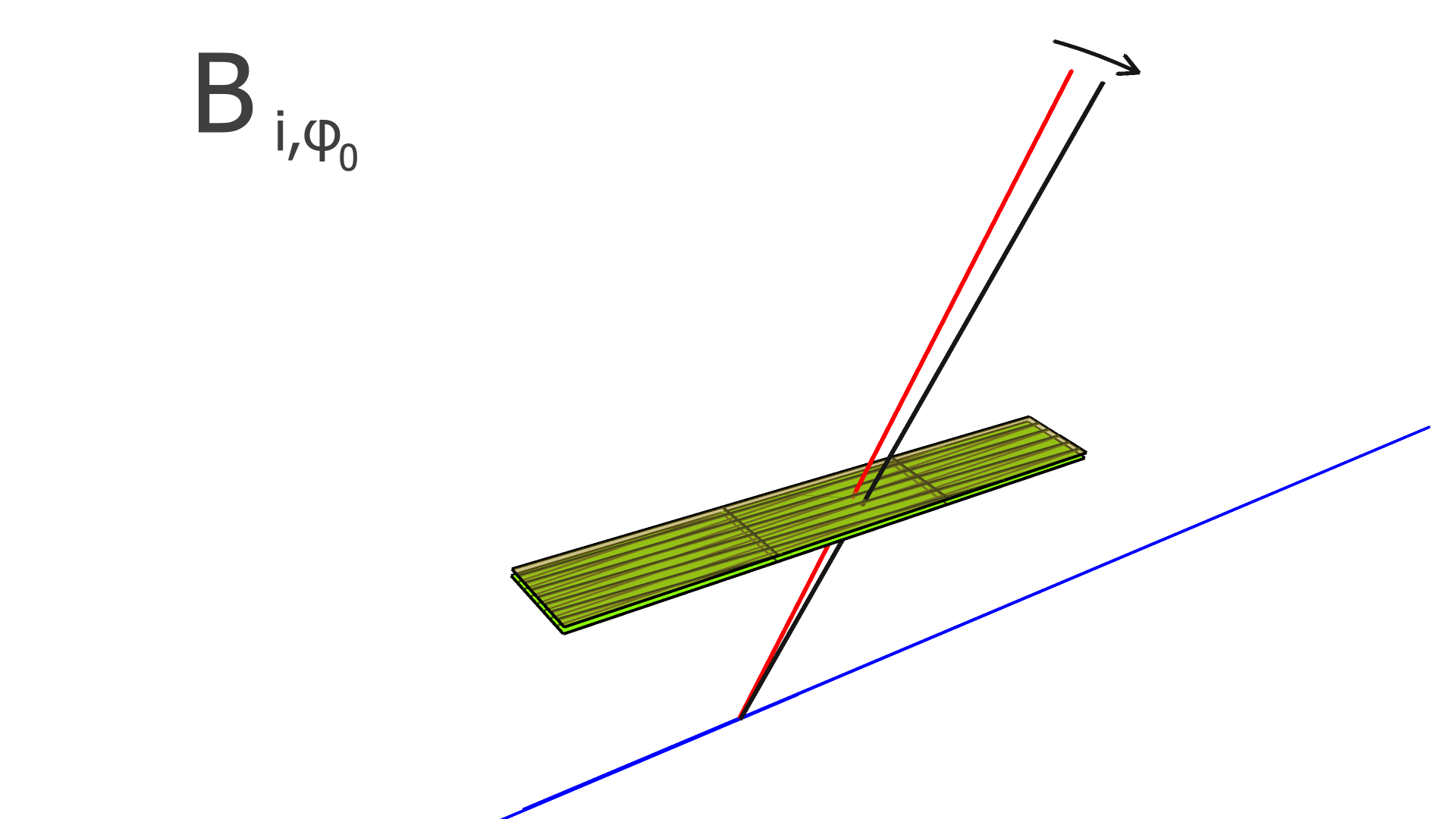}
    
    \includegraphics[width=\customwidth]{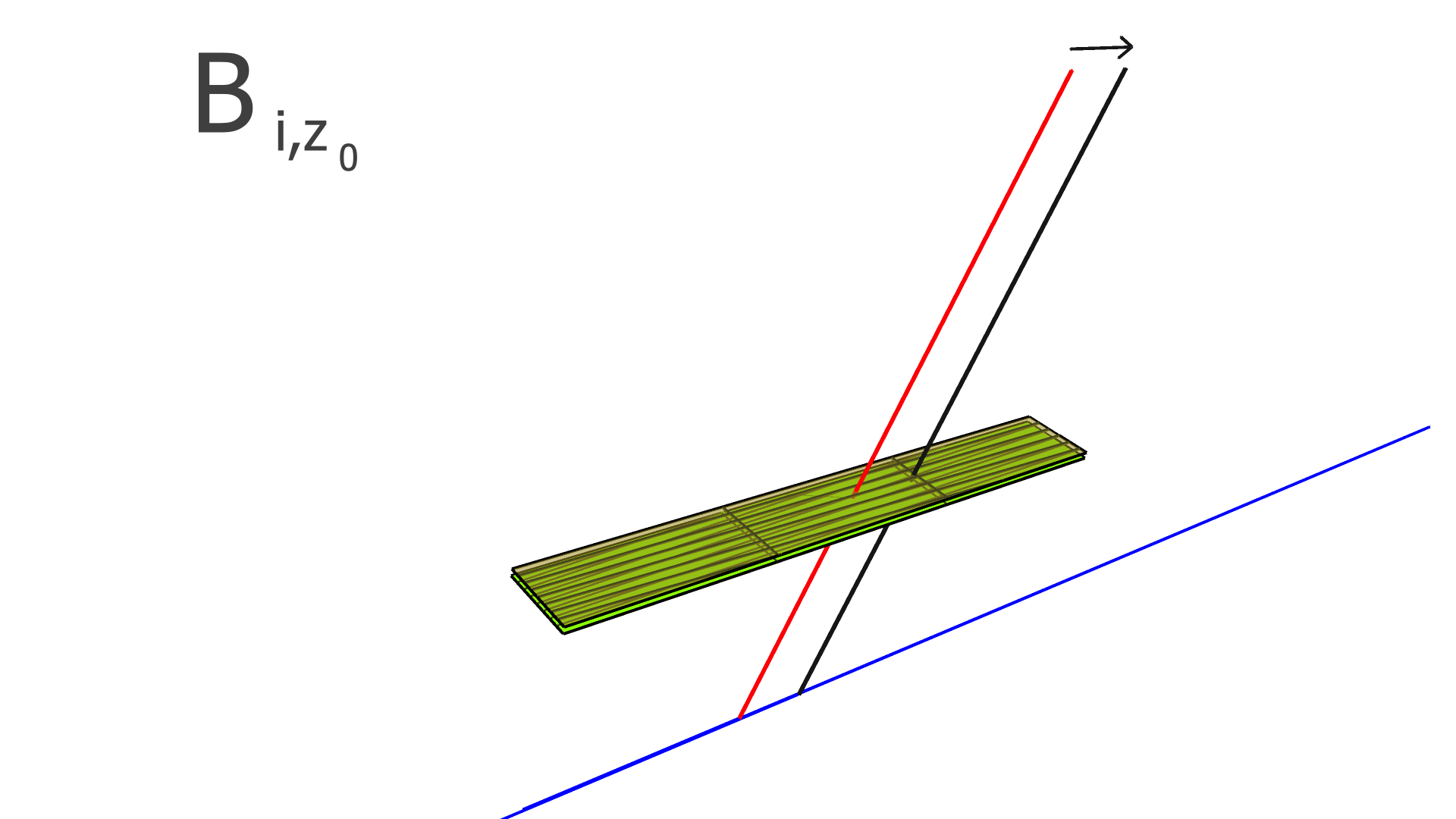}\includegraphics[width=\customwidth]{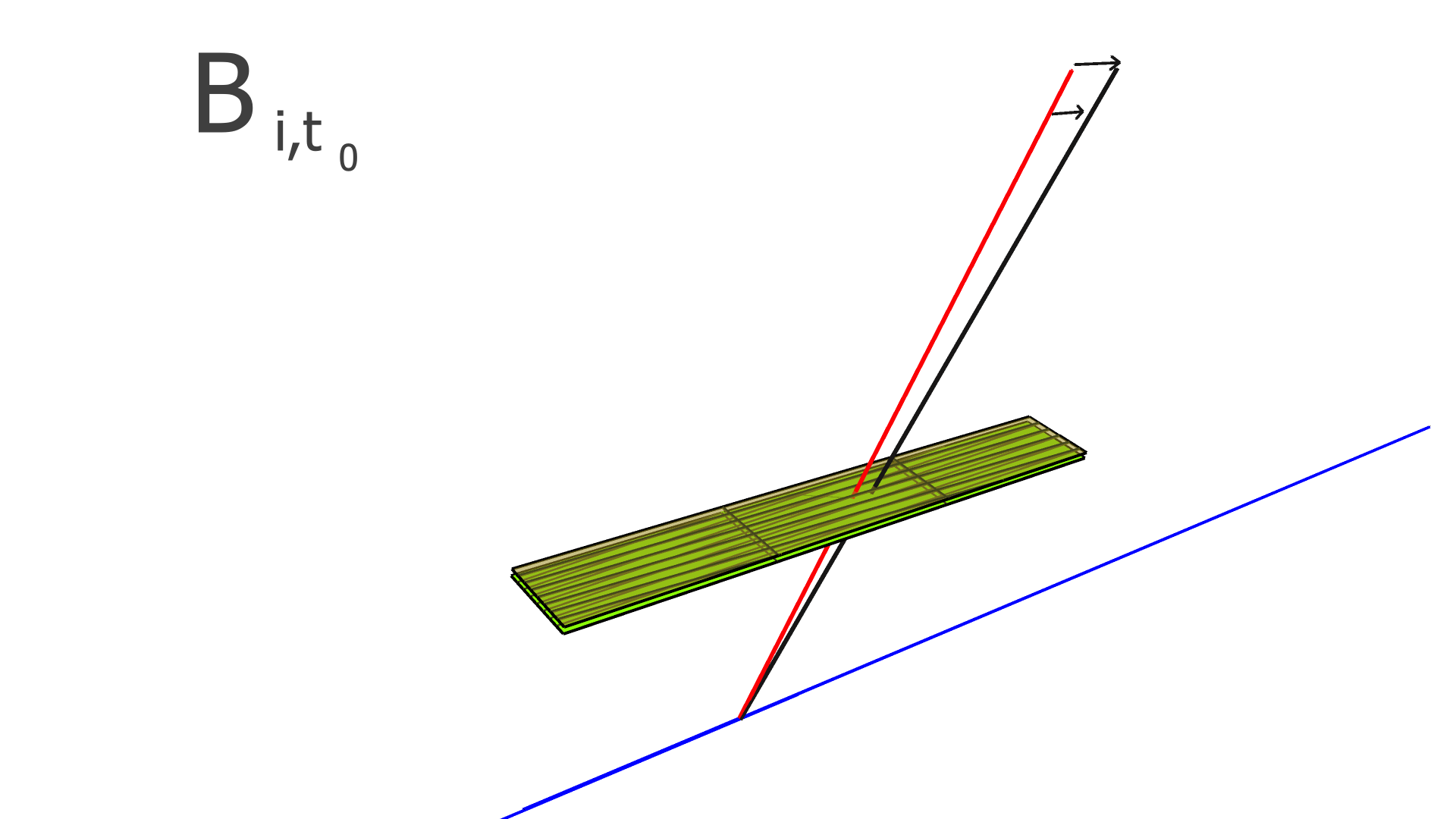}
    \caption{Illustration of matrix elements corresponding to translation degrees of freedom (top three panels, left column), rotation degrees of freedom (right column, top three panels), and variation in track parameters (bottom 2 rows) for one SVT module.  The reference trajectory is shown in red, and the beamline is shown in blue.  For the elements of the track-derivative matrix, $\mathbf B$, we show in black the trajectories with the indicated track parameter varied from the reference values.}
    \label{fig:deriv_SVT}
\end{figure*}

\begin{figure*}
    \centering
    \includegraphics[width=\customwidth]{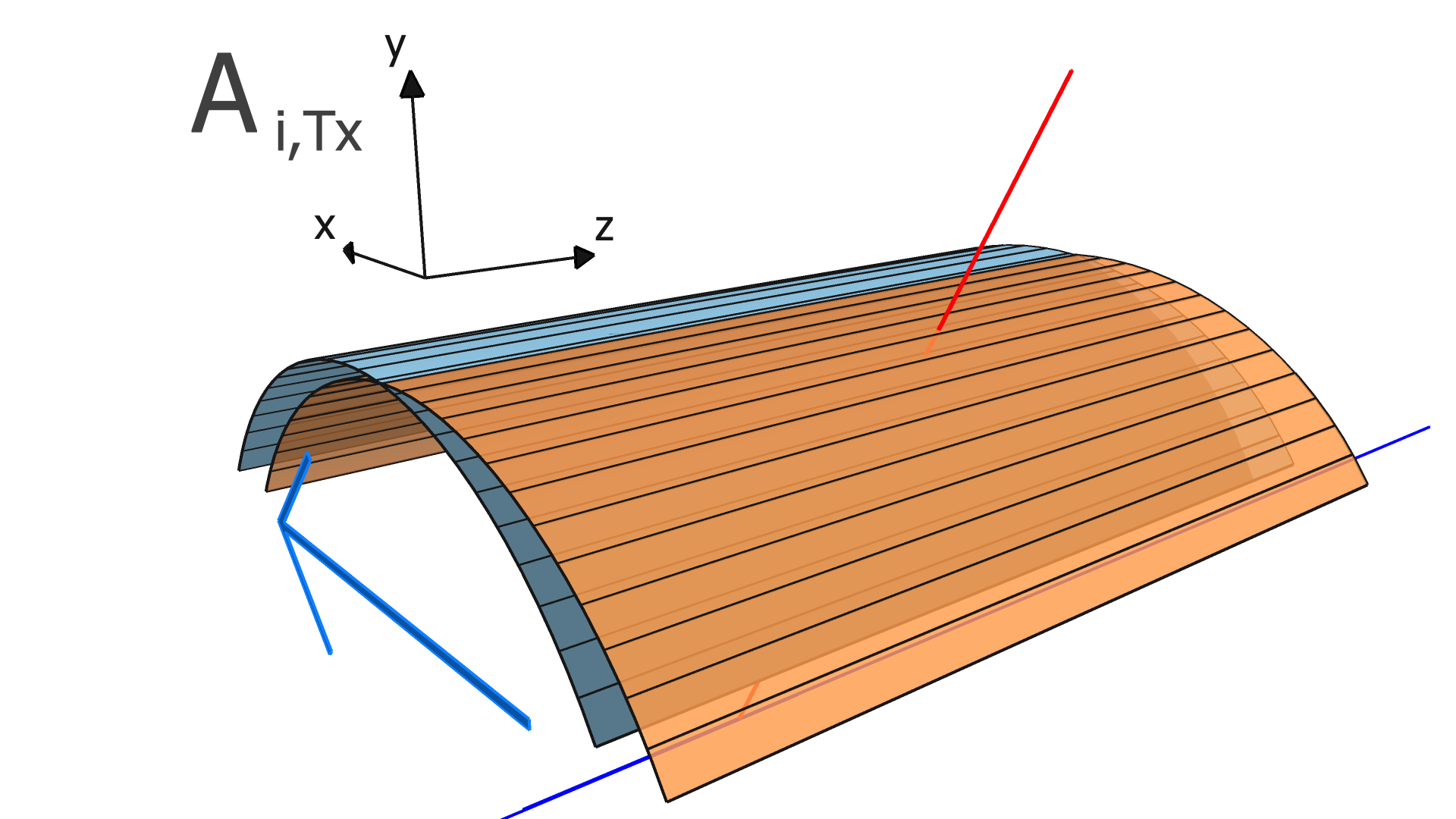}\includegraphics[width=\customwidth]{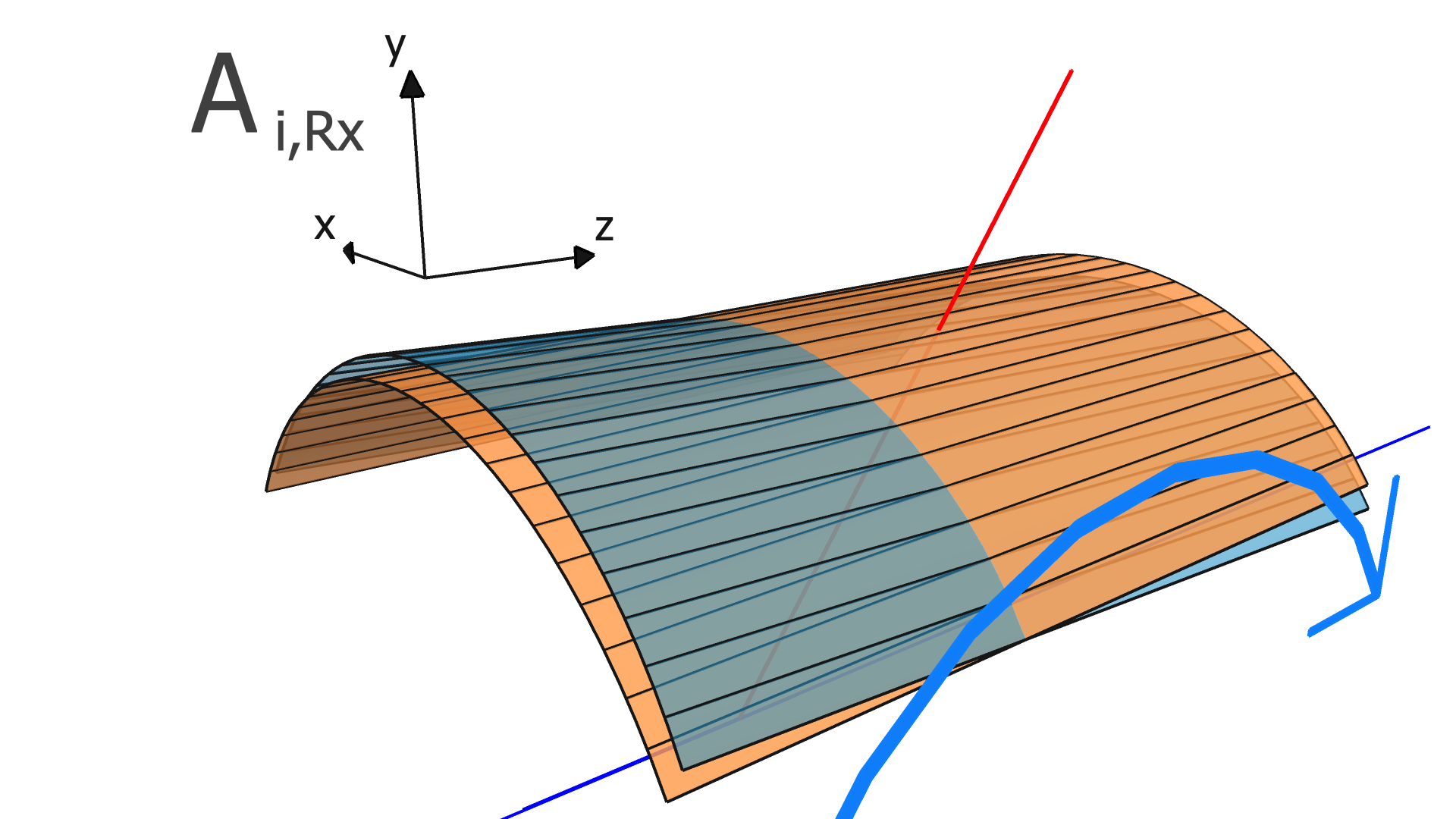}
    
    \includegraphics[width=\customwidth]{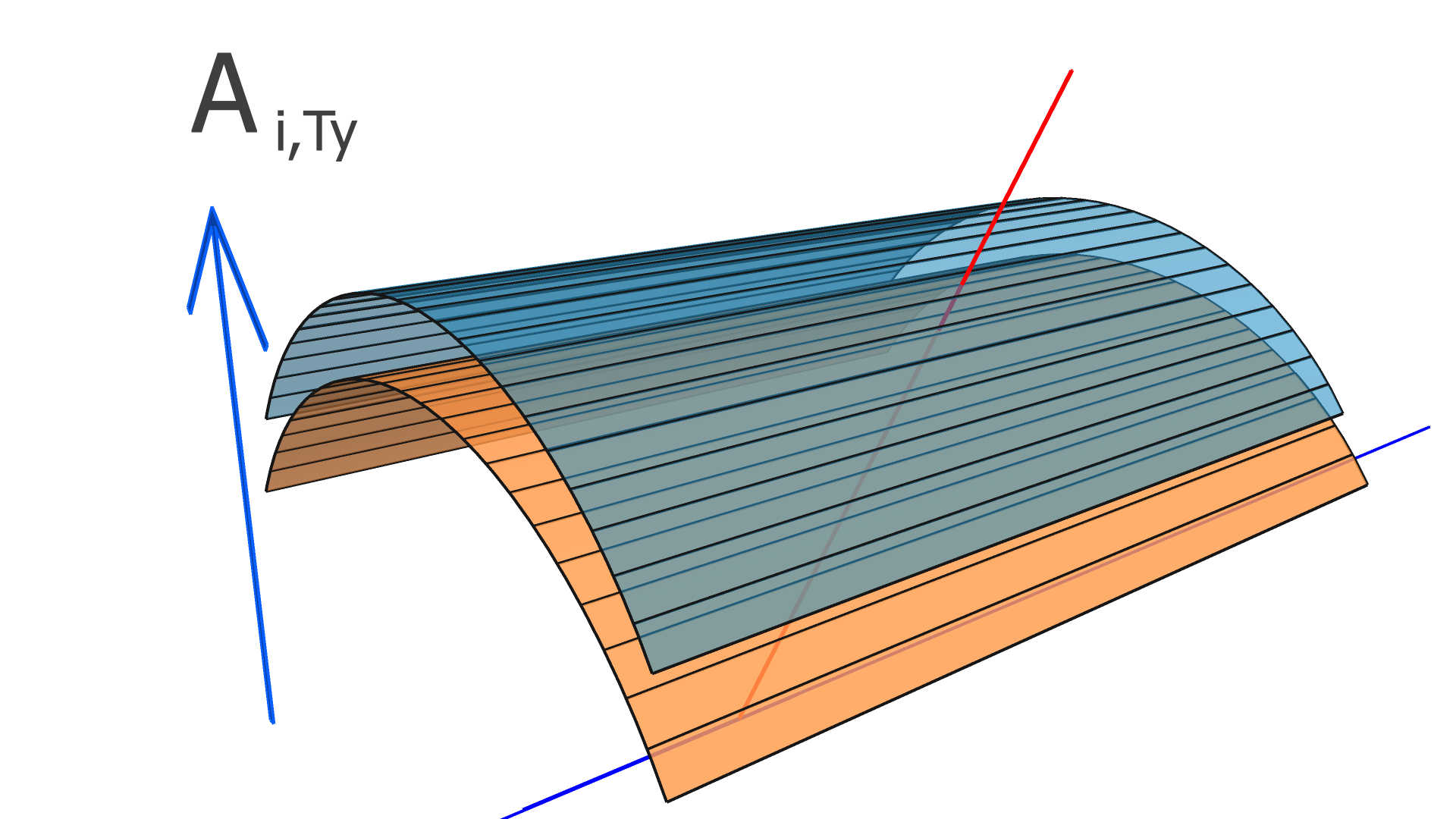}\includegraphics[width=\customwidth]{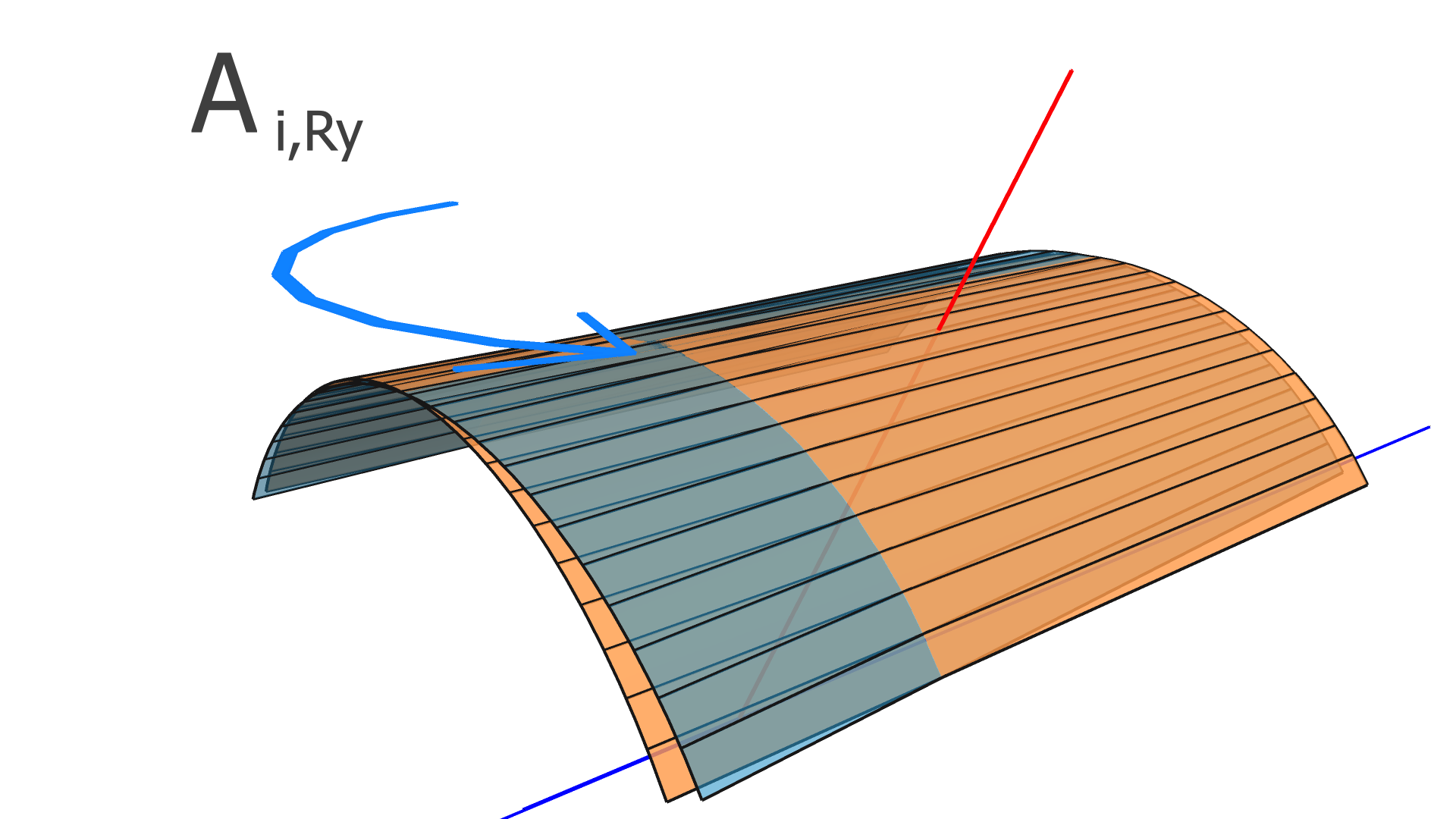}
    
    \includegraphics[width=\customwidth]{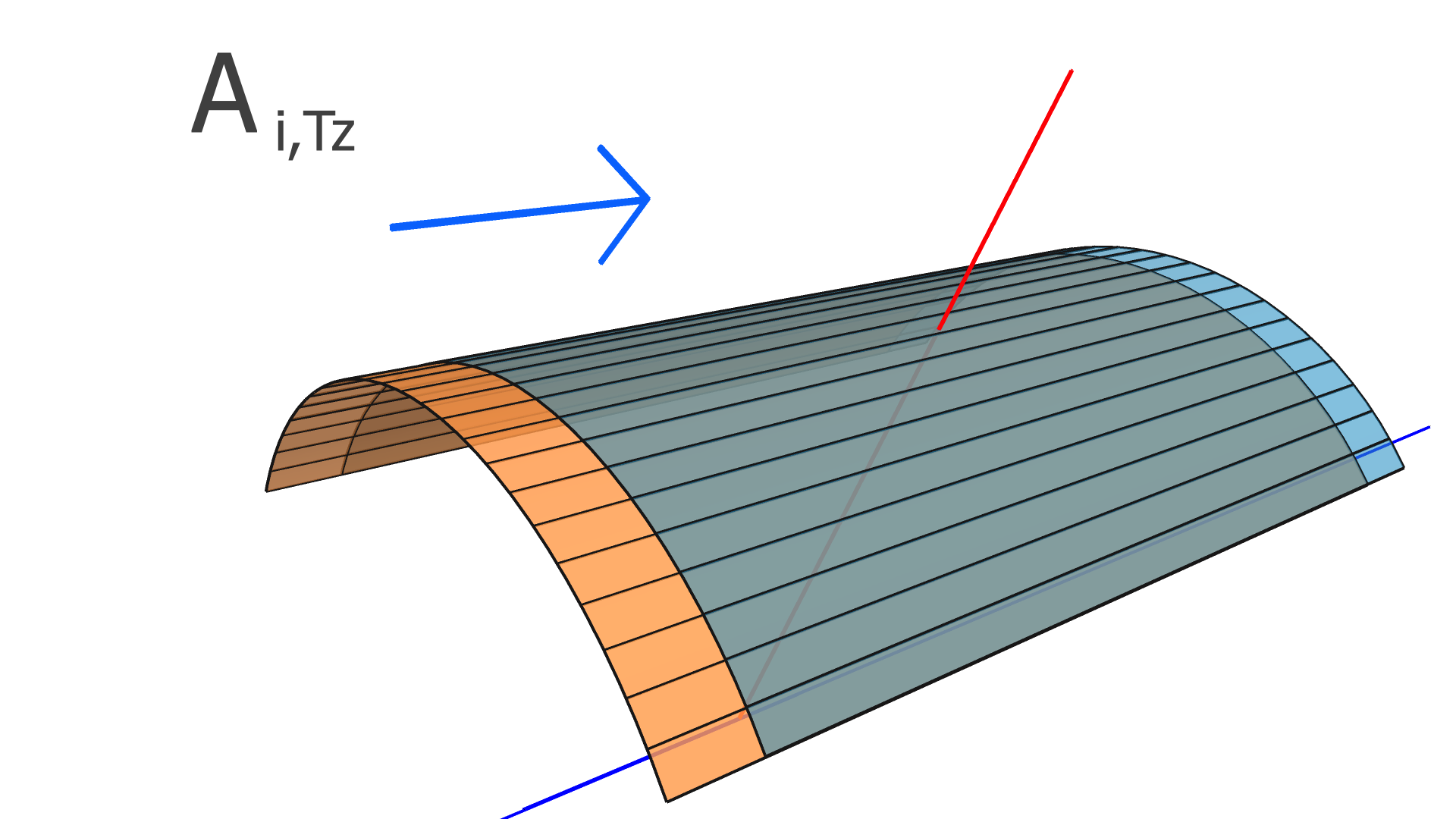}\includegraphics[width=\customwidth]{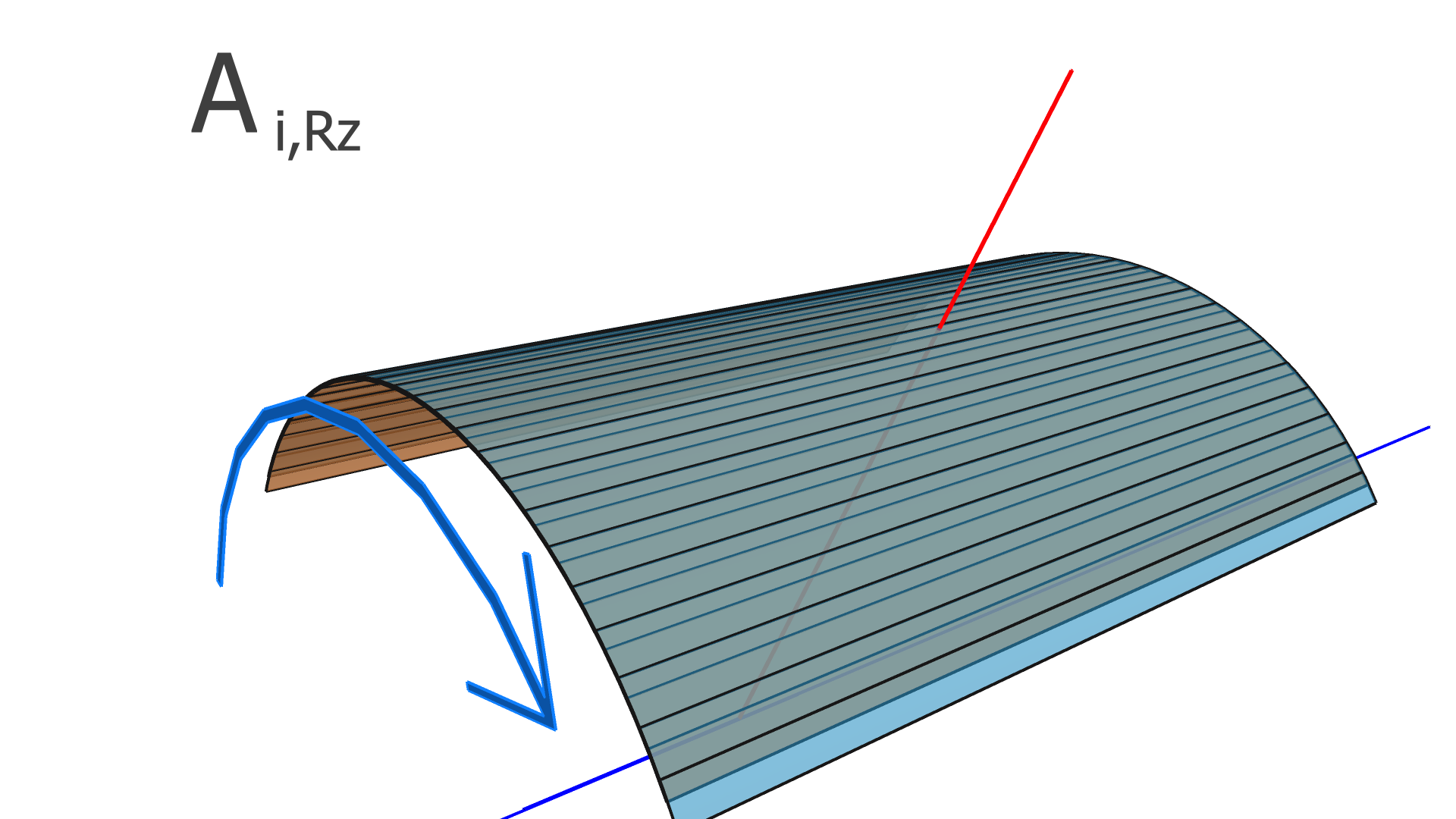}
    
    \includegraphics[width=\customwidth]{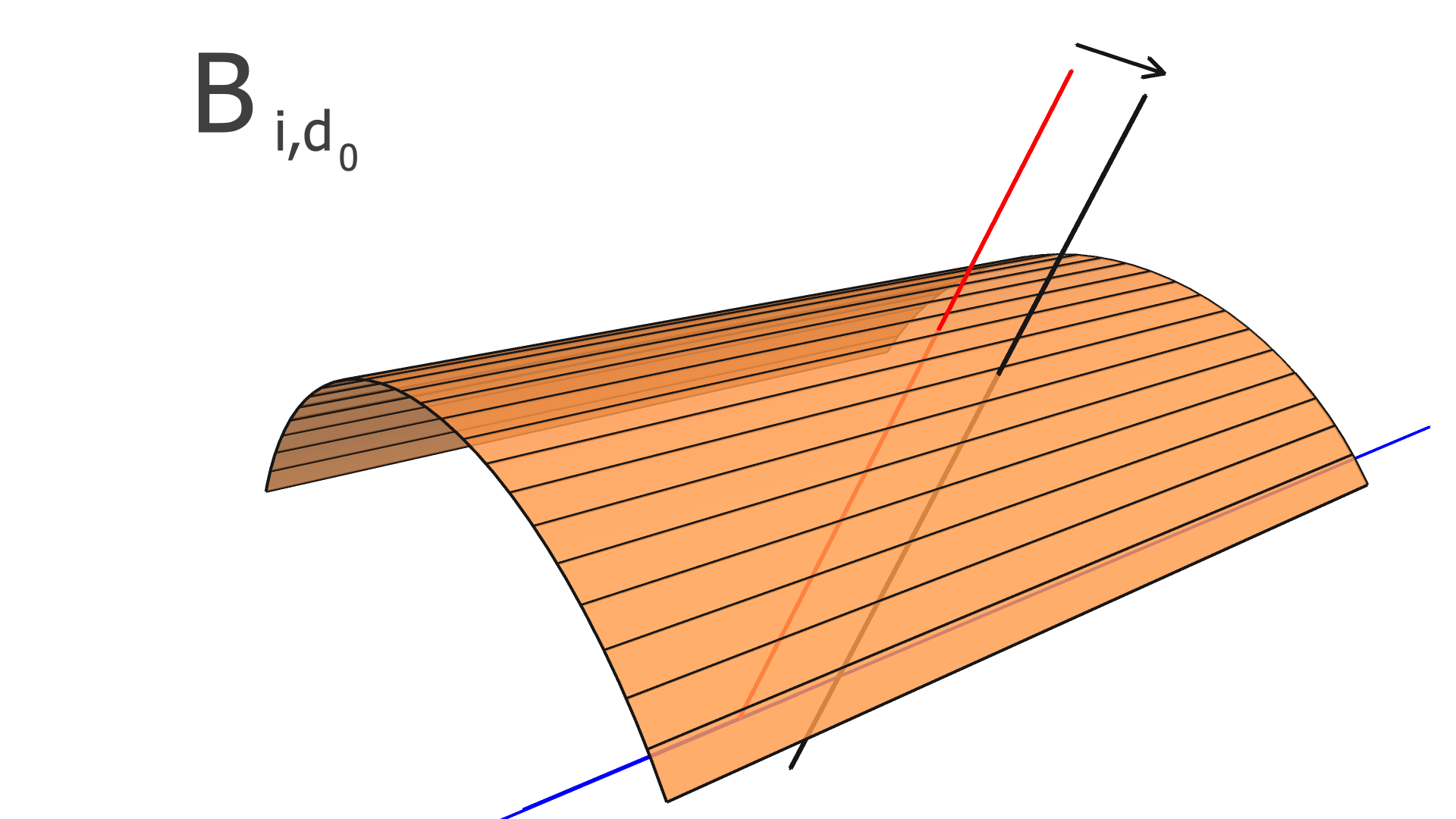}\includegraphics[width=\customwidth]{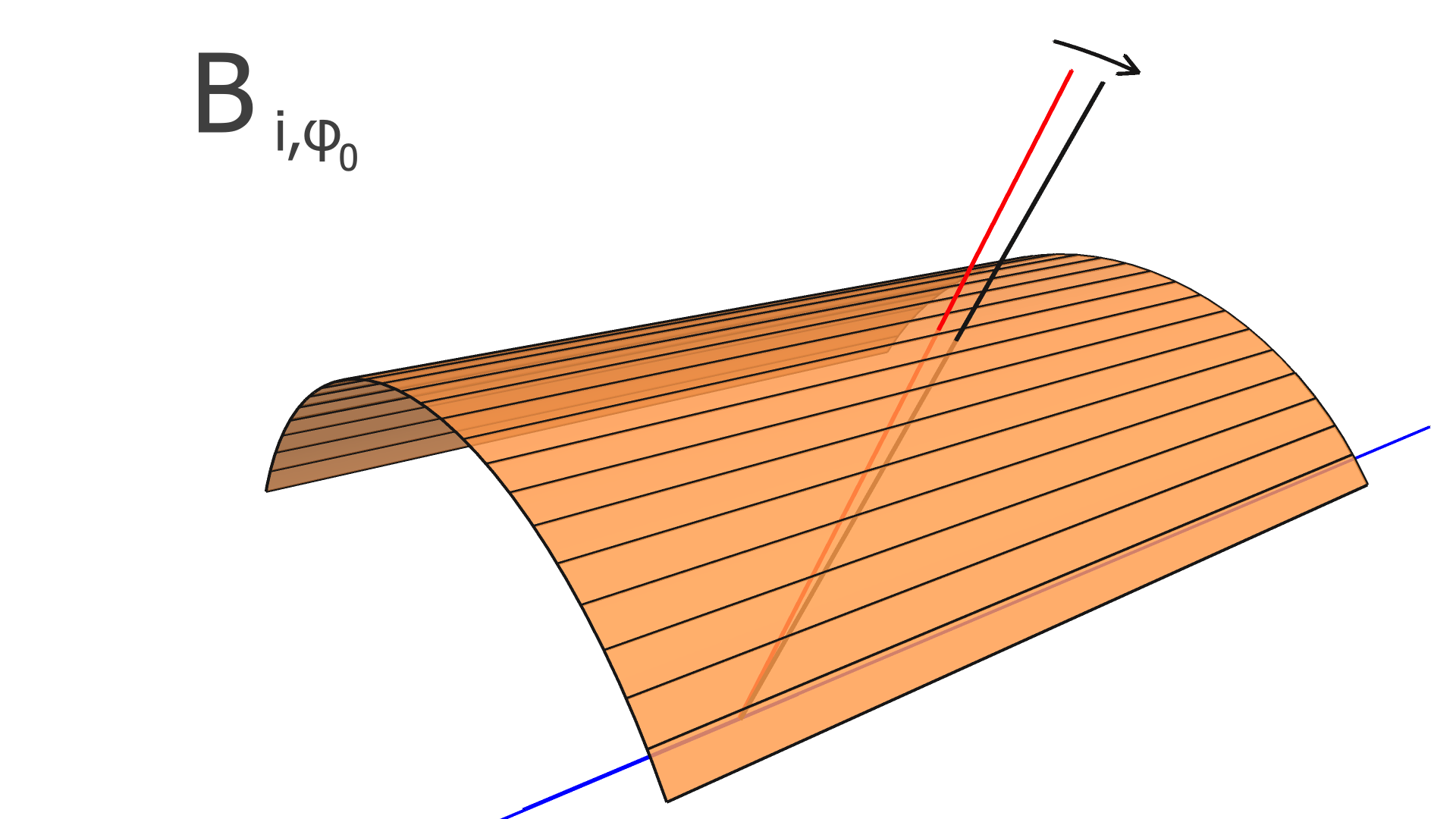}
    
    \includegraphics[width=\customwidth]{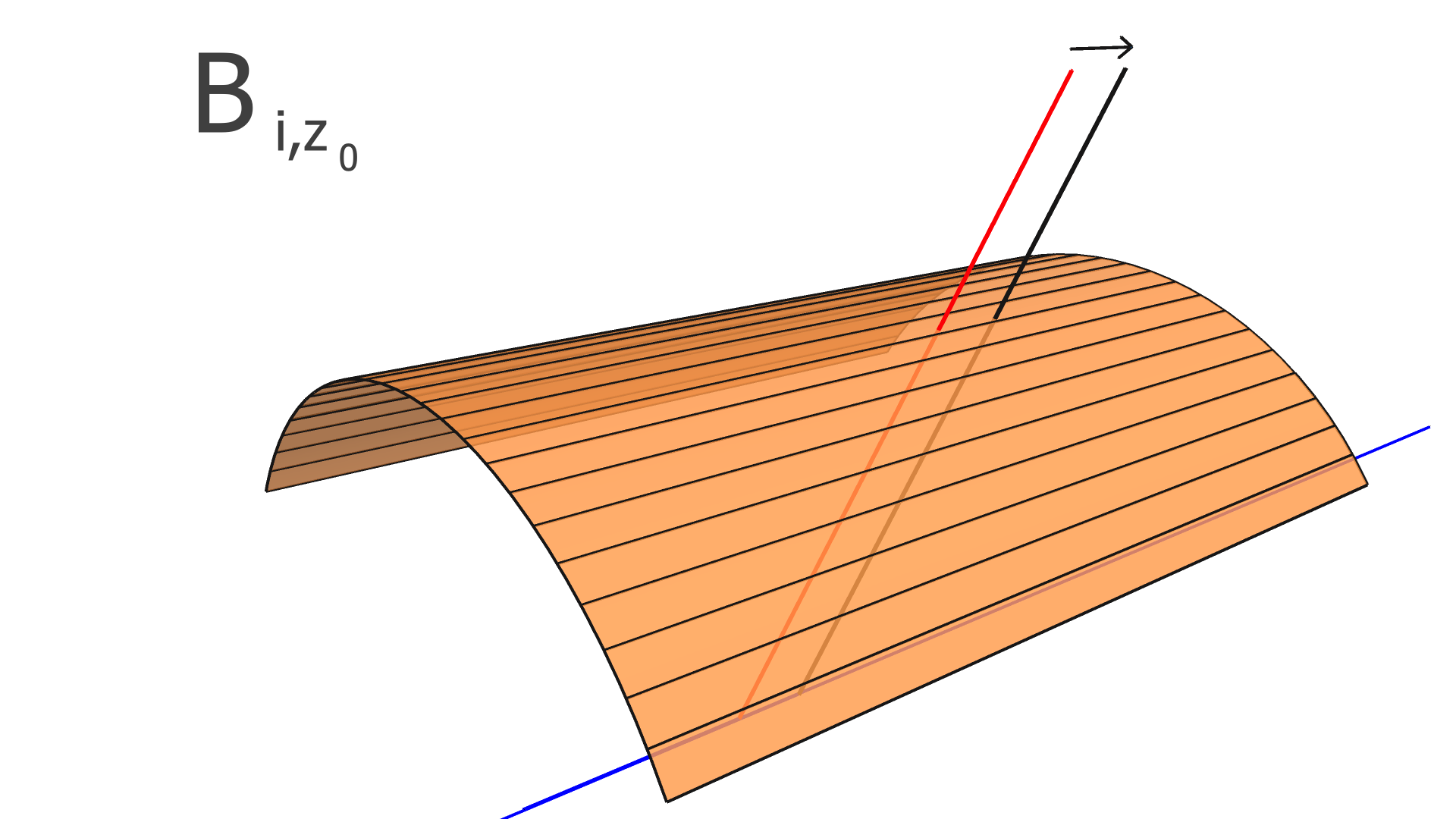}\includegraphics[width=\customwidth]{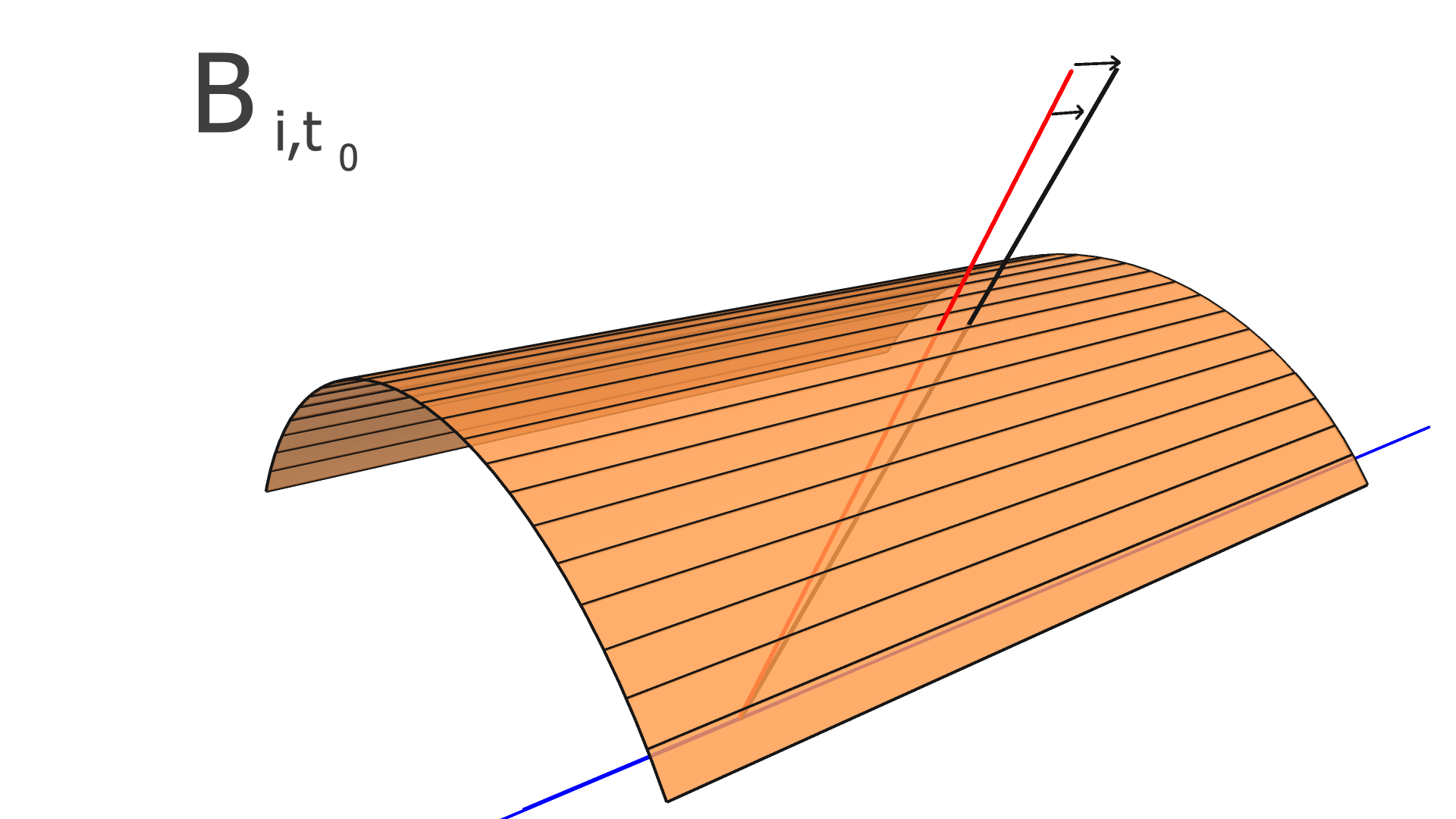}
    \caption{Illustration of matrix elements corresponding to translation degrees of freedom (top three panels, left column), rotation degrees of freedom (right column, top three panels), and variation in track parameters (bottom 2 rows) for one BMTZ module. The reference trajectory is shown in red, and the beamline is shown in blue.  For the elements of the track-derivative matrix, $\mathbf B$, we show in black the trajectories with the indicated track parameter varied from the reference values.}
    \label{fig:deriv_BMTZ}
\end{figure*}

\begin{figure*}
    \centering
    \includegraphics[width=\customwidth]{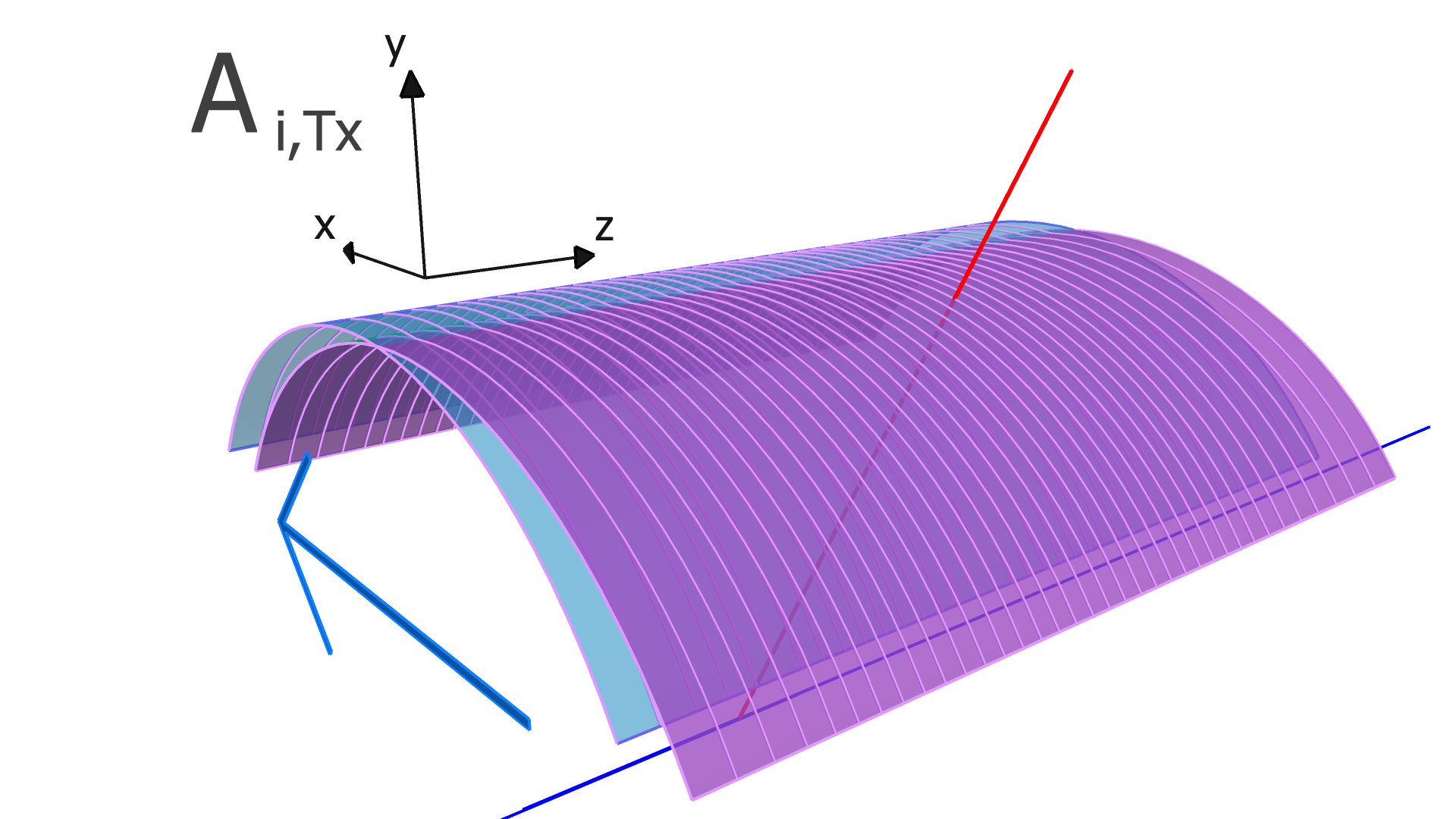}\includegraphics[width=\customwidth]{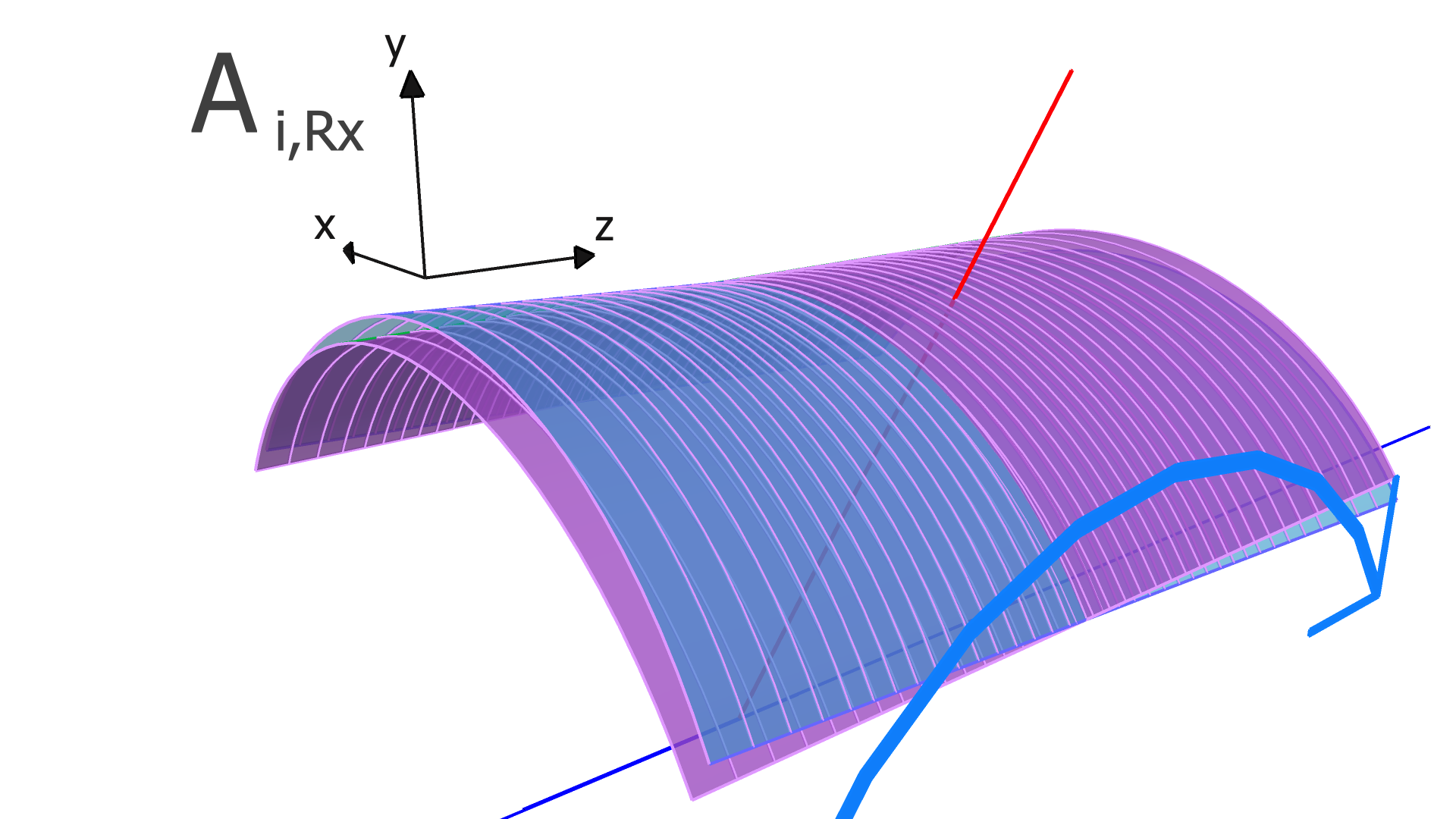}
    
    \includegraphics[width=\customwidth]{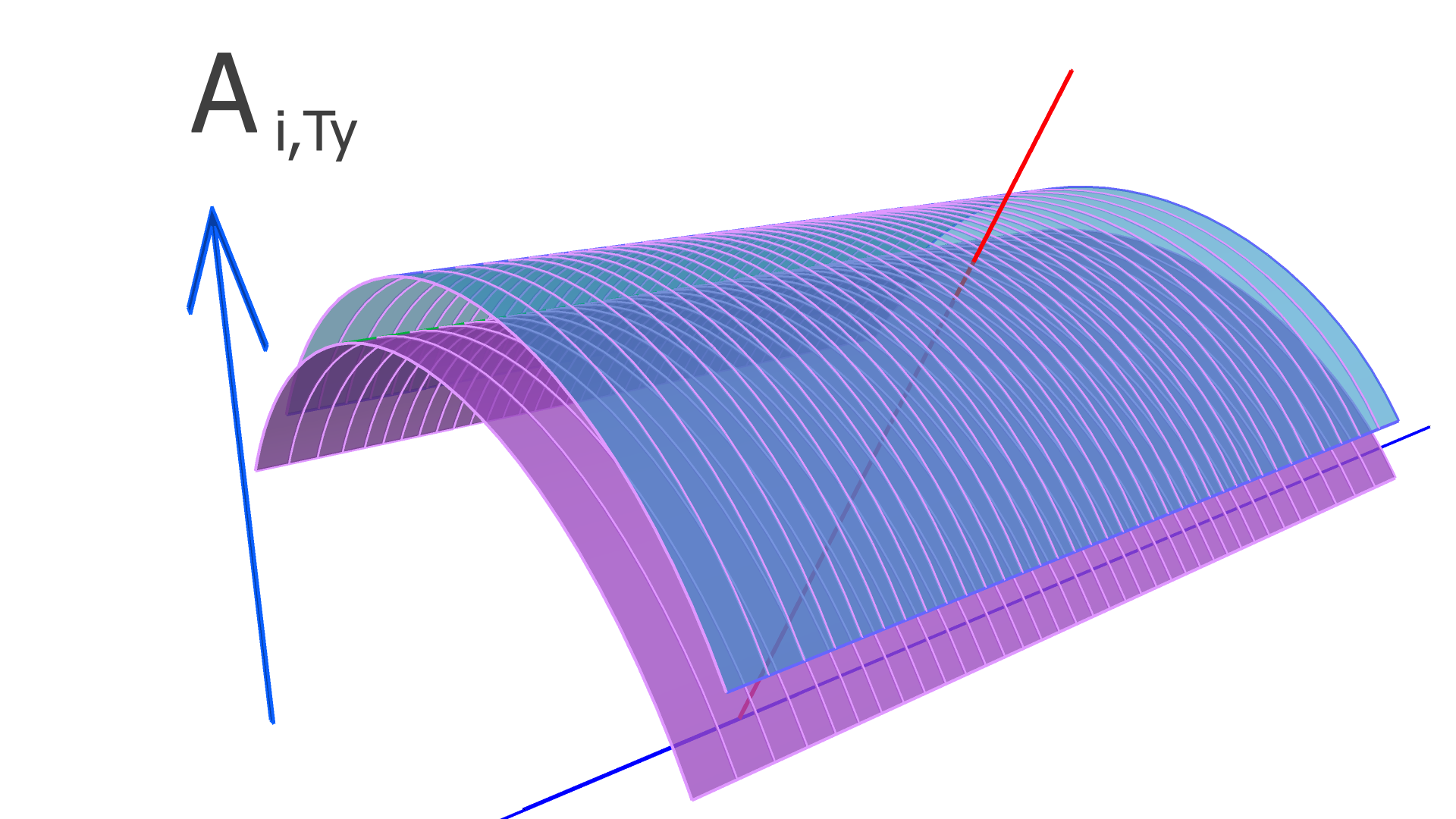}\includegraphics[width=\customwidth]{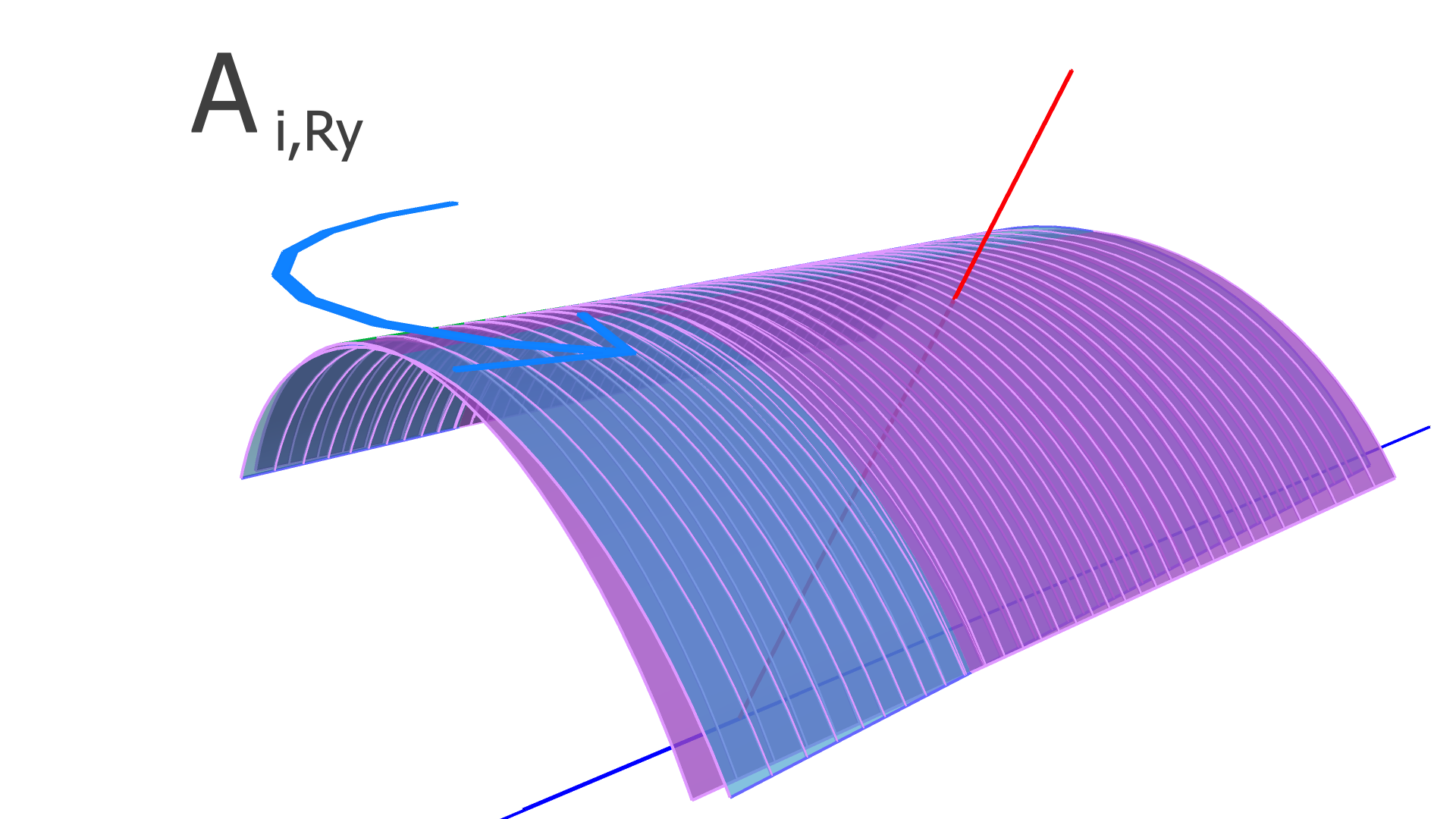}
    
    \includegraphics[width=\customwidth]{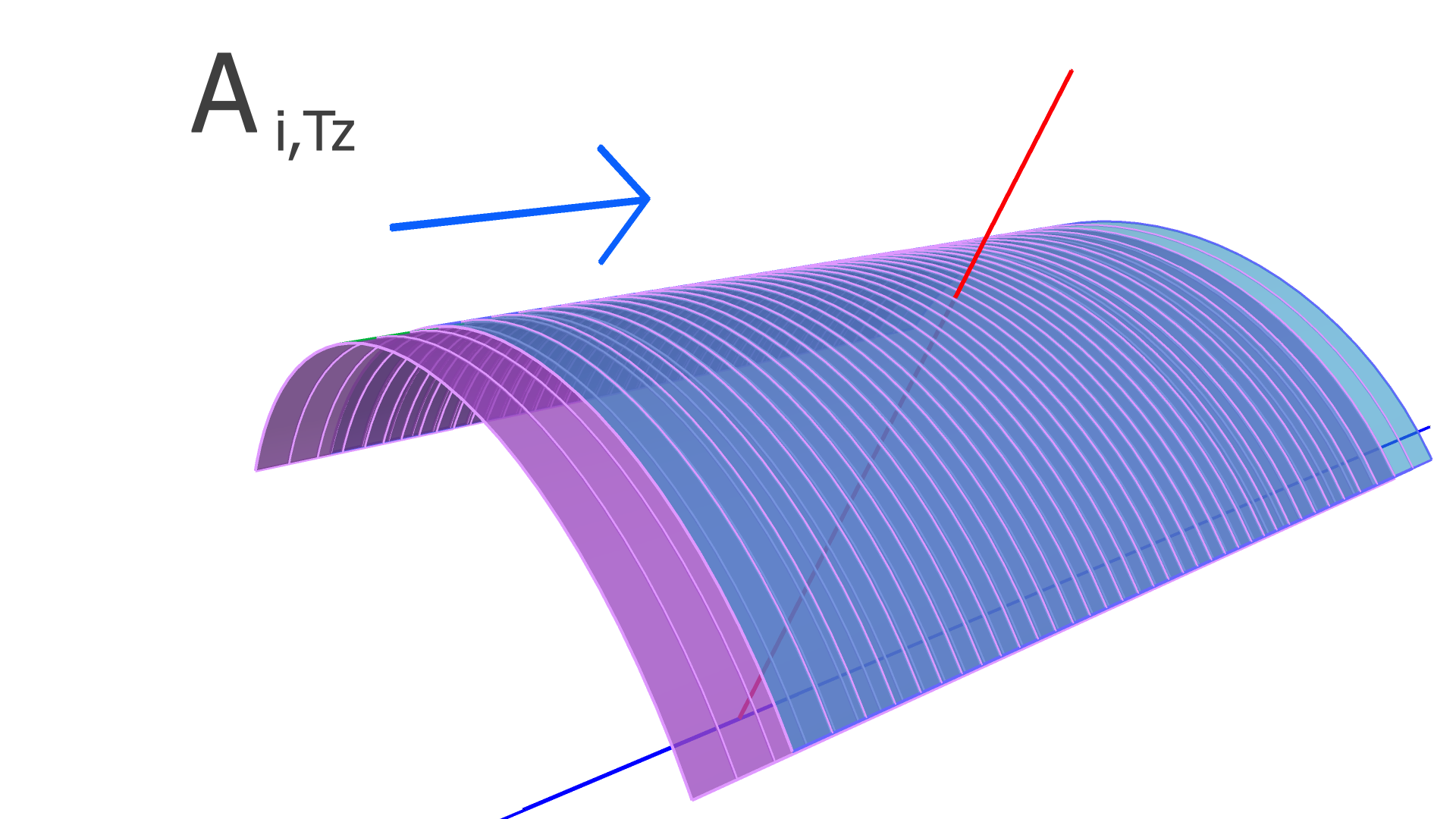}\includegraphics[width=\customwidth]{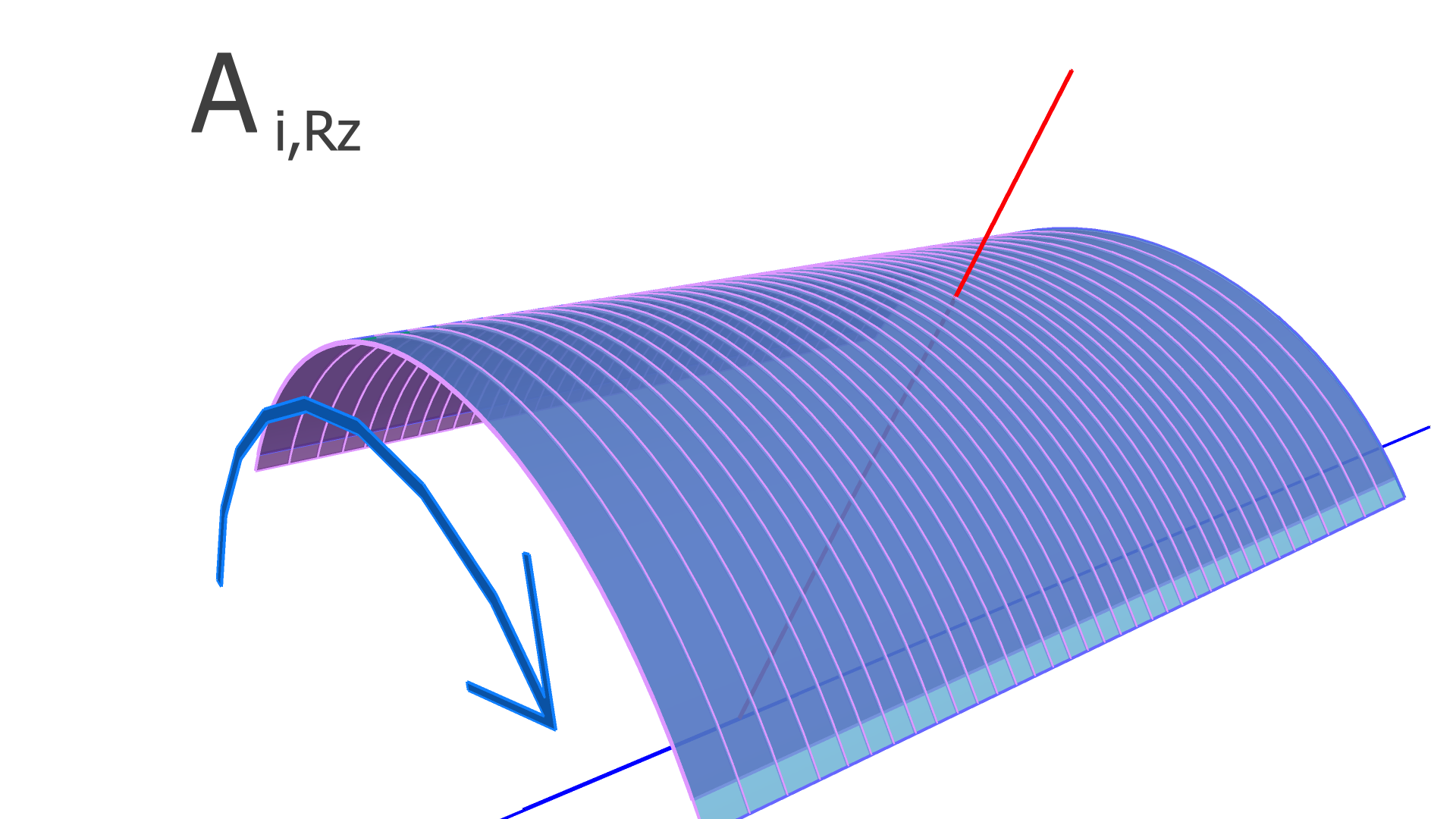}
    
    \includegraphics[width=\customwidth]{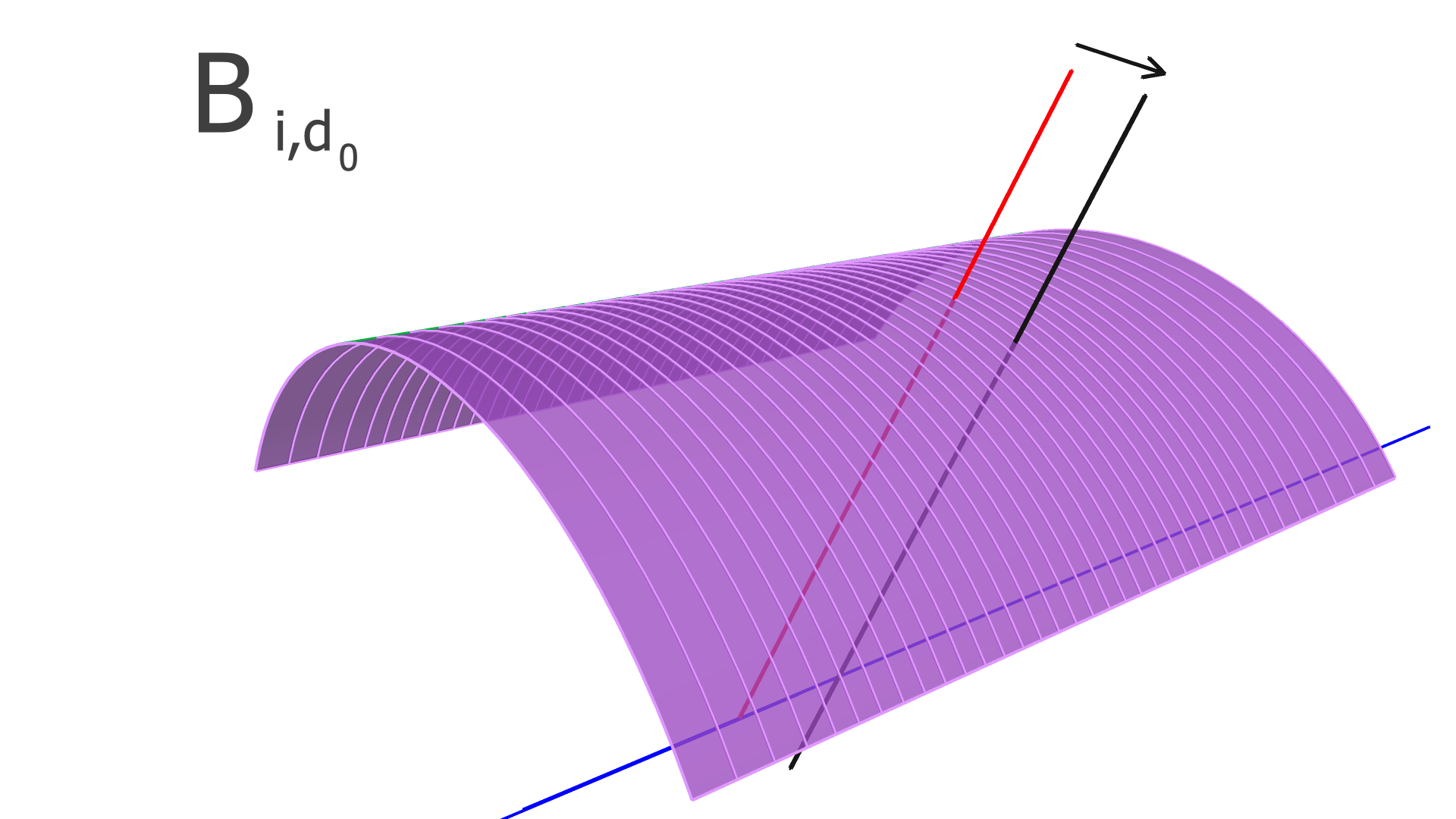}\includegraphics[width=\customwidth]{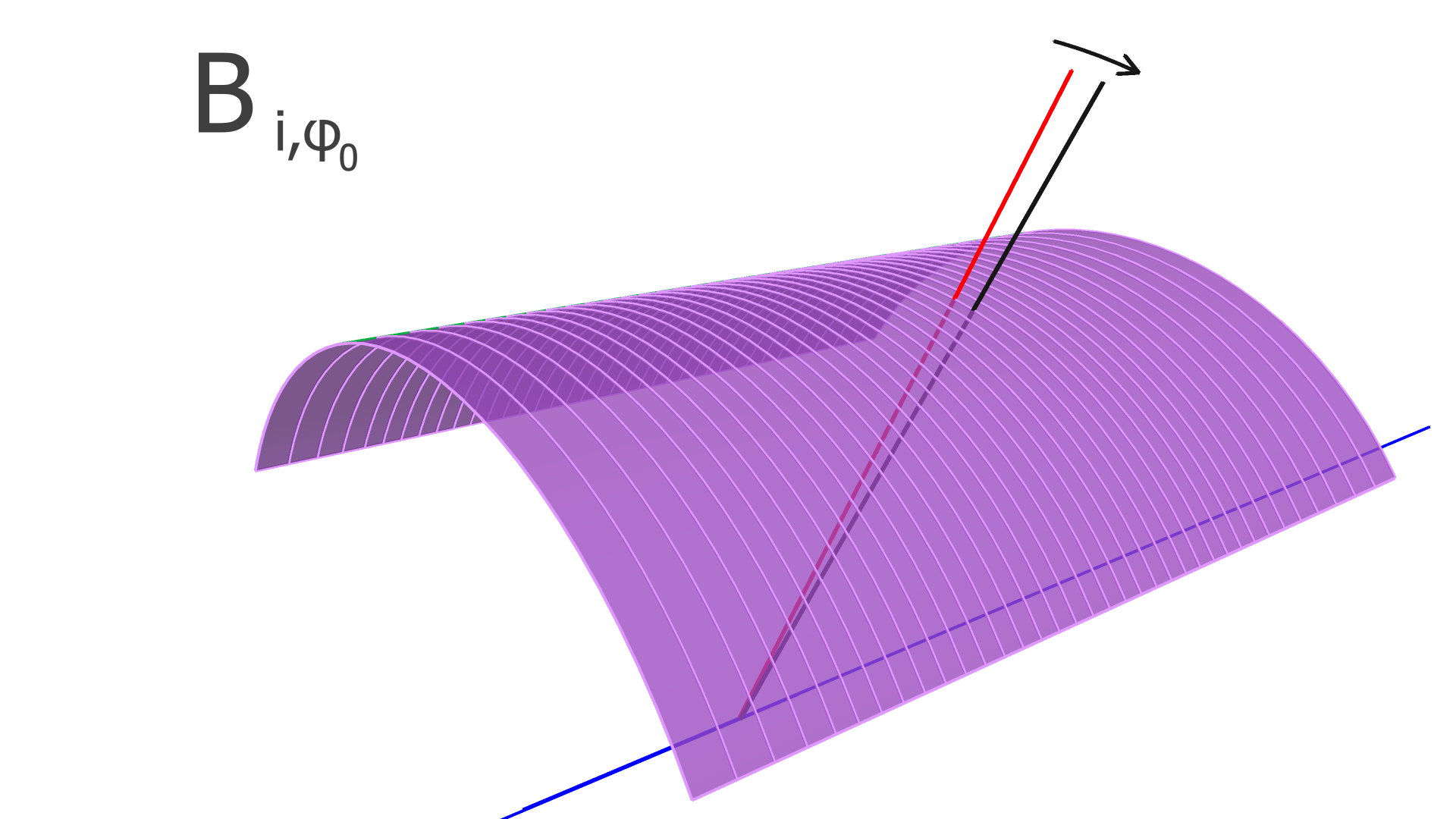}
    
    \includegraphics[width=\customwidth]{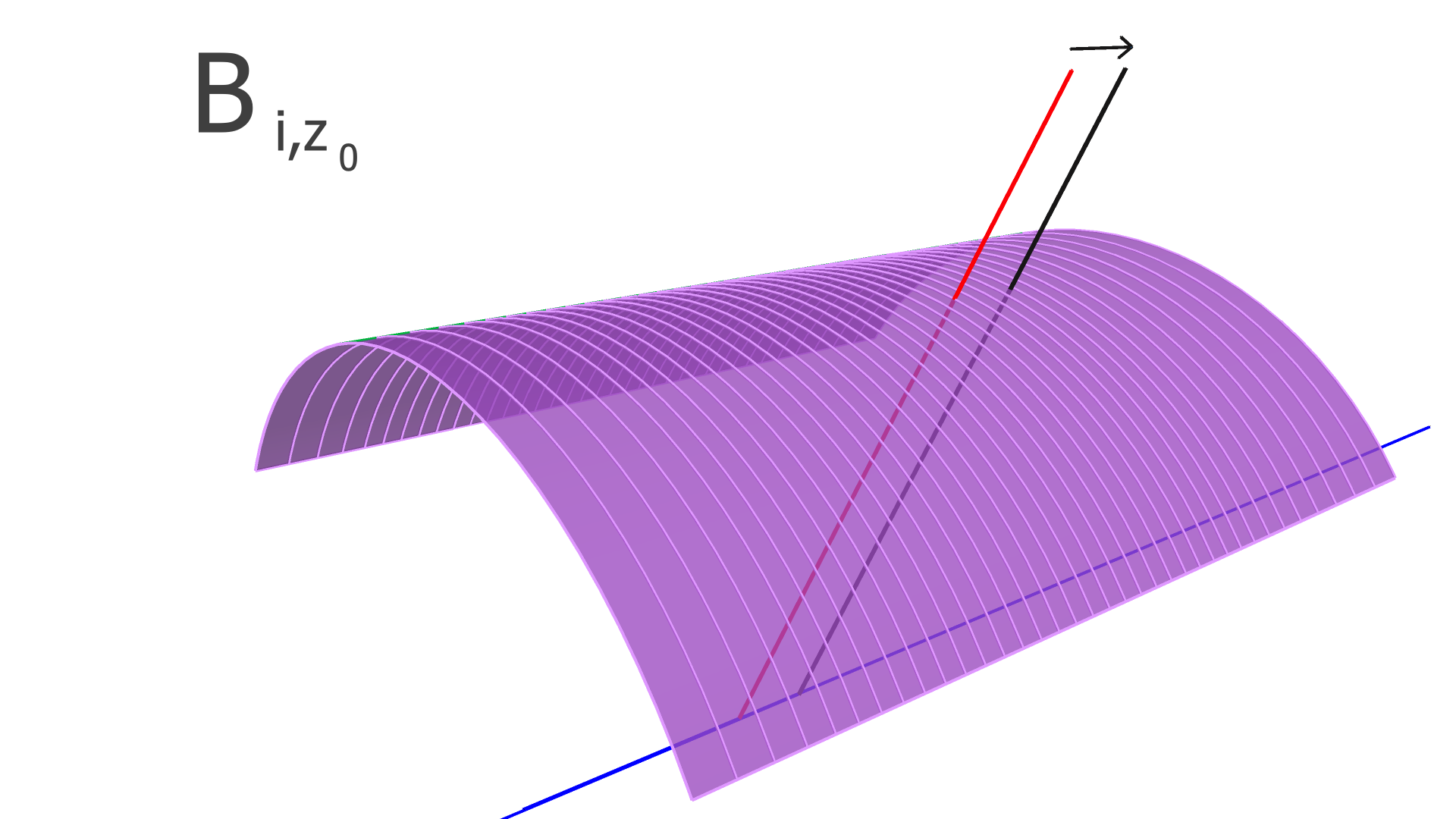}\includegraphics[width=\customwidth]{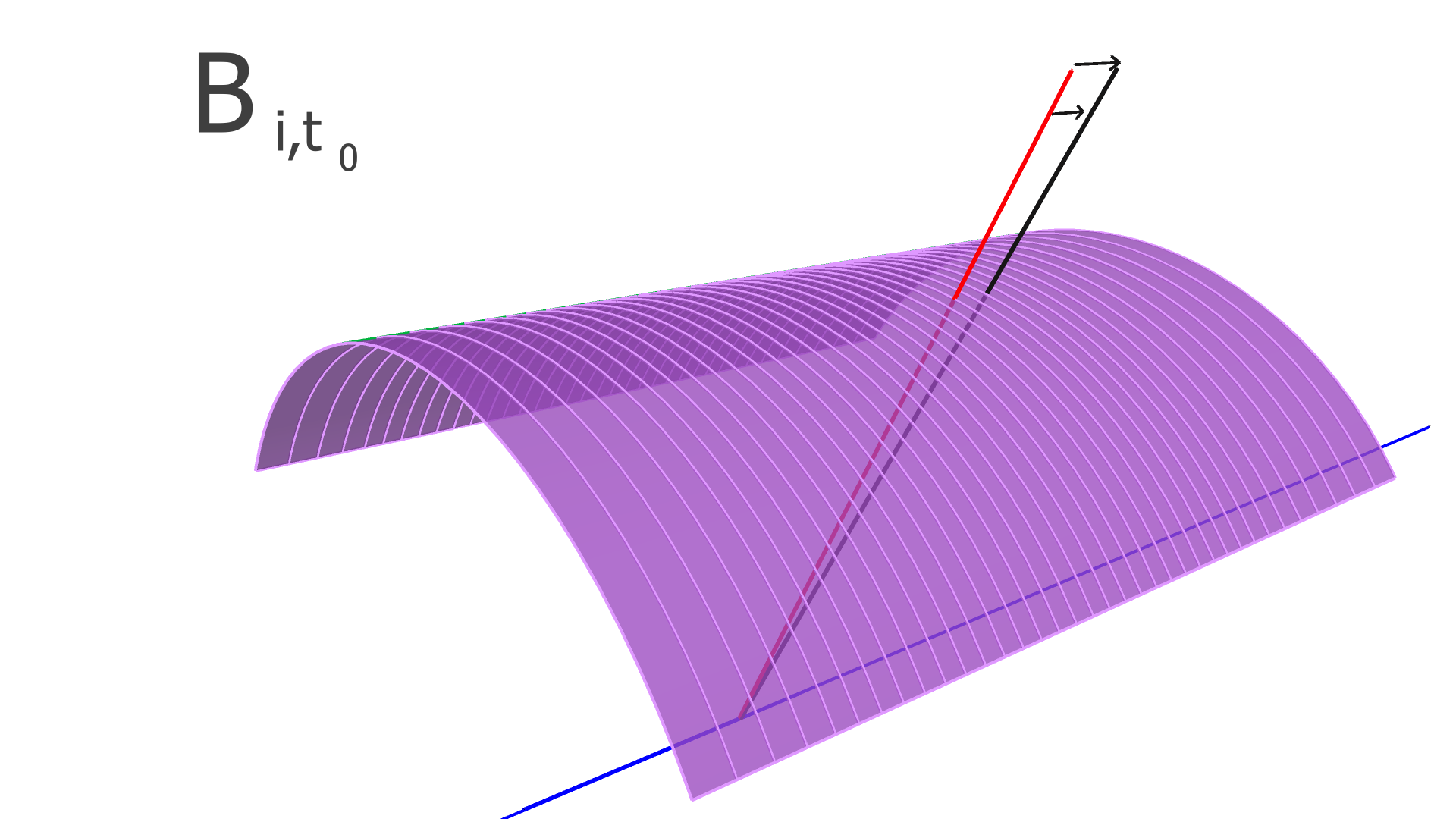}
    \caption{Illustration of matrix elements corresponding to translation degrees of freedom (top three panels, left column), rotation degrees of freedom (right column, top three panels), and variation in track parameters (bottom 2 rows) for one BMTC module.  The reference trajectory is shown in red, and the beamline is shown in blue.  For the elements of the track-derivative matrix, $\mathbf B$, we show in black the trajectories with the indicated track parameter varied from the reference values.}
    \label{fig:deriv_BMTC}
\end{figure*}

Following Refs.~\cite{Widl:2006mz,Widl:2008aqa}, KAA loops through all of the tracks, and updates $\mathbf d$ and $\mathbf{D}$ using Eqs.~\ref{eq:d_update}-\ref{eq:V_D_update} below (the derivations of these equations are beyond the scope of this article, and can be found in Refs.~\cite{Widl:2006mz,Widl:2008aqa}):
\begin{equation}
    \mathbf{d}' = \mathbf{d} + \mathbf{D}\mathbf{A}^{\rm T}\mathbf{G}\left(\mathbf{m}-\mathbf{c} -\mathbf{A}\mathbf{d}\right)
    \label{eq:d_update}
\end{equation}
and 
\begin{align}
    \mathbf{D}' = \left(\mathbf{I}-\mathbf{D}\mathbf{A}^{\rm T}\mathbf{G}\mathbf{A}\right)&\mathbf{D}\left(\mathbf{I}-\mathbf{A}^{\rm T}\mathbf{G}\mathbf{A}\mathbf{D}\right)\nonumber\\&+\mathbf{D}\mathbf{A}^{\rm T}\mathbf{GVGAD},
    \label{eq:D_update}
\end{align}
where 
\begin{equation}
    \mathbf{G} = \mathbf{V}^{-1}_D-\mathbf{V}^{-1}_D\mathbf{B}\left(\mathbf{B}^{\rm T}\mathbf{V}^{-1}_D \mathbf{B}\right)^{-1}\mathbf{B}^{\rm T}\mathbf{V}^{-1}_D
\end{equation}
and 
\begin{equation}
    \mathbf{V}_D = \mathbf{V}+\mathbf{ADA}^{\rm T},
    \label{eq:V_D_update}
\end{equation}
and $\mathbf I$ is the identity matrix of the same dimensions as $\mathbf D$.  The matrix $\mathbf V_D$ can be interpreted as sum of the covariance of the residuals due to measurement uncertainty and the covariance due to the alignment uncertainty.  $\mathbf G$ can be interpreted as a projection of the inverse of $\mathbf V_D$ such that ${\mathbf G}{\mathbf B} = {\mathbf B}^{\rm T}{\mathbf G}=0$, in order to remove bias.    

For some types of detector geometries, including that of the CVT, the residuals may depend non-linearly on the alignment parameters and/or the track parameters.  Such non-linearity can lead to a systematic bias in the alignment parameters obtained by the KAA.  We found that multiple iterations of the KAA, alternating with reiterations of the event reconstruction with the updated alignment parameters,  are necessary in order to converge on a non-biased set of alignment parameters.  This differs from the use of KAA in CMS, where the exclusive use of parallel strips and planar sensors cause the residuals to depend linearly on the alignment parameters.  For CMS, only a single pass of the KAA was necessary \cite{Widl_2010}.

\section{Datasets}
\label{sec:data}
\begin{figure*}[h!]
    \centering
    \includegraphics[width=\columnwidth]{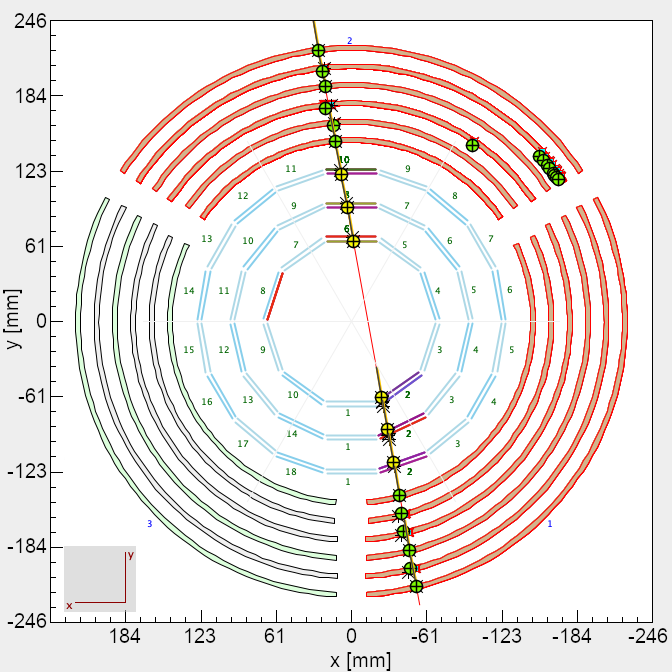}\includegraphics[width=\columnwidth]{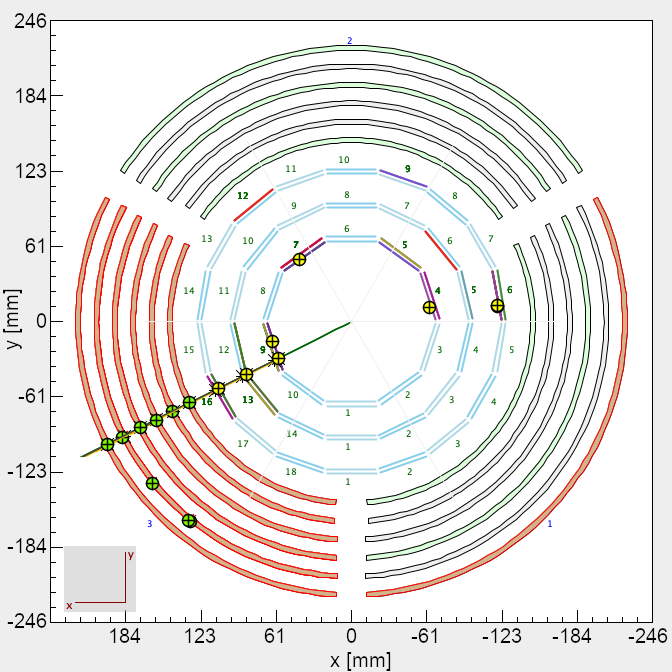}
    \caption{Example tracks from the ``cosmic'' (left) and ``field-off'' (right) configurations, as shown in the CLAS12 Event Display.  Units are mm.  BMT tiles that have been hit are outlined in red.  A yellow circle with a + represents a reconstructed crossing between pairs of clusters on the two sensors in the same SVT module.  The green circles represent the reconstructed position of BMT clusters (for BMTC, the azimuthal position is shown at the crossing of the track and the layer).  An asterisk is shown behind the circles at the position where the fitted track intersects the sensors.  The colors of the SVT sensors represent the ADC values of the hits on those sensors.}
    \label{fig:cosmic_and_field_off}
\end{figure*}

We used two special calibration runs taken in spring, 2019 during an experiment with a 10.6~GeV electron beam on a 5~cm liquid-deuterium target.  The first run was a ``cosmic run'', which was taken by turning off the beam and the spectrometer's magnetic field, and triggering on cosmic rays passing through the detector.  The second run was in the ``field-off'' configuration: the electron beam was on with 5~nA, and the target was in place\footnote{For this run, the target was in the ``empty'' configuration, \textit{i.e.} depressurized so that almost all of the scattering took place on the target windows, and only a small part of the data sample was from scattering from the residual gas.  This way, the longitudinal position of the target could be determined.}, but the magnetic field was turned off.  Example tracks from both runs are shown in Fig.~\ref{fig:cosmic_and_field_off}.  

For both configurations, the particles' trajectories are (neglecting multiple scattering) straight lines, which have several advantages over using helical tracks.  First, the straight tracks can be described with fewer parameters: four parameters rather than the five parameters for a helical track.  Second, no corrections need to be applied due to a possible non-homogeneity of the magnetic field.  Third, when the magnetic field is switched off, the Lorentz effect in the BMT is non-existent \cite{ACKER2020163423}, so no corrections are needed for this effect.  Finally, the formulas for the derivative matrices $\mathbf A$ and $\mathbf B$ (see Eqs.~\ref{eq:Adef}-\ref{eq:Bt}) are simpler for straight tracks than for helices.

The two data-taking configurations each have their strengths and weaknesses when used in alignment, so combining both in our sample takes advantage of both of their strengths.  Since the cosmic tracks pass through both the top half of the detector and the bottom half of the detector, they are useful for aligning the two halves together.  However, the cosmic tracks are less likely to pass through the SVT modules on the sides of the detector mounted vertically ($\phi$ near 0$^\circ$ or 180$^\circ$) and do not provide information about the alignment of the detector relative to the beamline.  The ``field-off'' tracks from the target have a nearly uniform distribution in $\phi$, and therefore have reasonable statistics in all of the SVT sectors. Since such tracks originate from the target, they can be used later on to constrain the alignment of the detector relative to the target and the beamline.  

The BMTC, in particular, cannot be aligned using only the tracks that originate from the beamline.  This is because each sector of the BMTC has a global weak mode\footnote{that is, a degree of freedom that is either entirely unconstrained or very poorly constrained.} in which the three BMTC layers within the sector are shifted along the beam direction by an amount proportional to their radii.  However, these weak modes can be constrained by using the cosmic tracks, which do not pass through the beamline.  By including both types of events in our sample, we remove the problematic weak modes and have sufficient statistics in all of the modules of the CVT.

Since the alignment procedure required rerunning the CLAS12 event reconstruction on each data set multiple times, we developed a procedure to create a sub-sample containing only the events with tracks that would be used in the KAA.  
First, we ran a preliminary event reconstruction using the CLAS12 reconstruction package \cite{ZIEGLER2020163472} with a pre-aligned version of the detector geometry, which was found using a detector survey followed by manual ad-hoc adjustments to individual alignment parameters.  We then filtered out events that did not have tracks in the CVT.  Events with more than two tracks were also removed, in order to get a cleaner sample.  If the angle between the reconstructed track direction and the normal of any sensor used in reconstructing the track was more than 75$^\circ$, or if the magnitude of the vector $\vec s\,'$ (see Eq.~\ref{eq:sp}) was greater than 10, then the whole track was rejected.  These cuts removed tracks that were difficult to accurately reconstruct with the detector.

To further improve the quality of our selected tracks, we required that all tracks had at least three BMTC clusters, two BMTZ clusters, and two pairs of clusters on paired sensors in the SVT. Further, we rejected tracks with very large residuals; these cuts were 7~mm for the BMTZ (which had the worst misalignments of the three subsystems before the alignment), and 2~mm for the BMTC and SVT.  
\FloatBarrier

\section{Aligning the CLAS12 CVT}
\label{sec:strategy}
The alignment procedure was comprised of several iterations of the following steps:
\begin{itemize}
    \item Running the CVT part of the CLAS12 reconstruction package~\cite{ZIEGLER2020163472} using the alignment parameters from the calibration-constants database (CCDB).  
    \item Running the KAA.  This is not part of the CLAS12 reconstruction, but rather a stand-alone software package, which takes as input from the reconstruction step a set of track measurements along with the alignment and track-derivative matrices (Eqs.~\ref{eq:Adef} and \ref{eq:Bdef}).   
    \item Adjusting the values in the CCDB based on the output of the KAA.
\end{itemize}

For the track fitting part of the reconstruction, we ignored the effects of multiple scattering, which were used in the standard variation of reconstruction.  We did this in order to avoid having an uneven weighting of hits in the outer layers during the fit, which would produce artificially large (small) residuals in the outer (inner) layers.  

Several cycles were necessary because the KAA operates using a linear expansion of the track residuals' dependence on the alignment parameters, as determined using the values of the alignment parameters at the time that the events were reconstructed, while the dependence in reality is non-linear, since the CLAS12 CVT contains non-parallel strips and curved sensors.  Therefore, the alignment values obtained from a single iteration may have some bias, which can be ameliorated by multiple iterations.  

We used an event sample that combines the cosmic and ``field-off'' event samples.  In order to avoid any bias from having all of the events of one of these two types at the beginning of the event sample and all of the other type at the end of the sample, we randomized the order of the events before starting the KAA.  

All three subsystems were fit simultaneously, rather than fitting them individually, since this takes into account the correlations between the the alignments of the different subsystems.  
At the beginning of each iteration, the diagonal elements of the covariance matrix $\mathbf D_{\rm init}$ were initialized to the following values:
\begin{itemize}
    \item $\epsilon^2=10^{-14}$ for the elements corresponding to fixed parameters.  This value is arbitrarily small, but non-zero in order to prevent $\mathbf D_{\rm init}$ from being singular.  This includes translations in $z$ for all BMTZ sensors and rotations in $z$ for all BMTC sensors.  We also chose to fix all parameters for one of the BMTZ modules (layer 5, sector 2), so that all alignment parameters would be relative to this sensor.  Since global translations in $z$ would otherwise be a weak mode, we also fixed the translations in $z$ for one of the BMTC modules (layer 6, sector 2).
    \item For the  non-fixed parameters, we used the following values:  $\Delta T^2 = (1.5$~mm$)^2$ for translations and $\Delta R^2=(0.005$~rad$)^2$ for rotations.  The values of $\Delta T$ and $\Delta R$ were chosen to be  bit larger than the maximum uncertainty of the precision of the preliminary survey\footnote{The survey had an estimated precision of a few hundred $\mu$m (in the global $x$ and $y$ directions) to 1~mm (in global $z$) for the BMT internal alignment, and about 100-150~$\mu$m for internal alignment of the SVT (due to the use of fiducials for every module), and 200~$\mu$m for the global SVT-BMT relative alignment.  Here, we define internal alignment of a detector subsystem as the relative alignment between modules in that subsystem.}.  It should be noted that due to the convergence of Kalman filter algorithms in general, overestimating the initial uncertainties has a very limited impact on the final results. 
\end{itemize}

Since the SVT sensors are rigidly attached back-to-back with one another (see Fig.~\ref{fig:svt_module}), the relative misalignment between paired sensors is much smaller than the alignment between different pairs or between the SVT modules and the BMT modules. We assumed that the alignments of the two sensors in a given SVT module only differ by translation transverse and longitudinal to the module (and had the same rotational alignment, as well as the same translation alignment normal to the sensors).  Therefore, we introduced parameters $\Delta T_\ell=$0.01~mm and $\Delta T_t=$0.01~mm as the uncertainty in the relative longitudinal and transverse alignment within the pair.  We then set the following off-diagonal elements:
\begin{itemize}
    \item $\Delta R^2-\epsilon^2/4$ for off-diagonal elements corresponding to the rotation about a given axis for one SVT sensor, and the rotation about the same axis for the other sensor in the same SVT module.   
    \item $\Delta T^2-(\epsilon^2\cos^2\phi+\Delta T_t^2\sin^2\phi)/4$, for the translations in $x$ of one sensor and the translation in $x$ of the other sensor in the pair.  Here, $\phi$ is the nominal azimuthal coordinate of the midplane of the SVT sector.  
    \item $\Delta T^2-(\epsilon^2\sin^2\phi+\Delta T_t^2\cos^2\phi)/4$, for the translations in $x$ of one sensor and the translation in $x$ of the other sensor in the back-to-back pair.  
    \item $(\epsilon^2-\Delta T_t^2)/4 \sin\phi\cos\phi$, for the translations in $y$ of one sensor and the translation in $x$ of the  other sensor in the back-to-back pair.  
    \item $-(\epsilon^2-\Delta T_t^2)/4 \sin\phi\cos\phi$, for the translations in $y$ of one sensor and the translation in $x$ of the same sensor.  
    \item $\Delta T^2-\Delta T^2_\ell/4$ for the translation in $z$ in one sensor and the translation in $z$ of the other sensor in the back-to-back pair.
\end{itemize}

Further, the inclusion of these constraints modifies the diagonal elements as well.  Instead of $\Delta R^2$ and $\Delta T^2$, the diagonal elements for the SVT are

\begin{itemize}
    \item $\Delta R^2+\epsilon^2/4$ for rotation parameters.
    \item $\Delta T^2 +(\epsilon^2\cos^2\phi+\Delta T_t^2\sin^2\phi)/4$ for translation parameters in $x$.
    \item $\Delta T^2 + (\epsilon^2\sin^2\phi+\Delta T_t^2\cos^2\phi)/4$ for translation parameters in $y$.
    \item $\Delta T^2 +T^2_\ell/4$ for translation parameters in $z$.
\end{itemize} 
All other elements of $\mathbf D_{\rm init}$, besides those listed above, were set to zero.

Since there are 6 parameters per module and 84 SVT sensors and 18 BMT sensors, there are $6\times(84+18)=612$ total parameters.  However, considering the fact that six parameters are fixed for global alignment, and four are fixed per SVT sensor pair, and one parameter is fixed for each BMT sensor, the remaining number of degrees of freedom is 420.

\section{Results}
\label{sec:results}
To align the detector using the cosmic-ray and ``field-off'' data from the Spring 2019 run, we followed the procedure detailed in Sec.~\ref{sec:strategy} for running the KAA with multiple iterations.
The KAA provides the alignment parameters needed to correct for errors in the reconstructed particle tracks, thus minimizing the residuals of the track reconstruction when those corrections are applied.  

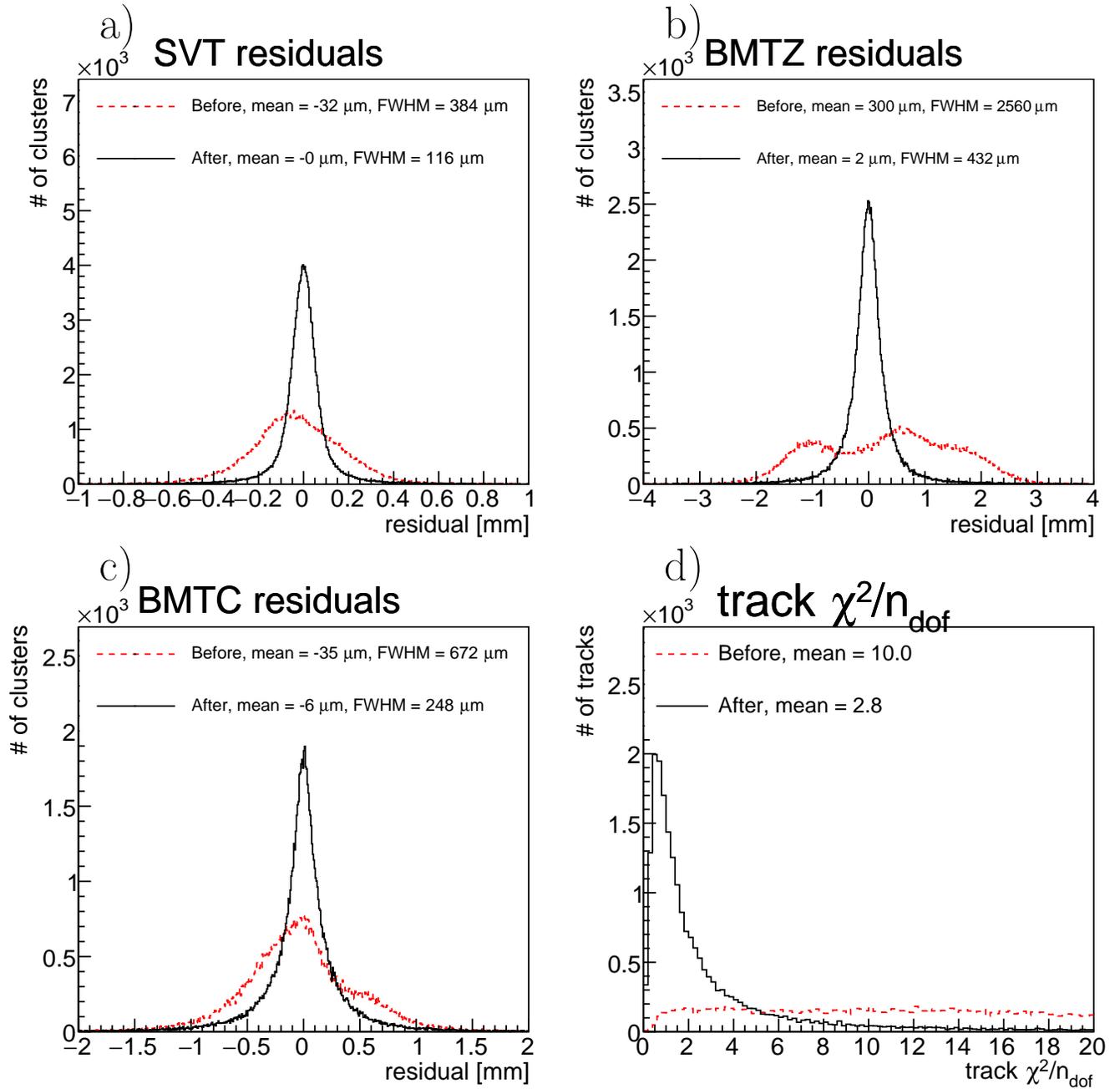
\begin{figure*}
    \centering
    \begin{overpic}[width=\textwidth]{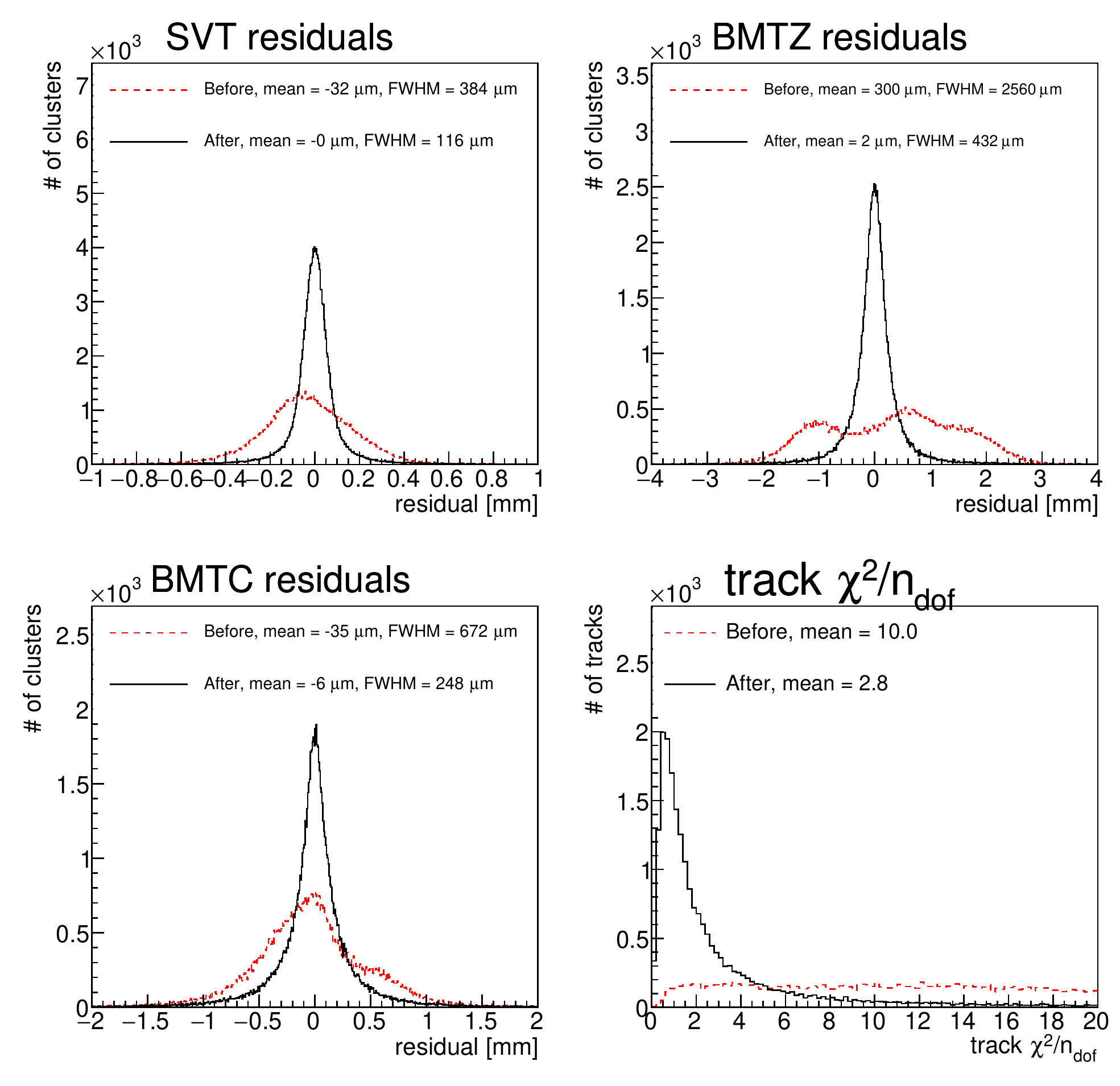}
    \put(10, 95) {\huge a)}
    \put(60, 95) {\huge b)}
    \put(10, 47) {\huge c)}
    \put(60, 47) {\huge d)}
    \end{overpic}
    \caption{Residuals distribution before (red, dashed) and after (black, solid) alignment for the SVT (a), BMTZ (b) and BMTC (c).  Panel (d) shows  the $\chi^2/n_{\rm dof}$ distribution (bottom right) for each reconstructed track.  Each cluster produces a single residual for a single track.}
    \label{fig:residuals_1D}
\end{figure*}

The distributions of residuals\footnote{As defined by Eq.~\ref{eq:rformula}.} of the sampled tracks before (red, dashed) and after (black, solid) alignment are shown in Fig.~\ref{fig:residuals_1D} for the SVT (a), BMTZ (b), and BMTC (c). 
In each detector, the residual distributions after alignment are much narrower than those before the alignment. We then determined the full widths at half maximum (FWHMs) of these distributions, which are 116~$\mu$m for the SVT, 432~$\mu$m for the BMTZ, and 248~$\mu$m for the BMTC.  Similarly, we also fit the cores of the distributions to Gaussian functions and obtained values that are about half of the values of the FWHMs\footnote{The ratio of the FWHM to the standard deviation of a distribution depends on its shape.  For reference, this ratio is $\approx 2.35$ for a Gaussian distribution.}: 57, 230 and 180~$\mu$m for the SVT, BMTZ, and BMTC, respectively.
These are comparable to the expected spatial resolutions of the SVT and BMT from Refs.~\cite{ANTONIOLI2020163701} and \cite{ACKER2020163423}, respectively.  The means of these distributions are on the order of a few $\mu$m, which is acceptable.  The measured resolutions are consistent with the system design goal of momentum resolution below 5\% for charged particles with momenta up to 1~GeV in stand-alone SVT reconstruction \cite{ANTONIOLI2020163701}. 

\begin{figure}
    \centering
    \begin{overpic}[width=\columnwidth]{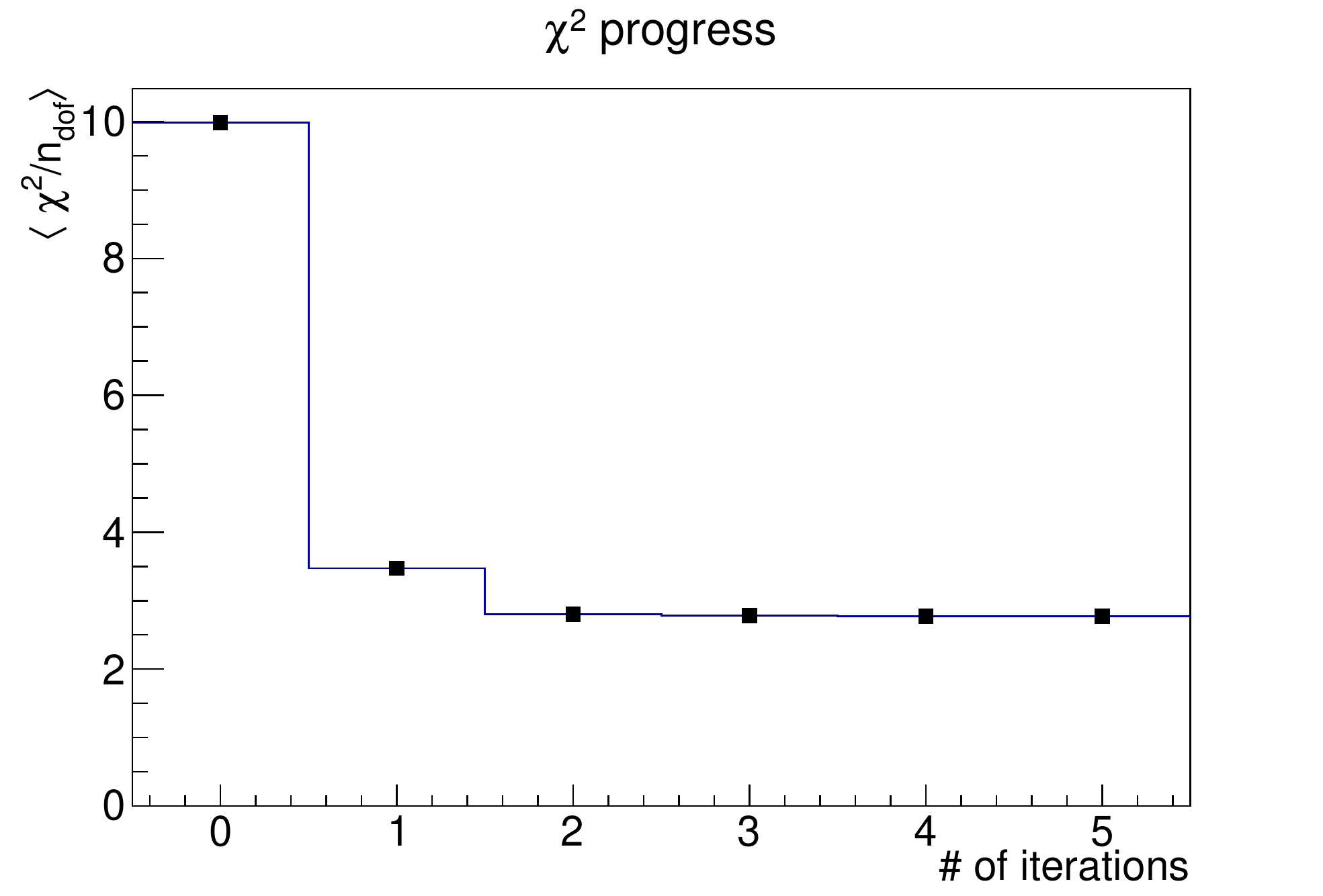}
    \end{overpic}
    \caption{Average track $\chi^2/n_{\rm dof}$ after the indicated number of iterations of the KAA.  The first point (at zero iterations) represents the average track $\chi^2/n_{\rm dof}$ before running any alignment with the KAA.}
    \label{fig:chi2_progress}
\end{figure}

We calculated the $\chi^2$ for each track as
\begin{equation}
\chi^2 = {\mathbf r}^{\rm T}{\mathbf V}^{-1}{\mathbf r},
\end{equation}
where the number of degrees of freedom, $n_{\rm dof}$, is the number of clusters in the track minus four (since there are four parameters for the track fit).  The distributions of the $\chi^2/n_{\rm dof}$ values before and after alignment are shown in  Fig.~\ref{fig:residuals_1D}(d).  As shown in Fig.~\ref{fig:chi2_progress}, the average $\chi^2/n_{\rm dof}$ goes down from 10.0 to about 3.4 after the first iteration, and down to about 2.8 after the second.  There is a very small improvement ($<$0.1) after the third iteration.  After the fourth and fifth iterations, there is no significant change to the average $\chi^2/n_{\rm dof}$.  Based on this assessment, there is no need to run the KAA for more than three iterations. 

\begin{figure*}
    \centering
    \begin{overpic}[width=0.97\textwidth]{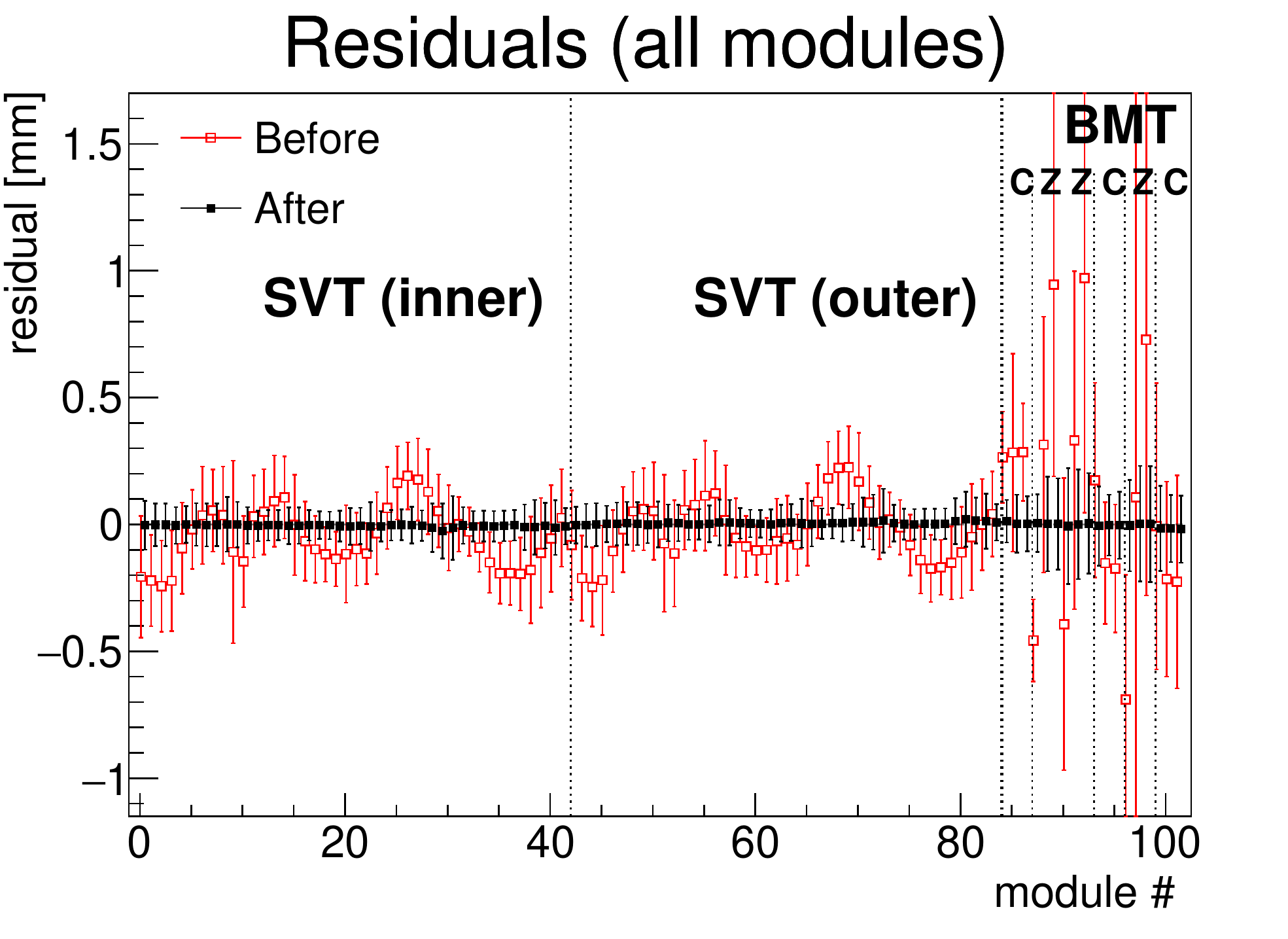}
    \end{overpic}
    \caption{Residuals for the each module, before (red, open symbols) and after (black, closed symbols) alignment.  The error bars for each point represent the FWHMs of the distributions, divided by two (so that the distance from the top of the upper error bar to the bottom of the lower error bar is one FWHM).  Module numbers 1-84 represent SVT sensors; numbers 85-102 represent BMT tiles.   
    Symbols are shifted horizontally slightly for clarity.}
    \label{fig:residuals_module}
\end{figure*}

Since the residual distributions in Fig.~\ref{fig:residuals_1D} are the sum over the residual distributions in all of the sensors of each given type, it does not provide information about the alignment of individual sensors.  Therefore, we determined the residual distributions of every sensor module individually in order to make sure that none of them had large misalignments.  We then determined the means and FWHMs of these distributions, which we show in Fig.~\ref{fig:residuals_module}.  
After fitting, the means of the residual distributions for all sensors are within 20~$\mu$m of zero, and the FWHMs are less than 170~(460)~$\mu$m for each of the SVT (BMT) sensors.  

\begin{figure*}
     \centering
    \begin{overpic}[width=0.83\textwidth]{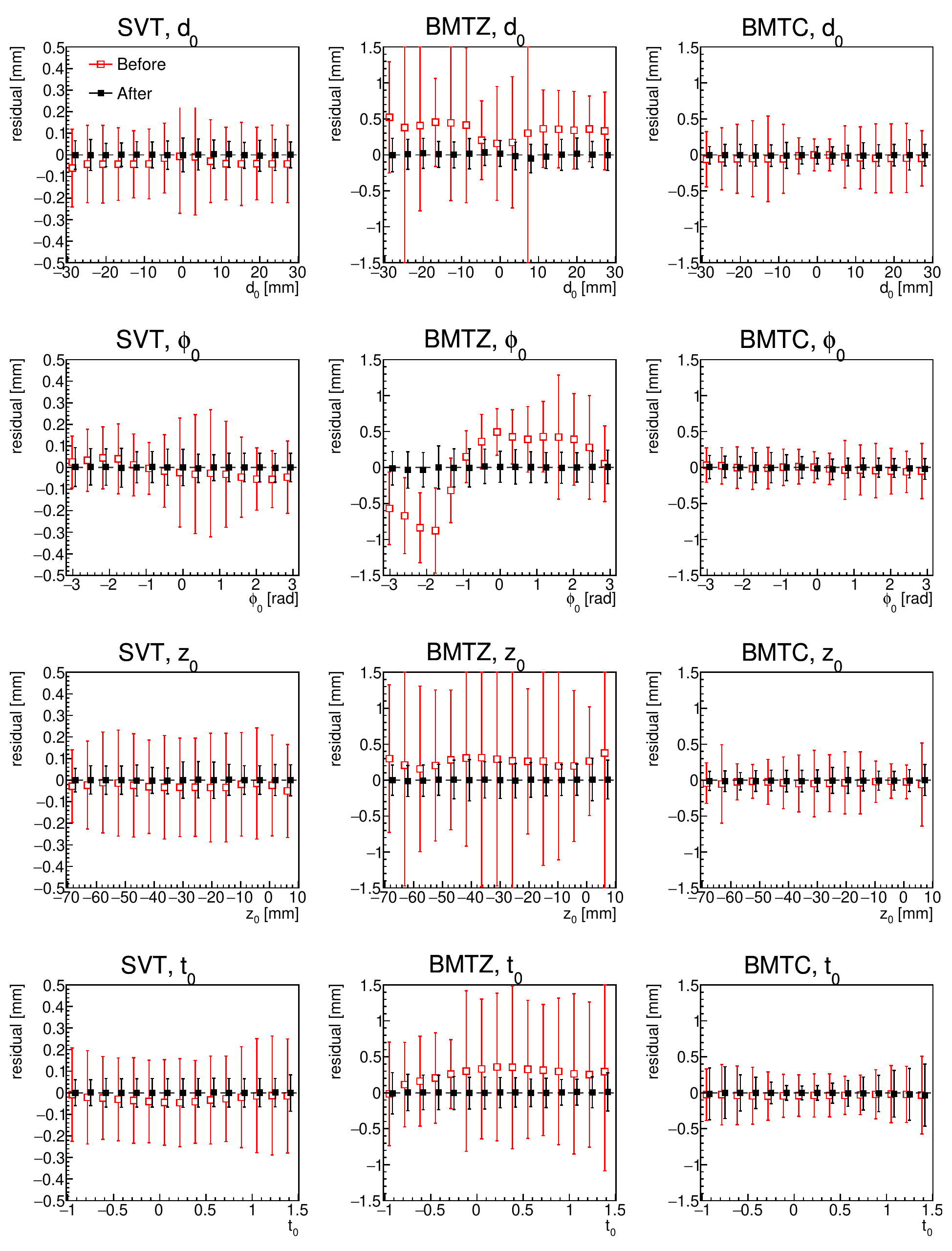}
    \end{overpic}
    \caption{Residuals before (red, open symbols) and after (black, closed symbols) alignment, as a function of the kinematic variables: from top to bottom, $d_0$, $\phi_0$, $z_0$ and $\theta_0$.  The error bars for each point represent the FWHMs of the distributions, divided by two (so that the distance from the top of the upper error bar to the bottom of the lower error bar is one FWHM).  From left to right, the results are shown for the SVT, BMTZ, and BMTC.  Symbols are shifted horizontally slightly for clarity.}
    \label{fig:residuals_vs_kinematics}
\end{figure*}

The alignment process can become biased to show lower performance for certain track locations due to data sampling and the specific algorithm implementation.  In order to show that there is no bias in the alignment, we studied the dependence of the residuals on the track parameters.  Figure~\ref{fig:residuals_vs_kinematics} shows the residuals for each of the three detector types as a function of the track kinematic variables $d_0$, $\phi_0$, $z_0$ and $t_0$.  The residual distributions after the alignment procedure are centered at zero, with no significant dependence on the kinematic variables.  

The KAA algorithm also yields the correlation among the alignment parameters.  The correlations are given by the matrix $\mathbf{C}$, where each element is given by 
\begin{equation}
    C_{ij}=D_{ij}/\sqrt{D_{ii}D_{jj}},
    \label{eq:correlation}
\end{equation}
where $\mathbf D$ is the covariance matrix.  By construction, the diagonal elements of $\mathbf C$ are equal to one.  Many of the parameters are strongly correlated with one another, leading to off-diagonal elements of $\mathbf C$ close to +1 (-1) when the correlations are strongly positive (negative).  We show plots of the values of the elements of $\mathbf C$ in \ref{sec:correlations} and discuss there in further detail which types of parameters are strongly or weakly correlated.

In order to see if the residual distributions depend on where the particles cross the sensors, we plot in \ref{sec:vs_coord} the distributions of the residuals versus the extrapolated lab-frame coordinates of the hits in the detectors, both before and after the alignment.  We also include the residuals versus the measured centroid strip number of the hits in each cluster.  We found that the residual distributions after alignment appear to be centered at zero regardless of the position of the hit in the detector.   

With an unaligned detector,the residuals in one sensor may be strongly correlated to those in another, whereas with a well-aligned detector, such correlations vanish. In \ref{sec:rvr}, we plot distributions of the residuals in one sensor versus those of another, for several different representative combinations of sensors.  The 2D residual distributions show strong correlation for some of these combinations before alignment, but there is no significant correlation between the residuals after alignment.  

To validate our results, we performed the same procedure on Monte-Carlo (MC) simulations, and present the results in  \ref{sec:results_MC}.  In the simulations, the means of the residual distributions are within about 15~$\mu$m of zero, which is comparable to the data.  However, the residual distributions are considerably narrower in the simulations than in the data, and as a result the $\chi^2/n_{\rm dof}$ distribution in the simulation has a smaller mean than in the data.  This could be due to a mis-modeling of the resolution effects in the detector, since the resolutions in the simulation were estimated using an idealized detector.

\begin{figure*}
    \centering
    \begin{overpic}[width=\textwidth]{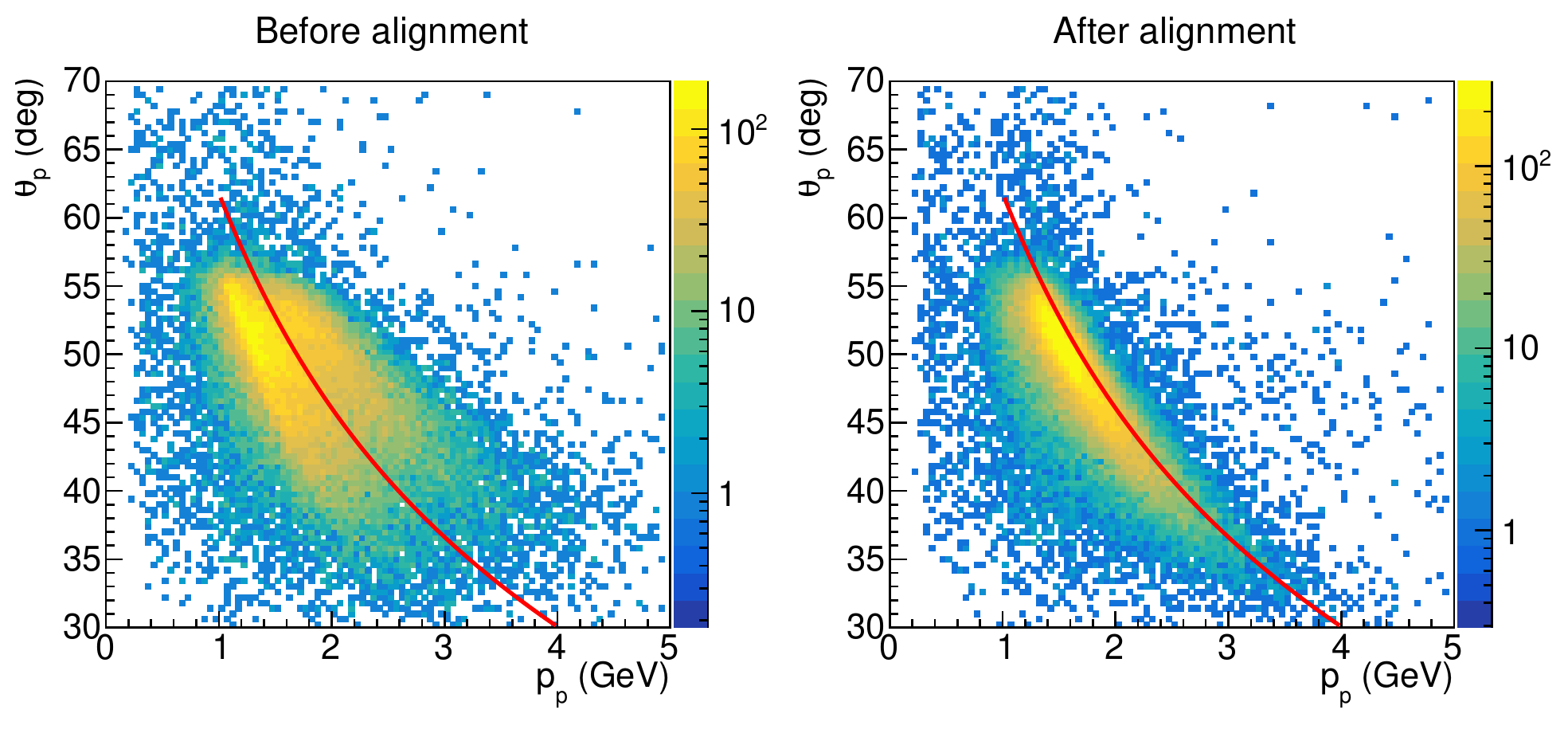}
    \end{overpic}
    \caption{Polar angle vs.~momentum distributions for elastically scattered protons in the CVT, before (left) and after (right) the alignment.  The curve shows the expected correlation between the two variables.}
    \label{fig:elastics}
\end{figure*}

Finally, we validated that the alignment works not only for straight tracks, but also for curved tracks (with the solenoid field turned on), using the following test.  Using a run configuration with 5~nA on liquid hydrogen at 10.2~GeV, we reconstructed events where electrons scattered elastically off a proton. These were selected by requiring one electron in the Forward Detector of CLAS12, with $W<1.1$~GeV\footnote{$W$ is defined as $\sqrt{2m_p\nu+m_p^2-Q^2}$, where $Q^2$ is the square of the four-momentum transfer of the reaction, $\nu$ is the energy transfer, and $m_p$ is the proton mass.}, \textit{i.e.}~in the elastic-peak region, and at least one positive track in CVT, which was assumed to be a proton.  We show the distribution of the reconstructed polar angle $\theta$ vs the reconstructed momentum $p$ of the protons in these reactions in Fig.~\ref{fig:elastics}, before (left) and after (right) the alignment procedure.  The expected relation between $\theta_p$ and $p_p$ for protons in elastic kinematics is:
\begin{equation}
    p_p=\frac{2E_b m_p(E_b+m_p)\cos\theta_p}{E_b^2\sin^2\theta_p+2E_b m_p+m_p^2},
\end{equation} 
where $m_p$ is the mass of a proton, and $E_b$=10.2~GeV is the beam energy; we show this as a curve overlaid on the distribution in Fig.~\ref{fig:elastics}.
The $\theta$~vs.~$p$ distribution obtained after the alignment follows the curve much more closely than the one obtained before the alignment.  

\section{Conclusions} \label{sec:conclusions}
We have adapted the KAA, originally developed for CMS, to align the CLAS12 CVT---a hybrid detector consisting of both silicon and micromegas tracking technologies, with both curved and non-parallel strips. 

Using a sample of cosmic-ray tracks and ``field-off'' data, we obtained residual distributions centered within 10~$\mu$m of zero for each of the silicon and micromegas sensors. In order to avoid significant bias from the non-linearity of the detector geometry, we had to run multiple iterations of the alignment, re-running the event reconstruction with the updated alignment parameters in between iterations.  

By adapting the algorithm to the CLAS12 CVT, we demonstrated the flexibility and power of the KAA~\cite{Widl:2006mz,Widl:2008aqa}. Future work will include extending these results to include the CLAS12 forward detectors or curved tracks as additional constraints.

The methodology and results detailed in this work could serve as reference for alignment of the CLAS12 CVT for upcoming experiments~\cite{doi:10.1146/annurev-nucl-101917-021129,Arrington:2021alx}, as well as for future experiments at the Electron-Ion Collider~\cite{AbdulKhalek:2021gbh}.  

\section*{Acknowledgements} \label{sec:acknowledgements}

We acknowledge the outstanding efforts of the staff of the Accelerator, the Nuclear Physics Division, Hall B, and the Detector Support Group at JLab that have contributed to the design, construction, installation, and operation of the CLAS12 detector.
We thank Maurizio Ungaro for his contributions in the CLAS12 simulations.  
We also thank the CLAS Collaboration for staffing shifts and taking high quality data.  
This work was supported by the United States Department of Energy under JSA/DOE Contract DE-AC05-06OR23177. 
This work was also supported in part by the U.S. National Science Foundation, the State Committee of Science of the Republic of Armenia, the Chilean Agencia Nacional de Investigaci\' on y Desarrollo, 
the Italian Istituto Nazionale di Fisica Nucleare, the French Centre National de la 
Recherche Scientifique, the French Commissariat a l'Energie Atomique, 
the Scottish Universities Physics Alliance (SUPA), the United Kingdom Science and Technology Facilities Council (STFC), the National Research Foundation of Korea, the Deutsche Forschungsgemeinschaft (DFG), and the Russian Science Foundation.  This research was funded in part by the French Agence Nationale de la Recherche contract no. 37. This work was supported as well by the EU Horizon 2020 Research and Innovation Program under the Marie Sklodowska-Curie Grant Agreement No. 101003460.

We also thank Martin Weber for developing a software implementation of the Kalman Alignment Algorithm and making it publicly available.

\bibliographystyle{elsarticle-num-etal} 
\bibliography{biblio.bib} 
\FloatBarrier
\appendix
\section{Correlations}\label{sec:correlations}
\begin{figure*}
    \centering
    \begin{overpic}[width=\textwidth]{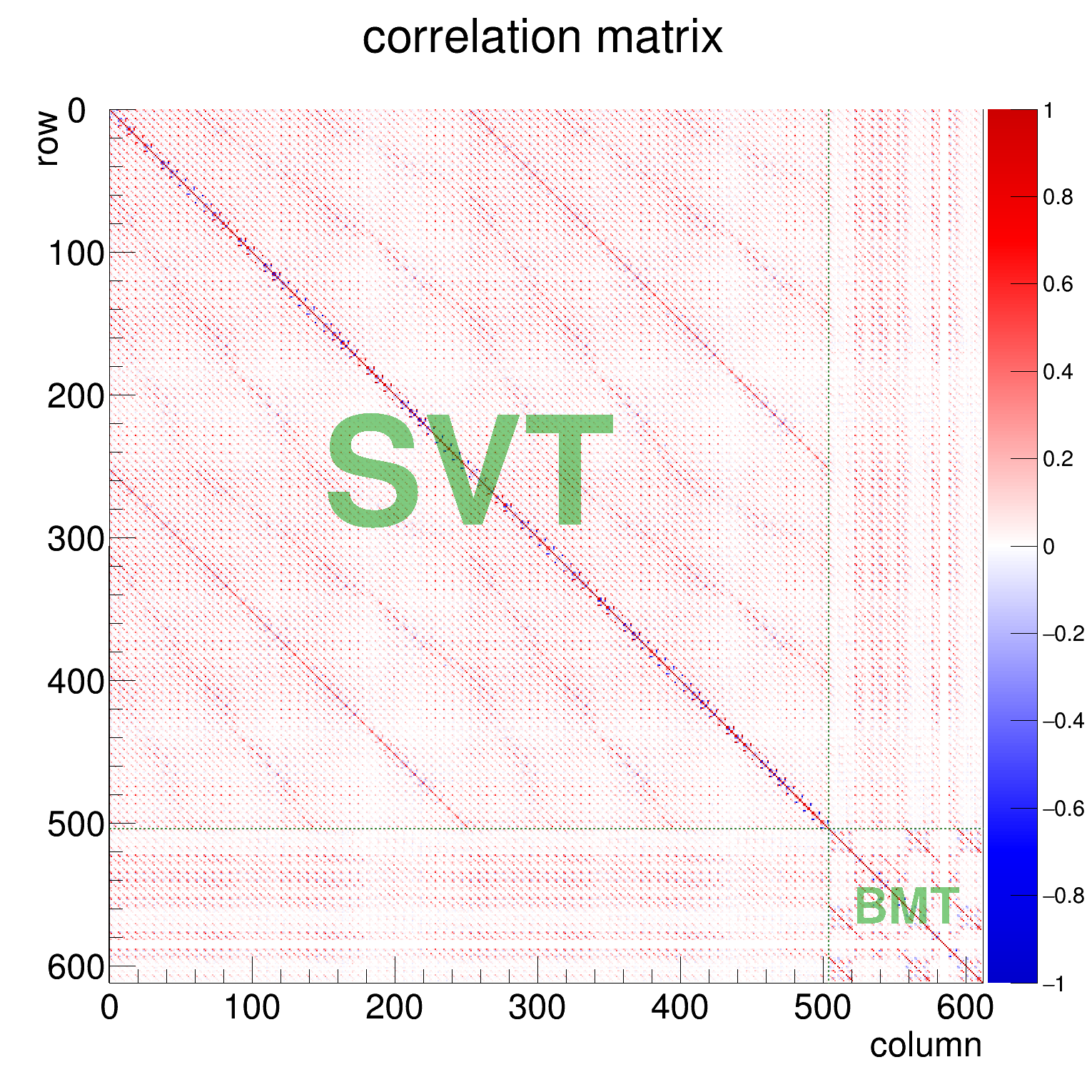}
    \end{overpic}
    \caption{Values of the matrix elements in the correlation matrix $\mathbf C$. Every group of 6 indices represent translations in $x$, $y$ and $z$ and rotations in $x$, $y$ and $z$ of a single sensor.  The first 84 of these groups represent the SVT, while the next 18 represent the BMT.  Positive values are shown in red and negative values are shown in blue.  The dark-green horizontal and vertical dotted lines demarcate the SVT and BMT regions.}
    \label{fig:correlation_matrix}
\end{figure*}

\begin{figure*}
    \centering
    \begin{overpic}[width=\textwidth]{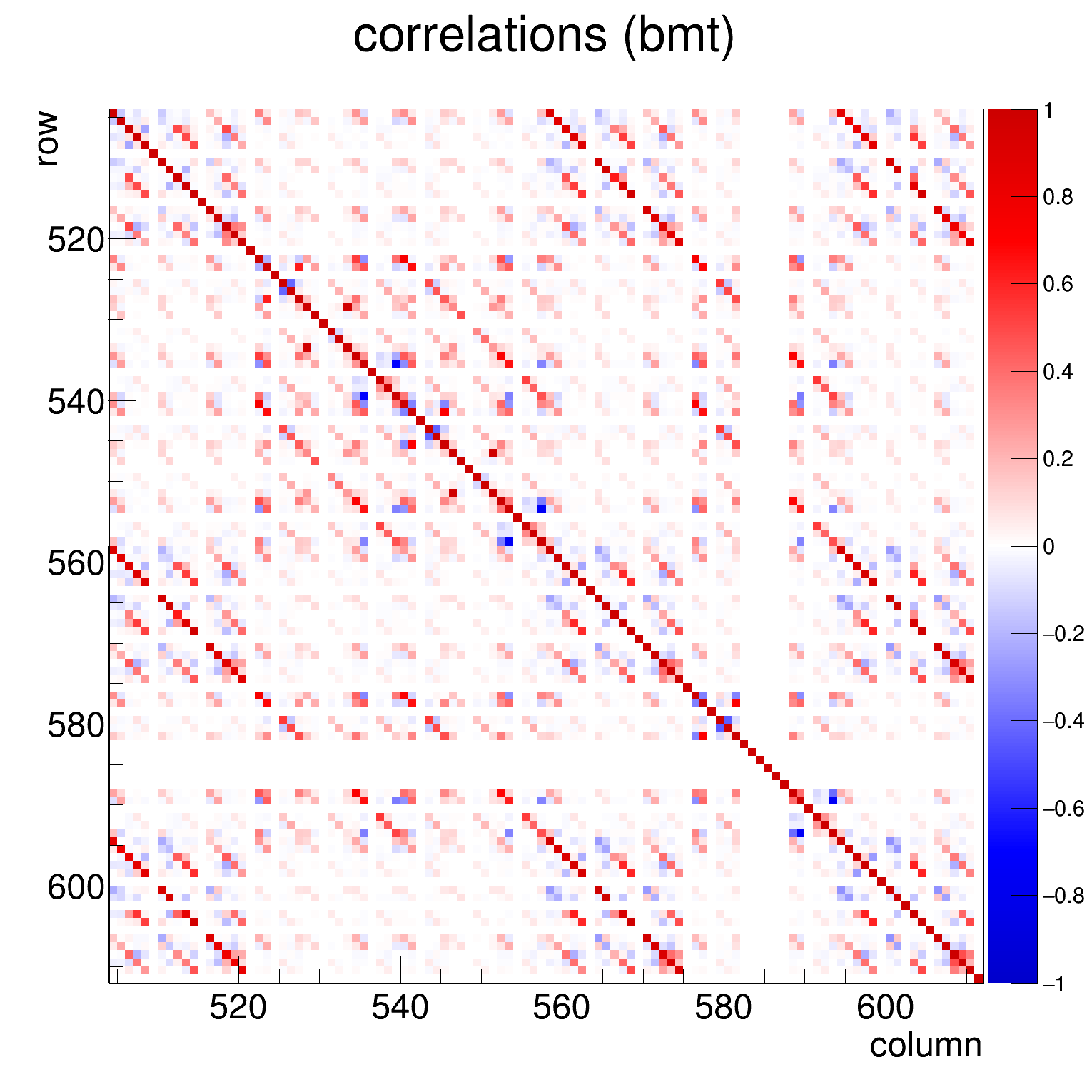}
    \put(17,17) {\rotatebox{-45}{\resizebox{0.5\textwidth}{0.1\textwidth}{\transparent{0.2}\circle{1}{\transparent{0}1}}}}
    \end{overpic}
    \caption{Zoom-in of Fig.~\ref{fig:correlation_matrix}, showing the submatrix of the correlation matrix $\mathbf C$ corresponding to correlations between BMT alignment variables.  Positive values are shown in red and negative values are shown in blue.  The ellipse indicates one of the diagonal ``streaks'' mentioned in the text.}
    \label{fig:correlation_matrix_bmt}
\end{figure*}

The values of the elements of the correlation matrix $\mathbf C$ (see Eq.~\ref{eq:correlation}), at the end of the final iteration, are shown in Fig.~\ref{fig:correlation_matrix}.  We also show a zoom-in of the BMT part of this matrix in Fig.~\ref{fig:correlation_matrix_bmt}.  

We find that the alignment parameters for any of the BMTC sensors are always very strongly positively correlated with the same parameters for the other BMTC sensors in the same sector.  This is apparent in Fig.~\ref{fig:correlation_matrix_bmt} as visible as dark red diagonal streaks, such as the one near the bottom right corner starting in row 594 (marked with an ellipse in Fig.~\ref{fig:correlation_matrix_bmt}), correlating BMTC layers 1 and 6.  The BMTC parameters are very weakly correlated with the parameters of the SVT and the BMTZ, which are apparent in Figs.~\ref{fig:correlation_matrix} and \ref{fig:correlation_matrix_bmt} as blocks of mostly white, suggesting that the internal alignment within the BMTC is much better than its alignment relative to the other components.  This is largely due to the fact that the BMTC measures the position of the track in $z$, which is a weak mode for the BMTZ.  Translations in $z$ are also less strongly constrained by the SVT than the BMTC, since the BMTC has much better precision on the $z$ positions of tracks than crosses between clusters in the SVT.  

Also, there is a strong correlation between the translations in $x$ and $y$ for a given SVT sensor and the rotation in $z$ for the same sensor.  This is because the rotations are defined around the CLAS12 origin, rather than the center of the sensors, and the widths of the sensors are much smaller than the distance between them and the beamline.  It is therefore difficult to distinguish between a rotation around the global $z$ axis and a translation of the sensor plane in the azimuthal direction.

\section{Additional plots}
\subsection{Residuals versus coordinates}
\label{sec:vs_coord}
\begin{figure*}
    \centering
    {\huge BEFORE}
    
    \vspace{0.3cm}
    \begin{overpic}[width=\textwidth]{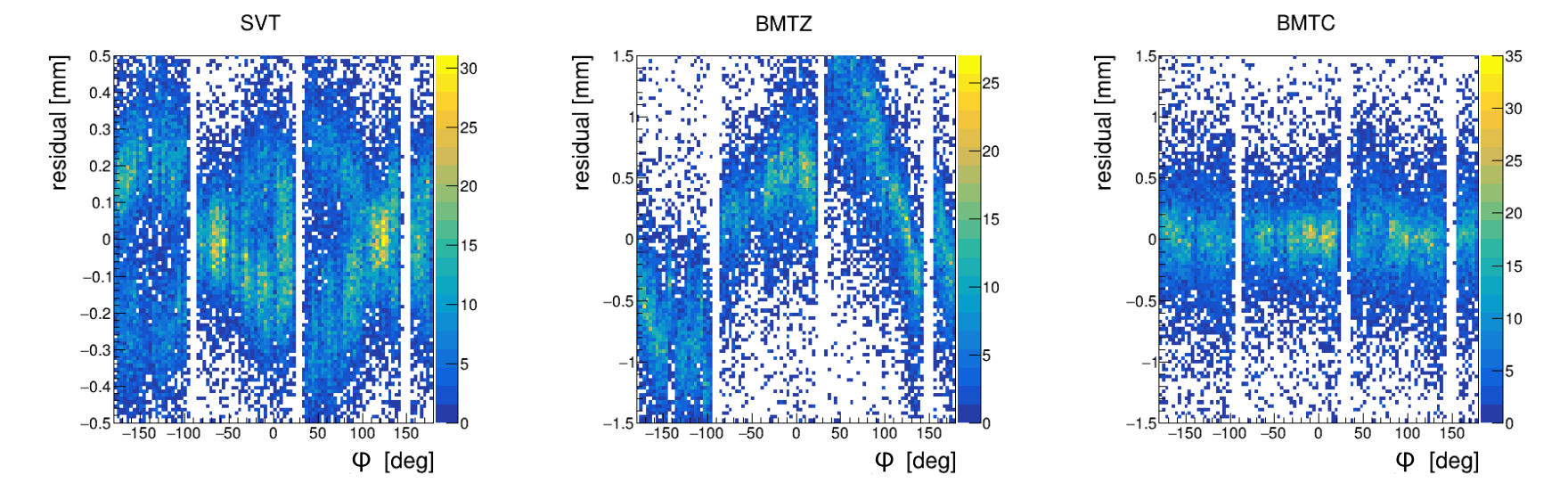}
    \end{overpic}
    
    \vspace{0.5cm}
    {\huge AFTER}
    
    \vspace{0.3cm}
     \begin{overpic}[width=\textwidth]{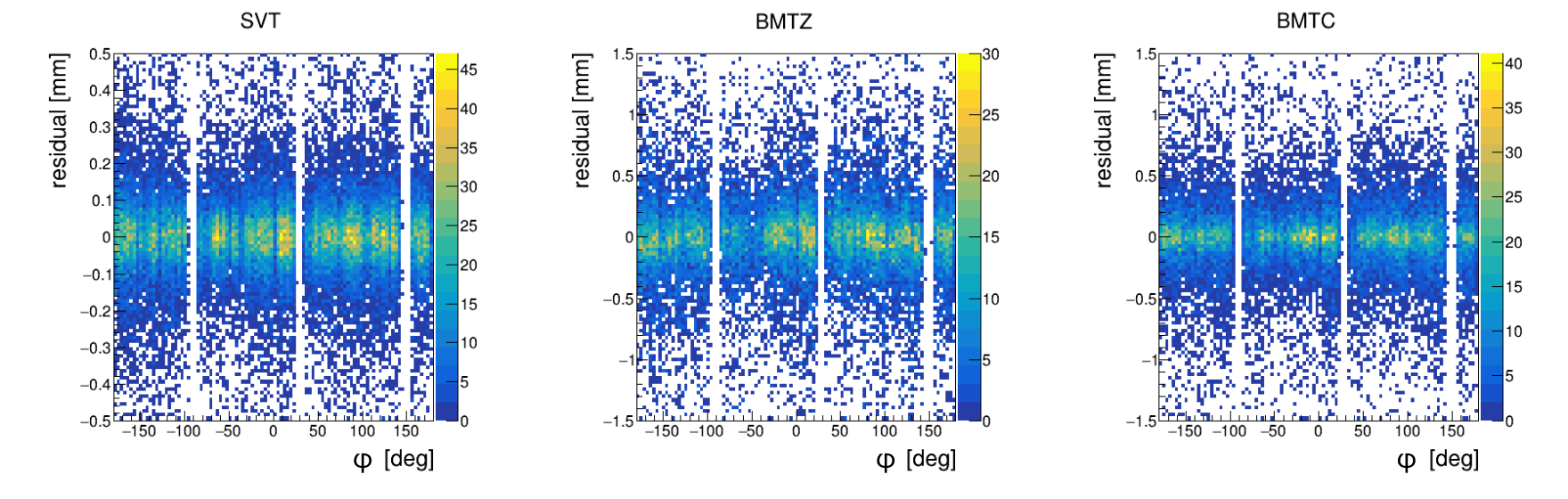}
    \end{overpic}
    \caption{Distributions of the residuals for SVT, BMTZ, and BMTC (left to right) vs.~the $\phi$ coordinate of the extrapolated hit positions before (top row) and after (bottom row) the alignment.}
    \label{fig:residuals_vs_phi}
\end{figure*}

\begin{figure*}
    \centering
    {\huge BEFORE}
    
    \vspace{0.3cm}
    \begin{overpic}[width=\textwidth]{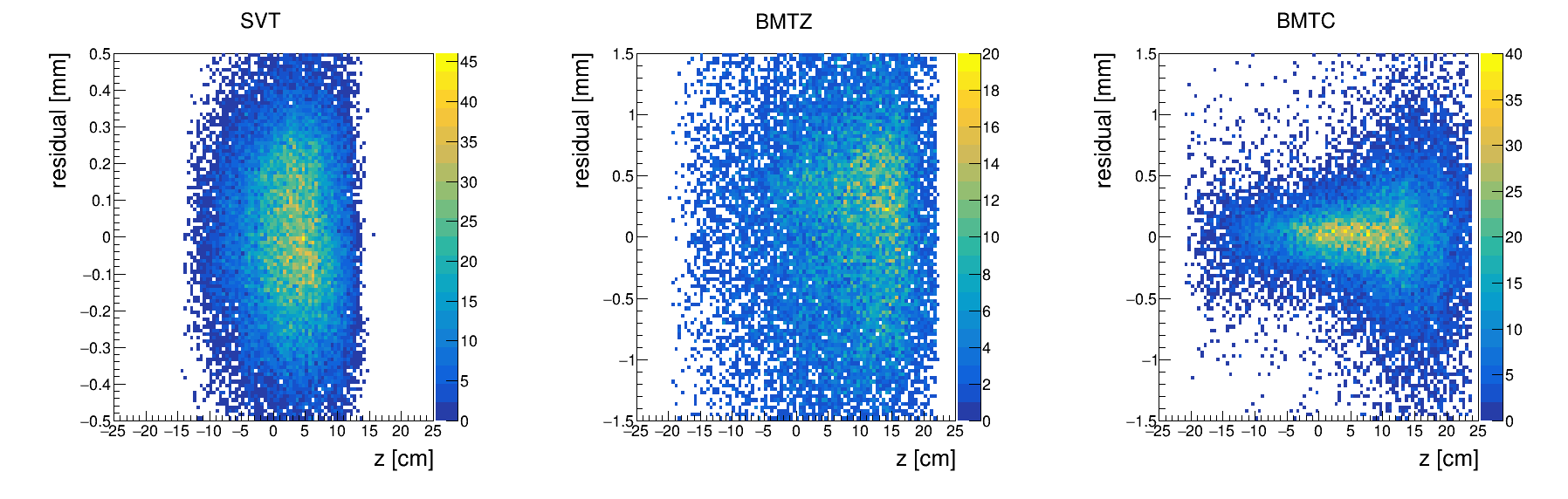}
    \end{overpic}
    
    \vspace{0.5cm}
    {\huge AFTER}
    
    \vspace{0.3cm}
     \begin{overpic}[width=\textwidth]{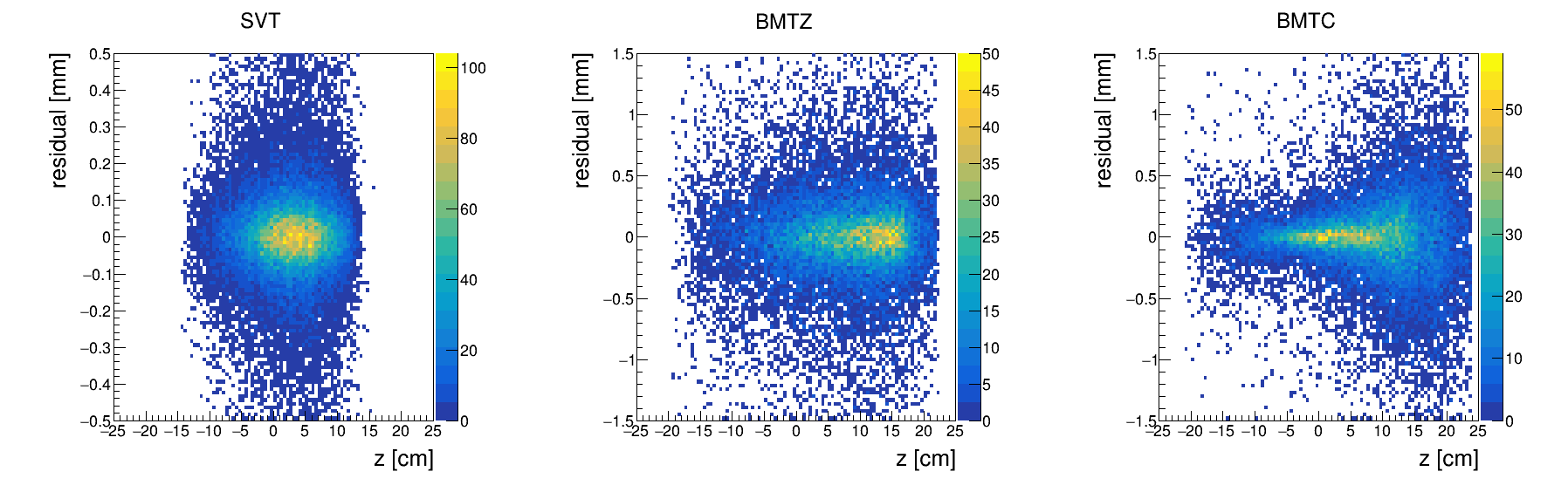}
    \end{overpic}
    \caption{Distributions of the residuals for SVT, BMTZ, and BMTC (left to right) vs.~the $z$ coordinate of the extrapolated hit positions before (top row) and after (bottom row) the alignment.}
    \label{fig:residuals_vs_z}
\end{figure*}

\begin{figure*}
    \centering
    {\huge BEFORE}
    
    \vspace{0.3cm}
    \begin{overpic}[width=\textwidth]{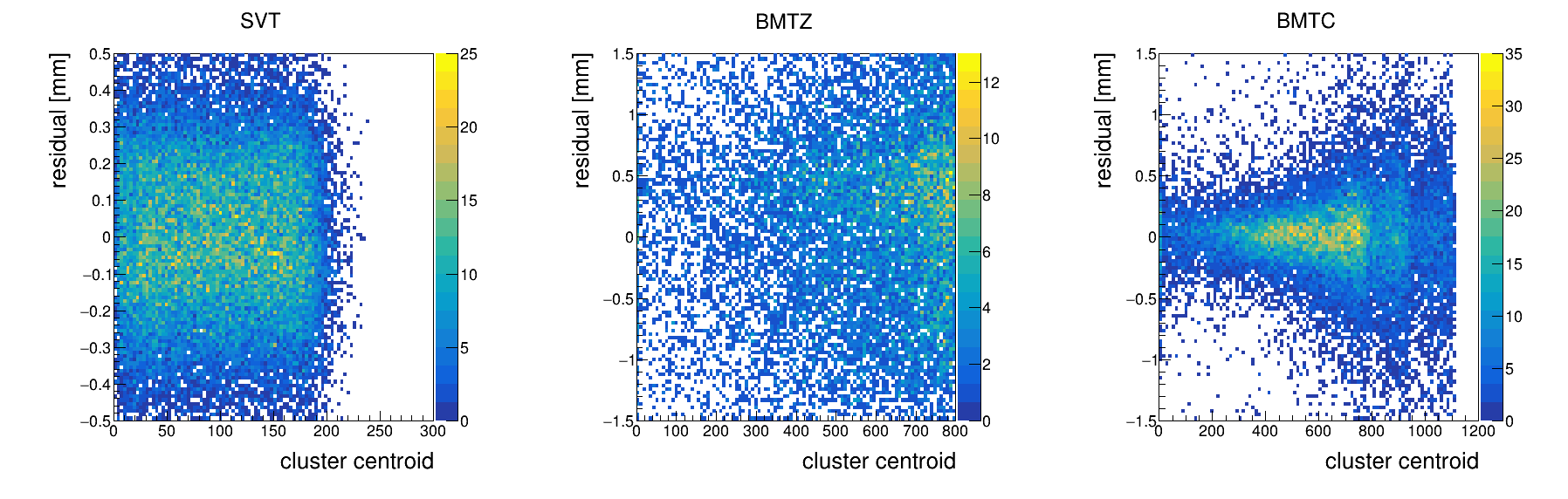}
    \end{overpic}
    
    \vspace{0.5cm}
    {\huge AFTER}

    \vspace{0.3cm}
     \begin{overpic}[width=\textwidth]{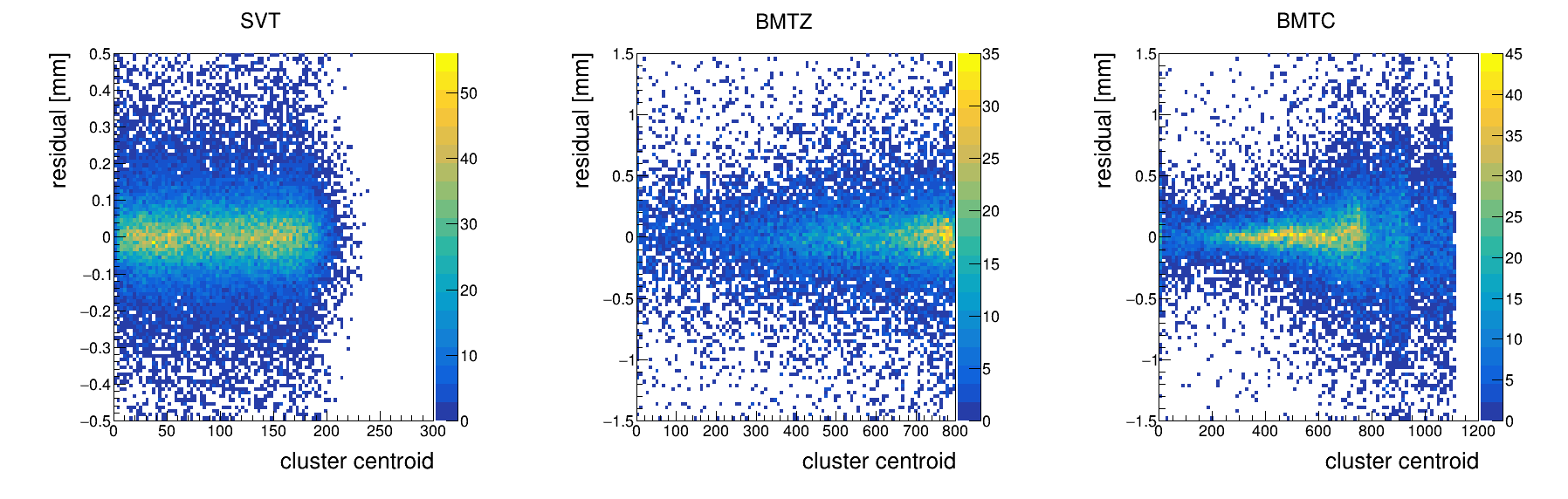}
    \end{overpic}
    \caption{Distributions of the residuals for SVT, BMTZ, and BMTC (left to right) vs.~the centroid strip numbers of the measured events  before (top row) and after (bottom row) the alignment.}
    \label{fig:residuals_vs_centroid}
\end{figure*}

To see if the residuals depend on the location where the particles cross the sensors, we include plots of the tracking residuals versus the global $\phi$ and $z$ coordinates of these intersection points in Figs.~\ref{fig:residuals_vs_phi} and \ref{fig:residuals_vs_z}, respectively.  We show residuals versus the measured centroid strip numbers in the sensors in Fig.~\ref{fig:residuals_vs_centroid}. The results before (after) alignment are shown in the top (bottom) row. There is a huge dependence of the residuals on the $\phi$ coordinate in the hit (see Fig.~\ref{fig:residuals_vs_phi}), but this vanishes after the alignment.  Since the centroid number correlates with $\phi$ in the BMTZ and the SVT, one would expect to see a similar dependence on the centroid number, but this would only be visible when looking at each sector individually.  

\subsection{Correlations between residuals in different sensors}
\label{sec:rvr}
To determine if the residuals in different sensors in the CVT are correlated to one another, we show in Fig.~\ref{fig:rvr1} and \ref{fig:rvr2} some 2D distributions of the residuals for various combinations of sensors, before (middle column) and after (right column) alignment.  For reference, the positions of the two sensors are shown to the left of the 2D residual plots.  The 2D residual distributions show strong correlations for some of these combinations before alignment, especially when the sensors' measurement directions, $\hat s$, are parallel or nearly parallel to one another, for instance between stereo pairs of SVT sensors (see first row of Fig.~\ref{fig:rvr1}).  In cases where the sensors' measurement directions are perpendicular to one another, such as one sensor in the BMTC and another in the BMTZ (see third row of Fig.~\ref{fig:rvr2}), there is no significant correlation between the residuals.  Moreover, in the ``after'' plots, there is almost no correlation between the residuals in one sensor versus those in another, except in the tails of the distributions. 

\begin{figure*}
    \centering
    \begin{overpic}[width=0.67\textwidth]{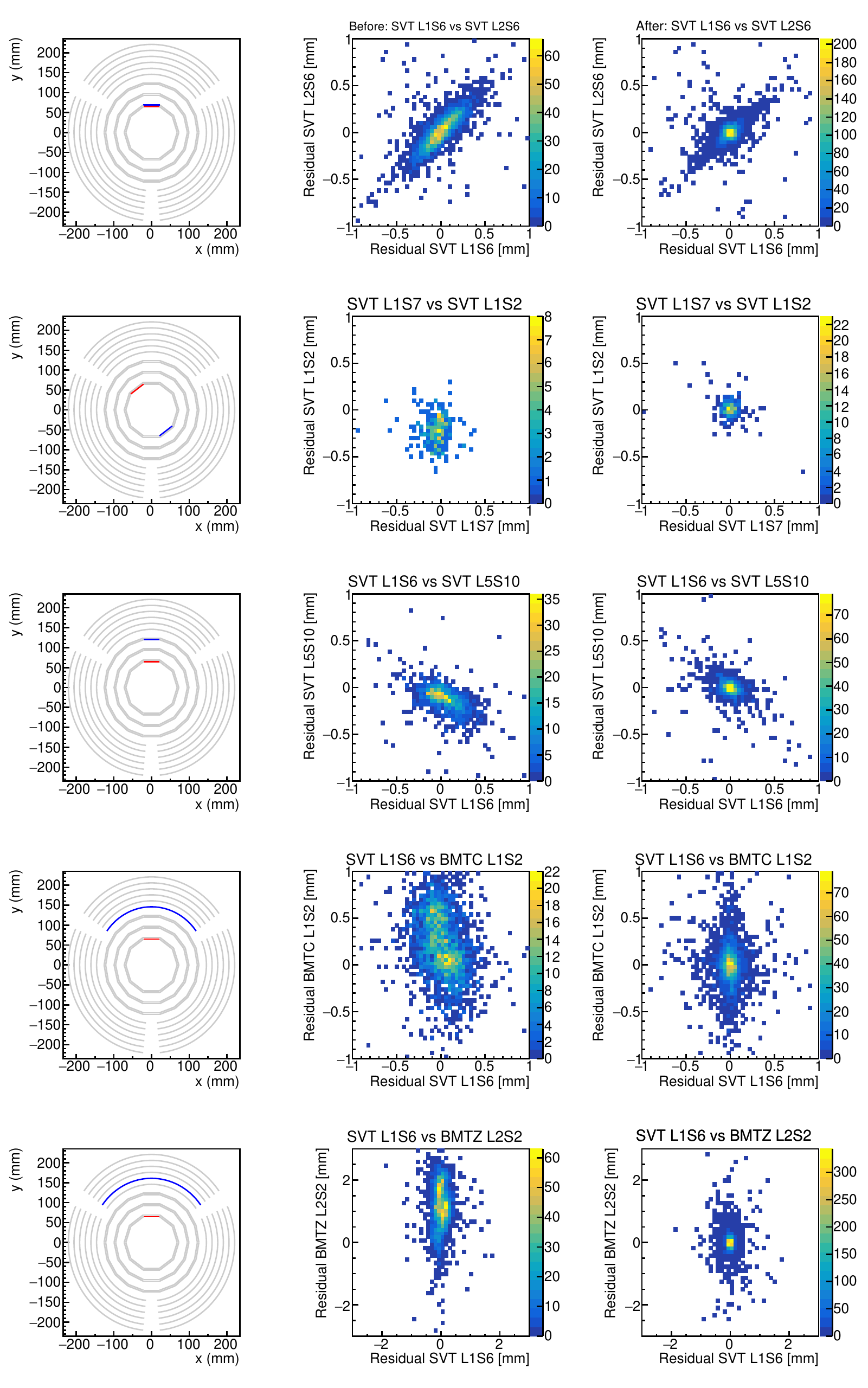}
    \end{overpic}
    \caption{Distributions of the residuals in one sensor versus those in another sensor within the same track before alignment (middle column) and after alignment (right column).  For reference, the positions of the two sensors are shown to the right of the 2D residual plots.  The combinations represent various topologies; from top to bottom, these represent: two SVT sensors in the same sector module, two SVT sensors in the same layer but azimuthally different sectors, two SVT sensors with overlapping sectors in different double-layers, an SVT sensor vs.~an overlapping BMTZ sensor, and an SVT sensor vs.~an overlapping BMTC sensor.}
    \label{fig:rvr1}
\end{figure*}

\begin{figure*}
    \centering
    \begin{overpic}[width=0.67\textwidth]{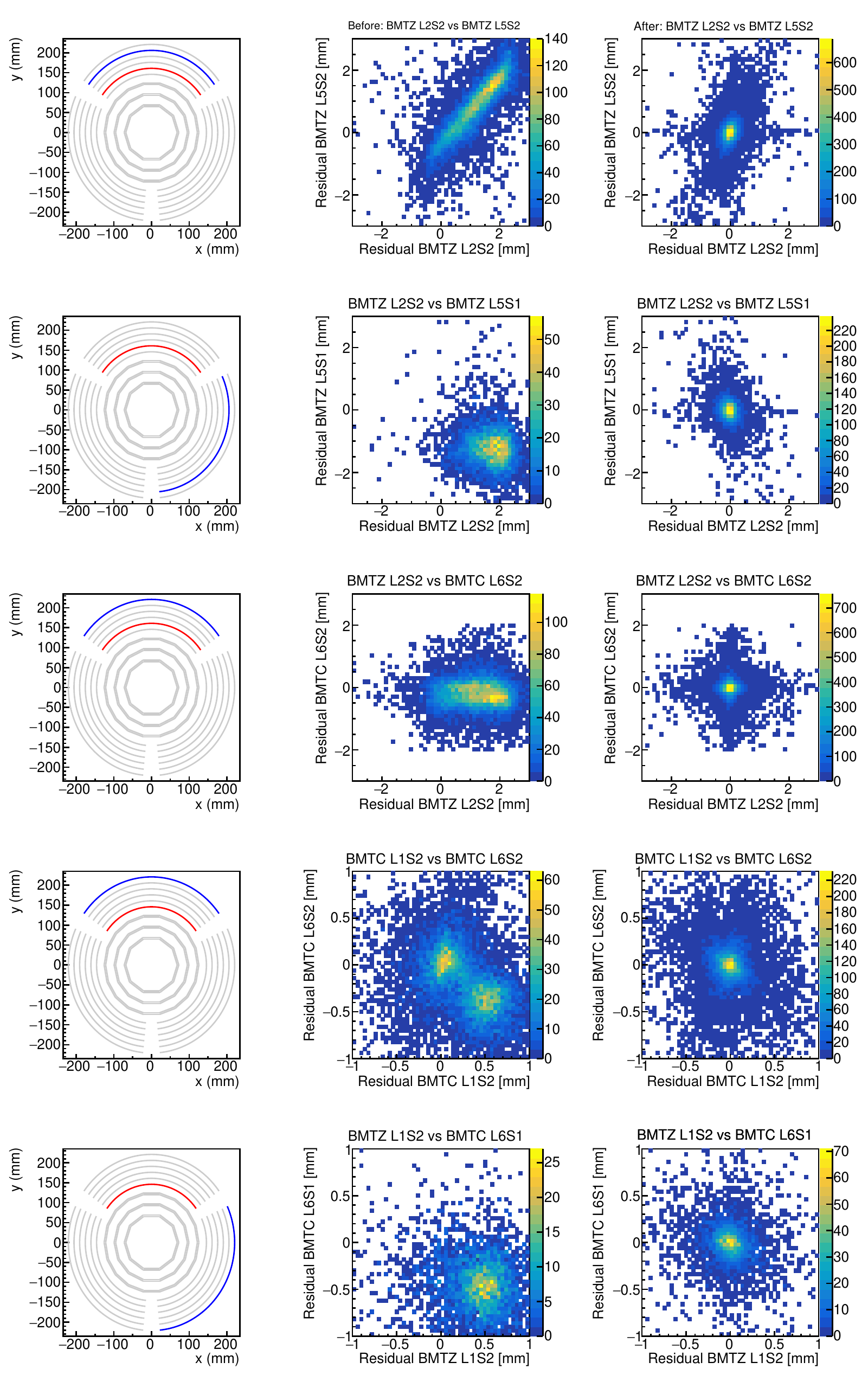}
    \end{overpic}
    \caption{Continued from Fig.~\ref{fig:rvr1}.   From top to bottom, the topologies of the combinations shown are: two BMTZ sensors in the same sector but different layers, two BMTZ sensors in differerent layers and different sectors, a BMTZ sensor and a BMTC sensor in different sectors and different layers, two BMTC sectors in different layers but the same sector, and  two BMTC sensors that are in different sectors and different layers.}
    \label{fig:rvr2}
\end{figure*}

\section{Validation through Simulations}
\label{sec:results_MC}
To validate the alignment process, we followed the procedure detailed in Sec.~\ref{sec:strategy} on MC simulations produced using the \textsc{GEMC} package \cite{Ungaro:2020xlc}, which is based on \textsc{Geant4}~\cite{GEANT4:2002zbu}.   The cosmic rays were simulated as $\sim$1~GeV muons, while the ``field-off'' tracks from the target were simulated as 0.4 to 5~GeV protons with polar angle $35^\circ<\theta<135^\circ$, and full azimuthal coverage.  

We performed three types of tests with simulations.  The first was to generate events with a misaligned geometry and to initialize the KAA with the nominal alignment parameters.  For this type of test, we only misaligned a few parameters at a time.  The second was to generate the events using the nominal alignment and to initialize the KAA using values other than the nominal ones.  The advantage of the second method is that multiple tests using different parameters could be performed for the same MC sample.  The third method is a hybrid of the first two, which included some misaligned parameters at the generator level, and non-nominal values for other parameters (we chose to use the survey values for these).  Only the results from the third type of test are included in this work, as it encapsulates the challenges from the other two tests; the other two were used only in the early stages of development of the analysis framework.  

\begin{figure*}
    \centering
    \begin{overpic}[width=\textwidth]{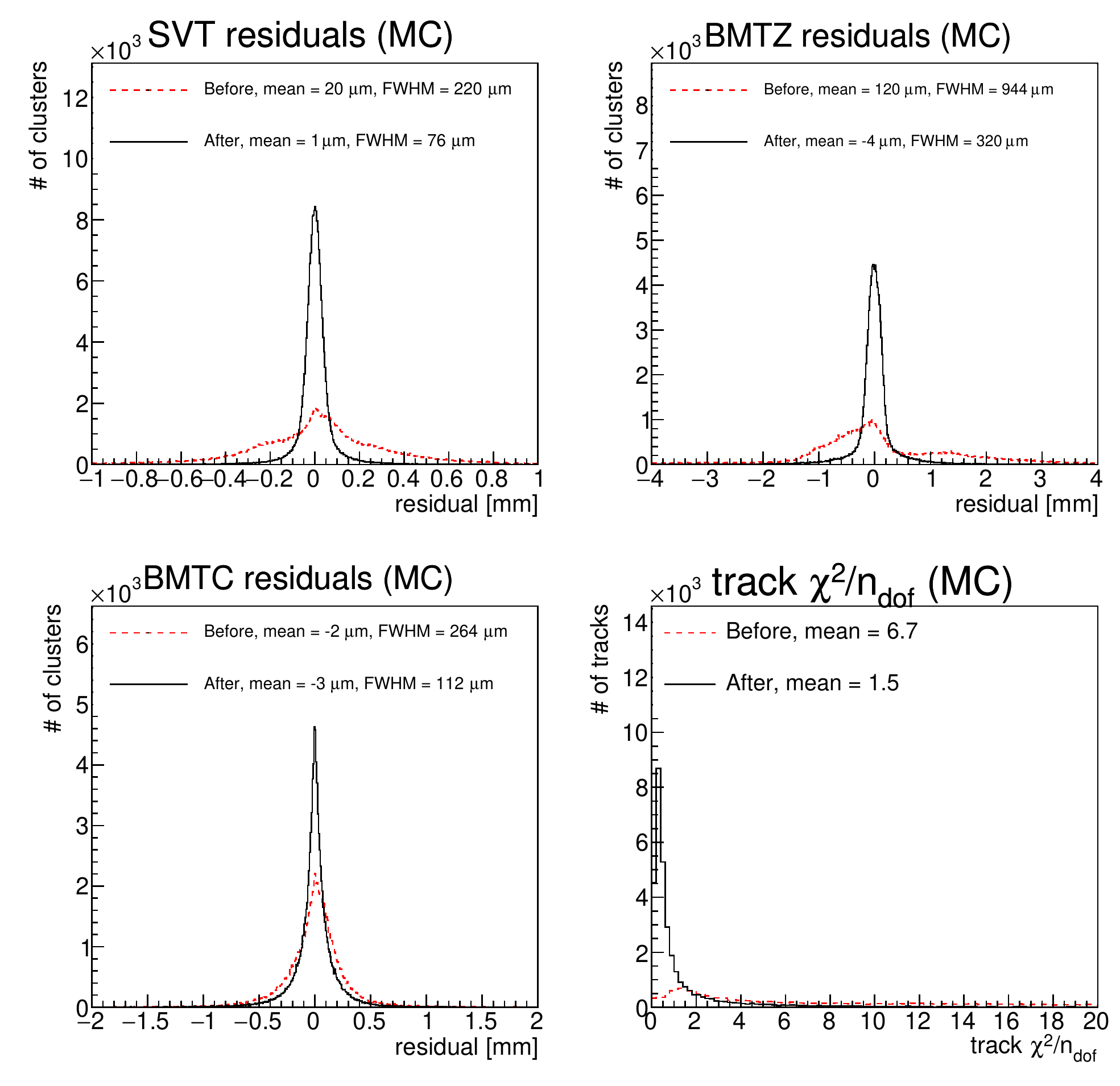}
    \put(10, 96) {\huge a)}
    \put(60, 96) {\huge b)}
    \put(10, 48) {\huge c)}
    \put(60, 48) {\huge d)}
    \end{overpic}
    \caption{Distributions of the residuals in the MC simulations before (red, dashed) and after (black, solid) alignment for the SVT (top left), BMTZ (top right) and BMTC (bottom left), and the $\chi^2$/dof distribution (bottom right) for each reconstructed track.  Each hit cluster produces a single residual in the track fit.}
    \label{fig:residuals_1D_MC}
\end{figure*}

The results with the third type of test are presented here in a similar format to Sec.~\ref{sec:results}.   The distributions of the residuals for each detector type and also the track $\chi^2$/dof are shown in Fig.~\ref{fig:residuals_1D_MC}.  The residual distributions are narrower in the simulations than in the data (see Fig.~\ref{fig:residuals_1D}), which may be attributed to a mis-modeling of the resolution effects in the simulations.  The estimates for the resolution effects in the simulations are based on an ideal version of the detectors, and can be adjusted to better match those of the real detectors.  

\begin{figure*}
    \centering
    \begin{overpic}[width=0.97\textwidth]{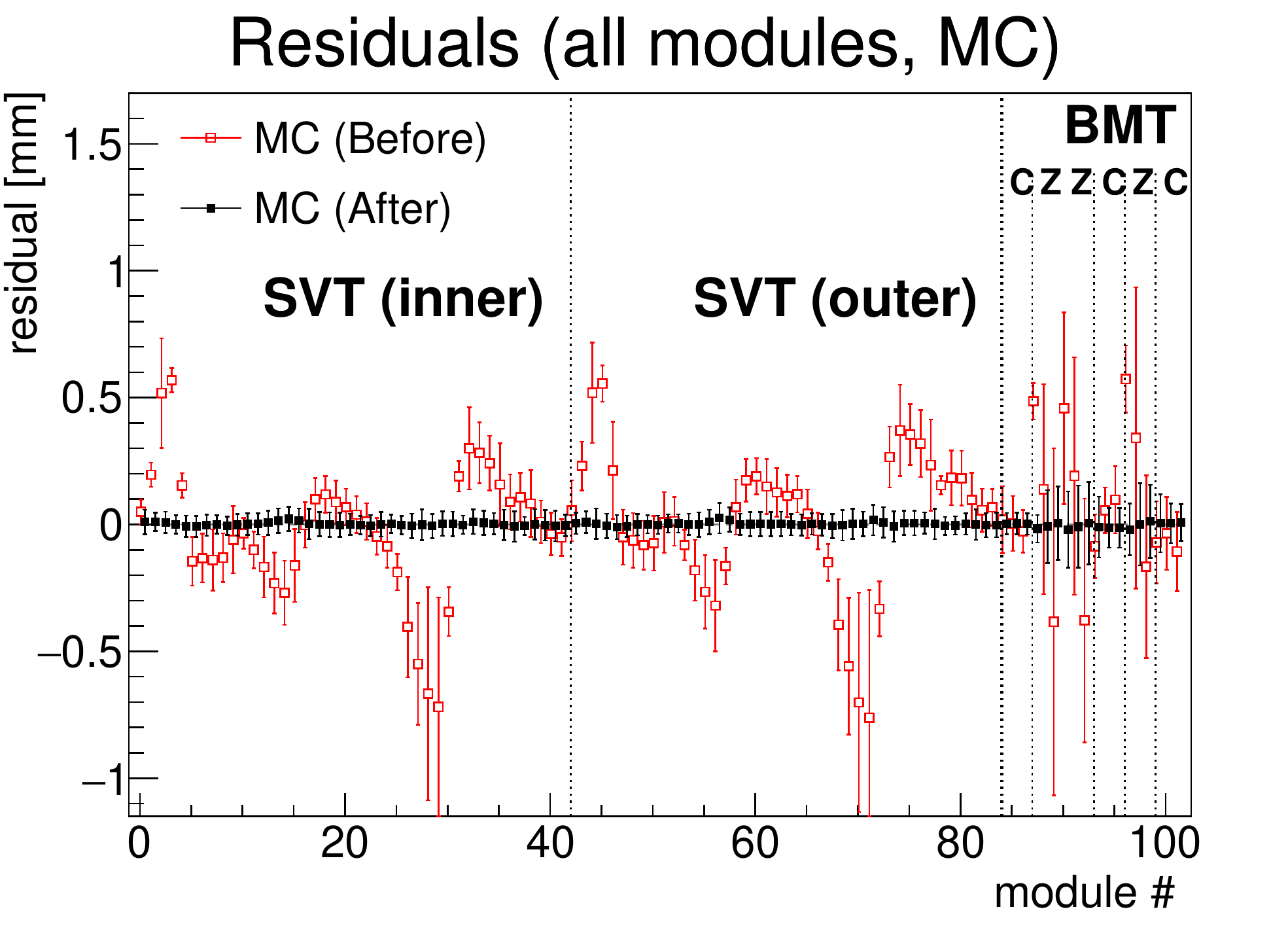}
    \end{overpic}
    \caption{Residuals in the MC simulations for the each module, before (red, open symbols) and after (black, closed symbols) alignment.  The error bars for each point represent the FWHMs of the distributions, divided by two (so that the distance from the top of the upper error to the bottom of the lower error bar is one FWHM).  Module numbers 1-84 represent SVT modules; numbers 85-102 represent BMT layers.  
    Symbols are shifted horizontally slightly for clarity.}
    \label{fig:residuals_module_MC}
\end{figure*}

\begin{figure*}
    \centering
    \begin{overpic}[width=0.83\textwidth]{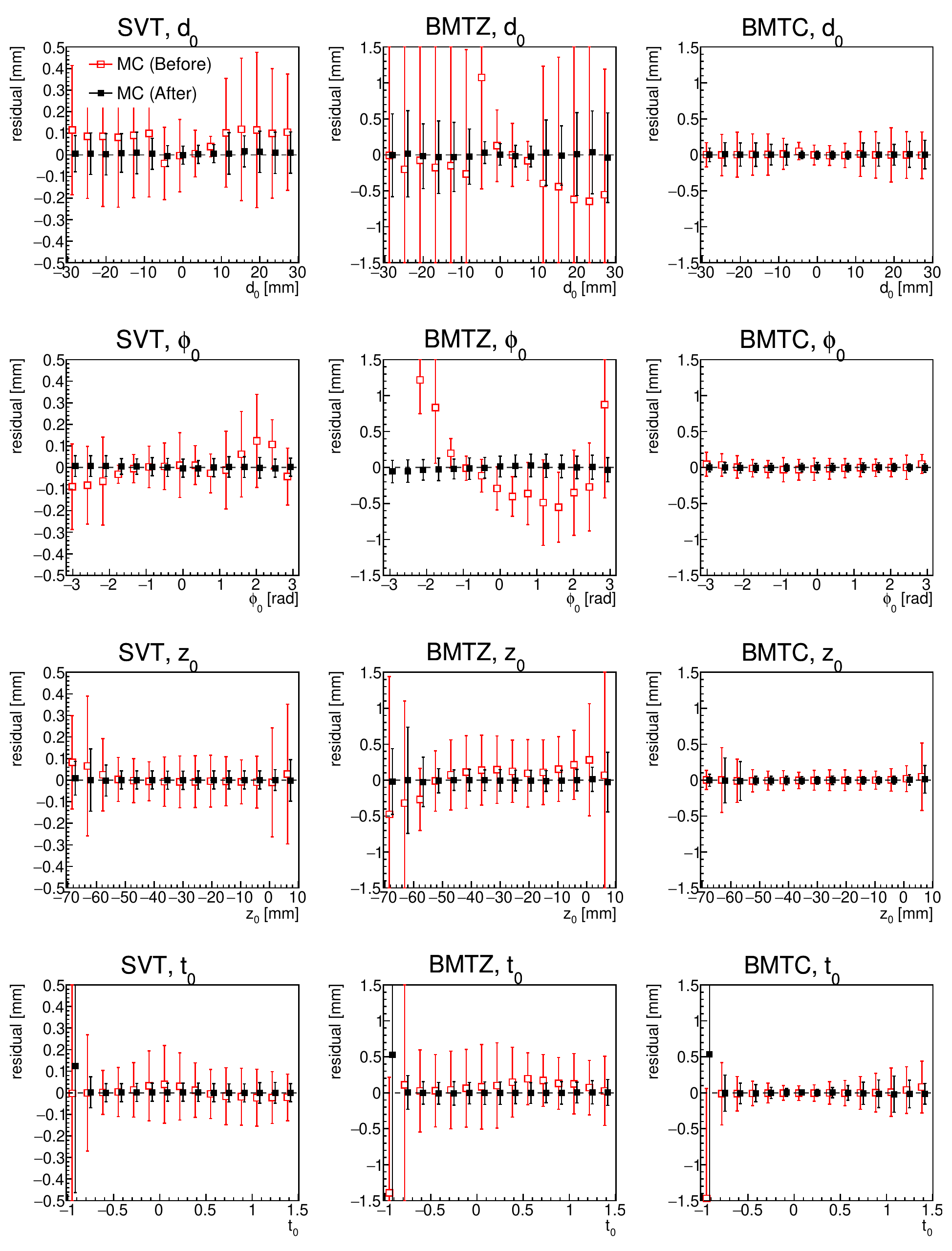}
    \end{overpic}
    \caption{Residuals in the MC simulations before (red, open symbols) and after (black, closed symbols) alignment, as a function of the kinematic variables: from top to bottom, $d_0$, $\phi_0$, $z_0$, and $\theta_0$.  The error bars for each point represent the FWHMs of the distributions, divided by two (so that the distance from the top of the upper error to the bottom of the lower error bar is one FWHM).  From left to right, the results are shown for the SVT, BMTZ, and BMTC.  Symbols are shifted horizontally slightly for clarity.  Note:  Some of the ``before'' points are missing due to being outside of the range of the plot.}
    \label{fig:residuals_vs_kinematics_MC}
\end{figure*}

We determined the mean and FWHMs of the residual distributions for each module.  These are shown in Fig.~\ref{fig:residuals_module_MC}. 
Finally, we show the residual distributions' means and FWHMs for the simulations as a function of the kinematics in Fig.~\ref{fig:residuals_vs_kinematics_MC}.  No trend is observed in the means of the distributions, however, the FWHMs in the BMTZ and SVT are considerably smaller for tracks with low $d_0$ (\textit{i.e.}, the  ``field-off'' configuration), than in tracks with large $d_0$ (\textit{i.e.}, cosmics).  The reason for this is that in the ``field-off'' configuration, the particles pass though the SVT and BMTZ detectors nearly perpendicular to the $\hat s$ direction, and therefore there is typically only a single hit in a cluster.  For the cosmic tracks, this is not necessarily the case, so there may be multiple hits in a given cluster, causing the resolution to be worse for such clusters.  

Overall, the MC simulations validate that our implementation of the algorithm works for the CLAS12 CVT. The FWHMs of the track-residual distributions are greatly reduced after the alignment (albeit to smaller values than those obtained in the data) and the average $\chi^2$ is reduced to near unity.  

\end{document}